\documentclass[%
reprint,
superscriptaddress,
longbibliography,
 amsmath,amssymb,
 aps,
prb,
citeautoscript,
]{revtex4-1}

\usepackage{dcolumn}
\usepackage[usenames,dvipsnames]{xcolor}
\usepackage{bm}
\usepackage{diagbox}

\usepackage{hyperref}
\hypersetup{ 
      colorlinks=true,
      linkcolor=blue,
      citecolor=black,
      urlcolor=blue
      }
      
\usepackage{tensor}
\usepackage{subfig}
\usepackage{braket}
\usepackage[export]{adjustbox}
\usepackage{array}
\usepackage{multirow}
\usepackage{booktabs}
\usepackage{ifthen}
\usepackage{tikz}

\usepackage{xspace}
\usepackage[english]{babel}
\usepackage{graphicx}
\usepackage[utf8]{inputenc}
\usepackage[normalem]{ulem}
\usepackage{placeins}

\usepackage{makecell}
\usepackage{nicefrac}
\newcommand{\myfont}[1]{\mbox{\textsc{#1}}}

\newcommand{\partd}[2]{\ifmmode \frac{\partial #1}{\partial #2} \else $\frac{\partial #1}{\partial #2}$\xspace\fi}
\newcommand{\partdd}[3]{\ifmmode \frac{\partial^2 #1}{\partial #2 \partial #3} \else $\frac{\partial^2 #1}{\partial #2 \partial #3}$\xspace\fi}

\newcommand{\fulld}{\mathrm{d}}
\newcommand{\totald}[2]{\ifmmode \frac{\fulld #1}{ \fulld #2} \else $\frac{\fulld #1}{ \fulld #2}$\xspace\fi}
\newcommand{\totaldd}[2]{\ifmmode \frac{\fulld^2 #1}{ \fulld #2^2} \else $\frac{\fulld^2 #1}{ \fulld #2^2}$\xspace\fi}

\newcommand{\funcd}[2]{\ifmmode \frac{\delta #1}{ \delta #2} \else $\frac{\delta #1}{ \delta #2}$\xspace\fi}
\newcommand{\funcdd}[2]{\ifmmode \frac{\delta^2 #1}{ \delta #2^2} \else $\frac{\delta^2 #1}{ \delta #2^2}$\xspace\fi}

\newcommand{\bfr}{\ifmmode  \mathbf{r}  \else $ \mathbf{r} $\xspace\fi}
\newcommand{\bfrp}{\ifmmode \mathbf{r}' \else $ \mathbf{r}' $\xspace\fi}
\newcommand{\bfR}{\ifmmode  \mathbf{R}  \else $ \mathbf{R} $\xspace\fi}
\newcommand{\bfa}[1]{\ifmmode \mathbf{a}_{#1} \else $\mathbf{a}_{#1}$\xspace\fi}
\newcommand{\bfb}{\ifmmode \mathbf{b} \else $\mathbf{b} $\xspace\fi}
\newcommand{\bfk}{\ifmmode  \mathbf{k}  \else $ \mathbf{k} $\xspace\fi}
\newcommand{\bfkp}{\ifmmode  \mathbf{k}' \else $ \mathbf{k}' $\xspace\fi}
\newcommand{\bfx}{\ifmmode \mathbf{x} \else $ \mathbf{x} $\xspace\fi}
\newcommand{\bfxp}{\ifmmode \mathbf{x}' \else $ \mathbf{x}' $\xspace\fi}
\newcommand{\bfp}{\ifmmode \mathbf{p} \else $ \mathbf{p} $\xspace\fi}
\newcommand{\bfpp}{\ifmmode \mathbf{p}' \else $ \mathbf{p}' $\xspace\fi}
\newcommand{\bfq}{\ifmmode \mathbf{q} \else $ \mathbf{q} $\xspace\fi}
\newcommand{\bfv}{\ifmmode \mathbf{v} \else $ \mathbf{v} $\xspace\fi}
\newcommand{\bfG}{\ifmmode \mathbf{G} \else $ \mathbf{G} $\xspace\fi}
\newcommand{\bGamma}{\ifmmode \boldsymbol{\Gamma} \else $ \boldsymbol{\Gamma} $\xspace\fi}
\newcommand{\btaup}{\ifmmode \boldsymbol{\tau}' \else $ \boldsymbol{\tau}' $\xspace\fi}
\newcommand{\bfO}{\ifmmode \mathbf{0} \else $ \mathbf{0} $\xspace\fi}

\newcommand{\bfe}{\ifmmode \mathbf{e} \else $\mathbf{e}$\xspace\fi}

\newcommand{\difft}{\ifmmode \fulld t \else $ \fulld t $\xspace\fi}
\newcommand{\diffr}{\ifmmode \fulld\bfr \else $ \fulld\bfr $\xspace\fi}
\newcommand{\diffrp}{\ifmmode  \fulld\bfrp \else $ \fulld\bfrp $\xspace\fi} 
\newcommand{\diffk}{\ifmmode \fulld\bfk \else $ \fulld\bfk $\xspace\fi}
\newcommand{\diffkp}{\ifmmode \fulld\bfkp \else $ \fulld\bfkp $\xspace\fi}
\newcommand{\diffx}{\ifmmode \fulld\bfx \else $ \fulld\bfx $\xspace\fi}
\newcommand{\diffp}{\ifmmode \fulld\bfp \else $ \fulld\bfp $\xspace\fi}
\newcommand{\diffpp}{\ifmmode \fulld\bfpp \else $ \fulld\bfpp $\xspace\fi}
\newcommand{\diffR}{\ifmmode \fulld\bfR \else $ \fulld\bfR $\xspace\fi}
\newcommand{\diffRN}{\ifmmode \fulld\bfR_1\cdots\fulld\bfR_N \else $ \fulld\bfR_1\cdots\fulld\bfR_N $\xspace\fi}
\newcommand{\diffP}{\ifmmode \fulld\bfP \else $ \fulld\bfP $\xspace\fi}
\newcommand{\diffPN}{\ifmmode \fulld\bfP_1\cdots\fulld\bfP_N \else $ \fulld\bfP_1\cdots\fulld\bfP_N $\xspace\fi}

\newcommand{\intBZ}[1]{\int_{\mathrm{BZ}}\!\!\!#1\,}

\newcommand{\norm}[1]{\ifmmode \| #1 \| \else $\| #1 \|$\xspace\fi}
\newcommand{\normsqd}[1]{\ifmmode \| #1 \|^2 \else $\| #1 \|^2$\xspace\fi}
\newcommand{\modulus}[1]{\ifmmode \left\vert #1 \right\vert \else $\left\vert #1 \right\vert$\xspace\fi}
\newcommand{\modulussqd}[1]{\ifmmode \left\vert #1 \right\vert^2 \else $\left\vert #1 \right\vert^2$\xspace\fi}

\newcommand{\psir}[1]{\ifmmode     \psi_{#1}(\bfr)       \else $ \psi_{#1}(\bfr)      $\xspace\fi}
\newcommand{\psirp}[1]{\ifmmode    \psi_{#1}(\bfrp)      \else $ \psi_{#1}(\bfrp)     $\xspace\fi}
\newcommand{\psiastpr}[1]{\ifmmode \psi_{#1}^\ast(\bfrp) \else $ \psi_{#1}^\ast(\bfrp) $\xspace\fi}
\newcommand{\phir}[1]{\ifmmode     \phi_{#1}(\bfr)       \else $ \phi_{#1}(\bfr)      $\xspace\fi}
\newcommand{\phirp}[1]{\ifmmode    \phi_{#1}(\bfrp)      \else $ \phi_{#1}(\bfrp)     $\xspace\fi}
\newcommand{\phiastrp}[1]{\ifmmode \phi_{#1}^\ast(\bfrp) \else $ \phi_{#1}^\ast(\bfrp) $\xspace\fi}
\newcommand{\phiastr}[1]{\ifmmode  \phi_{#1}^\ast(\bfr)  \else $ \phi_{#1}^\ast(\bfr) $\xspace\fi}
\newcommand{\phia}{\ifmmode \phi_\alpha \else $ \upphi_\alpha $\xspace\fi}
\newcommand{\phib}{\ifmmode \phi_\beta \else $ \upphi_\beta $\xspace\fi}
\newcommand{\psink}[1]{\ifmmode  \psi_{#1\bfk}(\bfr)  \else $\psi_{#1\bfk}(\bfr)$\xspace\fi}
\newcommand{\kunk}{\ifmmode \ket{u_{n\bfk}} \else $\ket{u_{n\bfk}}$\xspace\fi}
\newcommand{\kumk}{\ifmmode \ket{u_{m\bfk}} \else $\ket{u_{m\bfk}}$\xspace\fi}
\newcommand{\unk}{\ifmmode u_{n\bfk} \else $u_{n\bfk} $\xspace\fi}
\newcommand{\umk}{\ifmmode u_{m\bfk} \else $u_{m\bfk} $\xspace\fi}
\newcommand{\unkp}{\ifmmode u_{n\bfkp}(\bfr) \else $u_{n\bfkp}(\bfr)$\xspace\fi}
\newcommand{\unkt}{\ifmmode \widetilde{u}_{n\bfk}(\bfr) \else $\widetilde{u}_{n\bfk}(\bfr)$\xspace\fi}
\newcommand{\unkR}{\ifmmode u_{n\bfk}(\bfr+\bfR) \else $u_{n\bfk}(\bfr+\bfR)$\xspace\fi}
\newcommand{\wannr}[2]{\ifmmode w_{#1#2}(\bfr) \else $ w_{#1#2}(\bfr) $\xspace\fi}
\newcommand{\wann}[2]{\ifmmode w_{#1#2} \else $ w_{#1#2} $\xspace\fi}
\newcommand{\kwann}[2]{\ifmmode \Ket{w_{#1 #2}} \else $ \Ket{w_{#1 #2}} $\xspace\fi}
\newcommand{\kpsi}[2]{\ifmmode \Ket{\psi_{#1 #2}} \else $ \Ket{\psi_{#1 #2}} $\xspace\fi}

\newcommand{\omi}{\Omega_{\mathrm{I}}}
\newcommand{\omt}{\widetilde{\Omega}}
\newcommand{\omd}{\Omega_{\mathrm{D}}}

\newcommand{\operator}[1]{\ifmmode \hat{#1} \else $ \hat{#1} $\xspace\fi}
\newcommand{\adjointop}[1]{\ifmmode \operator{#1}^\dag \else $ \operator{#1}^\dag $\xspace\fi}
\newcommand{\ketbra}[2]{\ifmmode \ket{#1}\bra{#2} \else $ \ket{#1}\bra{#2} $\xspace\fi}
\newcommand{\Ketbra}[2]{\ifmmode \Ket{#1}\Bra{#2} \else $ \Ket{#1}\Bra{#2} $\xspace\fi}

\newcommand{\commutator}[2]{\ifmmode \left[#1,#2\right] \else $ \left[#1,#2\right] $\xspace\fi}

\newcommand{\vecspace}[1]{\ifmmode \mathcal{#1} \else $ \mathcal{#1} $\xspace\fi}
\newcommand{\cspace}{\ifmmode \mathbb{C} \else $ \mathbb{C} $\xspace\fi}
\newcommand{\realspace}{\ifmmode \mathbb{R} \else $ \mathbb{R} $\xspace\fi}

\newcommand{\elm}[3]{\ifmmode \braket{#1 \vert #2 \vert #3} \else $ \braket{ #1 \vert #2 \vert #3 } $\xspace\fi }
\newcommand{\Elm}[3]{\ifmmode \Braket{#1 \vert #2 \vert #3} \else $ \Braket{ #1 \vert #2 \vert #3 } $\xspace\fi }
\newcommand{\inprod}[2]{\ifmmode \Braket{ #1 | #2 } \else $ \Braket{ #1 \vert #2 } $\xspace\fi} 
\newcommand{\overlap}[2]{\ifmmode \inprod{\phir{#1}}{\phir{#2}} \else $\inprod{\phir{#1}}{\phir{#2}} $\xspace\fi} 

\newcommand{\sub}[1]{\ifmmode _{\mathrm{#1}} \else $_{\mathrm{#1}} $\xspace\fi}
\newcommand{\super}[1]{\ifmmode ^{\mathrm{#1}} \else $ ^{\mathrm{#1}} $\xspace\fi}
\newcommand{\mrm}[1]{\ifmmode   \mathrm{#1} \else $ \mathrm{#1} $\xspace\fi}
\newcommand{\tinysub}[1]{\ifmmode _{\mbox{\scriptsize{#1}}} \else $ _{\mbox{\scriptsize{#1}}} $\xspace\fi}
\newcommand{\tinysup}[1]{\ifmmode ^{\mbox{\scriptsize{#1}}} \else $ ^{\mbox{\scriptsize{#1}}} $\xspace\fi}

\newcommand{\orthomat}[1]{\ifmmode \mathbf{#1}^{\perp} \else $ \mathbf{#1}^{\perp} $\xspace\fi}
\newcommand{\trans}{\ifmmode \mathsf{T} \else $ \mathsf{T} $\xspace\fi}
\newcommand{\transpose}[1]{\ifmmode {#1}^\trans \else $ {#1}^\trans $\xspace\fi}
\newcommand{\mmn}{\ifmmode M_{mn}^{(\mathbf{k,b})} \else  $ M_{mn}^{(\mathbf{k,b})} $\xspace\fi}
\newcommand{\mnn}{\ifmmode M_{nn}^{(\mathbf{k,b})} \else  $ M_{nn}^{(\mathbf{k,b})} $\xspace\fi}
\newcommand{\amn}{\ifmmode A_{mn}^{(\mathbf{k})} \else $ A_{mn}^{(\mathbf{k})} $\xspace\fi}
\newcommand{\umn}{\ifmmode U_{mn}^{(\mathbf{k})} \else $ U_{mn}^{(\mathbf{k})} $\xspace\fi}
\newcommand{\zmn}{\ifmmode Z_{mn} \else $ Z_{mn} $\xspace\fi}
\newcommand{\matUk}{\ifmmode \mathbf{U}^{(\bfk)} \else $ \mathbf{U}^{(\bfk)} $\xspace\fi}
\newcommand{\matMbk}{\ifmmode \mathbf{M}^{(\bfk,\bfb)} \else $ \mathbf{M}^{(\bfk,\bfb)} $\xspace\fi}

\newcommand{\trace}{\ifmmode \mathrm{tr} \else $ \mathrm{tr} $\xspace\fi}
\newcommand{\Trace}{\ifmmode \mathrm{Tr} \else $ \mathrm{Tr} $\xspace\fi}

\newcommand{\ON}[1]{\ifmmode \mathcal{O}\left(N^{#1}\right) \else $\mathcal{O}\left(N^{#1}\right)$\xspace\fi}

\newcommand{\overbar}[1]{\mkern1.5mu\overline{\mkern-1.5mu#1\mkern-1.5mu}\mkern 1.5mu}

\newcommand{\averagern}{\ifmmode \overbar{\bfr}_n \else $ \overbar{\bfr}_n $\xspace\fi}
\newcommand{\eg}{e.g.}
\newcommand{\ie}{i.e.}

\newcommand{\etal}{\textit{et al.}}

\newcommand{\Wannier}{\myfont{Wannier90}}
\newcommand{\QE}{\myfont{Quantum ESPRESSO}}

\newcommand{\pwtowannier}{\myfont{pw2wannier90}}

\newcommand{\MLWFs}{\mbox{MLWFs}}

\newcommand{\abinitio}{\mbox{\textit{ab initio}}}

\newcommand{\angstrom}{~\AA}

\newcommand{\simpleangstrom}{\AA}

\newcommand\corr{Corresponding author vvitale@ic.ac.uk}
\newcommand{\titlefig}[1]{{\bf#1}.}

\begin{document}

\title{Automated high-throughput Wannierisation}

\author{Valerio Vitale}
\altaffiliation{\corr}
 \affiliation{Cavendish Laboratory, Department of Physics, University of Cambridge, 19 JJ Thomson Avenue Cambridge UK}
\affiliation{Departments of Materials and Physics, and the Thomas Young Centre for Theory and Simulation of Materials, Imperial College London, London SW7 2AZ, UK} 

\author{Giovanni Pizzi}
\affiliation{Theory and Simulation of Materials (THEOS) and National Centre for Computational Design and Discovery of Novel Materials (MARVEL), \'Ecole Polytechnique F\'ed\'erale de Lausanne, Lausanne, Switzerland}

\author{Antimo Marrazzo}
\affiliation{Theory and Simulation of Materials (THEOS) and National Centre for Computational Design and Discovery of Novel Materials (MARVEL), \'Ecole Polytechnique F\'ed\'erale de Lausanne, Lausanne, Switzerland}

\author{Jonathan R. Yates}
\affiliation{Department of Materials, University of Oxford, Parks Road, Oxford OX1 3PH, UK}

\author{Nicola Marzari}
\affiliation{Theory and Simulation of Materials (THEOS) and National Centre for Computational Design and Discovery of Novel Materials (MARVEL), \'Ecole Polytechnique F\'ed\'erale de Lausanne, Lausanne, Switzerland}

\author{Arash A. Mostofi}
\affiliation{Departments of Materials and Physics, and the Thomas Young Centre for Theory and Simulation of Materials, Imperial College London, London SW7 2AZ, UK} 
\date{\today}

\begin{abstract}
Maximally-localised Wannier functions (MLWFs) are routinely used to compute from first-principles advanced materials properties that require very dense Brillouin zone integration and to build accurate tight-binding models for scale-bridging simulations. At the same time, high-throughput (HT) computational materials design is an emergent field that promises to accelerate the reliable and cost-effective design and optimisation of new materials with target properties. The use of MLWFs in HT workflows has been hampered by the fact that generating MLWFs automatically and robustly without any user intervention and for arbitrary materials is, in general, very challenging. We address this problem directly by proposing a procedure for automatically generating MLWFs for HT frameworks. Our approach is based on the selected columns of the density matrix method (SCDM)
and we present the details of its implementation in an AiiDA workflow. We apply our approach to a dataset of $200$ bulk crystalline materials that span a wide structural and chemical space. We assess the quality of our MLWFs in terms of the accuracy of the band-structure interpolation that they provide as compared to the band-structure obtained via full first-principles calculations. 
Finally, we provide a downloadable virtual machine that can be used to reproduce the results of this paper, including all first-principles and atomistic simulations as well as the computational workflows.
\newline
\newline
NOTE: In addition to the main manuscript and supplemental materials, we have added in the Materials Cloud entry a dataset with the Wannierized band structures for all 200 materials (which can be downloaded from \url{https://archive.materialscloud.org/record/file?file_id=22842ba6-5528-48d7-9005-daa8d6a32d9d&record_id=425&filename=Vitale-2020-all-bands.pdf}).
\end{abstract}

\maketitle

\section*{Introduction}\label{sec1:intro}
The combination of modern high-performance computing, robust and scalable software for first-principles electronic structure calculations, and the development of computational workflow management platforms, has the potential to accelerate the design and discovery of materials with tailored properties using first-principles high-throughput (HT) calculations\cite{Curtarolo2013,Oba_2018,Marzari2016,Mounet_2018}.

Wannier functions (WFs) play a key role in contemporary state-of-the-art first-principles electronic structure calculations. First, they provide a means by which to bridge lengthscales by enabling the transfer of information from the atomic scale (e.g., density-functional theory and many-body perturbation theory calculations) to mesoscopic scales at the level of functional nano-devices (e.g., tight-binding calculations with a first-principles-derived WF basis)\cite{Calzolari_PRB69,Gresch_PRM2}. Second, the compact WF representation provides a means by which advanced materials properties that require very fine sampling of electronic states in the Brillouin zone (BZ) may be computed at much lower computational cost, yet without any loss of accuracy, via Wannier interpolation\cite{Yates_PRB75}.

Among several variants of WFs\cite{MMYSV_RMP84}, maximally-localised Wannier functions (MLWFs), based on the minimisation of the Marzari--Vanderbilt quadratic spread functional $\Omega$, are those most employed in actual calculations in the solid state\cite{MMYSV_RMP84}.
One ingredient in the canonical minimisation procedure is the specification of a set of initial guesses for the MLWFs. These are typically trial functions localised in real-space that are specified by the user, based on their experience and chemical intuition. As shall be described in more detail later, in the case of an isolated manifold of bands, the final result for the MLWFs is almost always found to be independent of the choice of initial guess\cite{MV_PRB56}. In the case of entangled bands\cite{SMV_PRB65}, however, this tends not to be the case and the choice of initial guess strongly affects the quality of the final MLWFs, presenting a challenge to the development of a general-purpose approach to generating MLWFs automatically without user intervention.

Several approaches have been put forward to remove the necessity for user-intervention in generating MLWFs, including the iterative projection method of Mustafa \etal{}\cite{Mustafa_PRB92}, the smooth orthonormal Bloch frames of Levitt \etal{}\cite{Levitt_PRB95}, and the automated construction of pseudo-atomic orbitals rather than WFs as the local basis to represent the target space, as described by Agapito \etal{}\cite{Agapito_PRB_88,Agapito_PRB_93,Agapito_PRB97}. In addition, some {\it ad hoc} solutions have been proposed, whose range of applicability is focused onto specific classes of materials\cite{Coh_PRMat2,Zhang_JPCL_2018,Olsen_PRMat3_2019,Gresch_PRMat2_2018}.

A recently proposed algorithm by Damle \etal{}  \cite{DL_2015_SCDM,DL_2018_SIAM}, known as the selected columns of the density matrix (SCDM) method, has shown great promise in avoiding the need for user intervention in obtaining MLWFs. Based on QR factorisation with column pivoting (QRCP) of the reduced single-particle density matrix, SCDM can be used without the need for an initial guess, making the approach ideally suited for HT calculations. The method is robust, being based on standard linear-algebra routines rather than on iterative minimisation. Moreover, the authors have proposed an efficient algorithm for the QRCP factorisation that operates on a smaller and numerically more tractable matrix than the full density matrix. Finally, SCDM is parameter-free for an isolated set of composite bands, and requires only two parameters in the case of entangled bands together with the choice of the target dimensionality for the disentangled subspace (i.e., the number of MLWFs required).
We emphasize here that the SCDM method can be seen as an extension to solid-state periodic systems of the Cholesky orbitals approach of Aquilante \emph{et al.}\cite{Aquilante_2006}, that has been developed from a quantum-chemistry molecular perspective for finite systems. SCDM focuses instead on periodic systems, and it is based on a real-space grid discretisation of the wavefunctions. We discuss in more detail this equivalence in the~\nameref{sec:SCDM} section and in the Methods section of the Supplementary material.

In this article, we present a fully-automated protocol based on the SCDM algorithm for the construction of \MLWFs{}, in which the two free parameters are determined automatically (in our HT approach the dimensionality of the disentangled space is fixed by the total number of states used to generate the pseudopotentials in the DFT calculations). We have implemented the SCDM algorithm in the \pwtowannier{} interface code between the \QE{} software package\cite{giannozzi_qe_2017} and the \Wannier{} code\cite{MOSTOFI20142309}. We have used our implementation as the basis for a complete computational workflow for obtaining \MLWFs{} and electronic properties based on Wannier interpolation of the BZ, starting only from the specification of the initial crystal structure. We have implemented our workflow within the AiiDA\cite{Pizzi_AiiDA} materials informatics platform, and we used it to perform a HT study on a dataset of 200 materials. 

We anticipate here that our scheme works extremely well for our purposes, i.e. band-structure interpolation of both insulating and metallic systems with Wannier functions, but is less suitable for other applications where, for instance, a specific symmetry character of the WFs is required. It is worth mentioning that there are other approaches for constructing Wannier functions, which are based on a minimisation procedure and therefore require an initial guess~\cite{Thygesen_POWFs_PRL,Thygesen_POWFs_PRB, Damle_SIAM_2019} and which could also be automated in a similar fashion. In this work however, we focus only on the automatic generation of maximally-localised Wannier functions. We also note that there exist efficient non-Wannier-based techniques for band-structure interpolation, e.g., Shirley interpolation~\cite{Shirley_PRB,Prendergast_PRB}. Whilst these approaches have their own advantages, they do not provide the same insight afforded by a real-space, localized description of the electronic structure, which can often be very helpful for understanding and computing advanced properties.

The manuscript is organised as follows. First, we present a summary of the background theory, starting with MLWFs for isolated and entangled bands followed by the SCDM algorithm, where we focus in particular on providing a physical interpretation of the method. In the~\nameref{sec2:results} section we first provide a preliminary comparison, for a few well-known materials, between MLWFs obtained via the conventional method (i.e., with user-defined initial guesses) and those obtained from SCDM. We then proceed to show the validation of the SCDM method and our workflow for the valence bands of 81 insulating materials. We then discuss our automated protocol to determine the free parameters in the case of entangled bands and validate it on a dataset of 200 semiconducting and metallic materials. Finally, details on the implementation of the SCDM method in \pwtowannier{} and of the AiiDA workflow are presented in the~\nameref{sec6:implementation} section.

%

We summarise in this section the main concepts and notations related to maximally-localised Wannier functions that will be useful in the rest of the paper, following the notation of Ref.~[\onlinecite{MMYSV_RMP84}].

A Wannier function associated to a band $n$ can be obtained via a unitary transformation of the Bloch state \kpsi{n}{\bfk}, known as Wannier transform\cite{W_PR52}
\begin{equation}
\kwann{\bfR}{n} =\frac{V}{(2\pi)^3} \intBZ{\diffk}
\kpsi{n}{\bfk}e^{-i\bfk\cdot\bfR},
\label{eq4:wannier-home}
\end{equation}
where $V$ is the \mbox{real-space} primitive cell volume, $\bfR$ is a Bravais lattice vector, and the integral is
over the first BZ. For clarity of notation, we assume spin-degeneracy unless otherwise specified. 

The gauge freedom of the Bloch state under multiplication by a $k$-dependent phase $e^{{i\varphi_n(\bfk)}}$ results in a non-uniqueness in the definition of the Wannier function. Maximally-localised Wannier functions represent the choice of gauge in which the real-space quadratic spread of the Wannier function is minimised\cite{MV_PRB56,MMYSV_RMP84}.
In order to obtain a minimal TB basis set it is therefore beneficial to select the optimal phases that minimise the total spread, so that overlaps and Hamiltonian matrix elements between different Wannier functions decay rapidly to zero as a function of the distance between their centres.
Since the integral transformation in Eq.~(\ref{eq4:wannier-home}) is still a unitary transformation, the resulting $\{\kwann{\bfR}{n}\}$ span the same Hilbert space as the original Bloch states $\{\kpsi{n}{\bfk}\}$. Moreover, from the orthogonality of the $\kpsi{n}{\bfk}$ readily follows the orthogonality of the $\kwann{\bfR}{n}$, since unitary transformations preserve inner products. Finally,
two WFs $\kwann{\bfR}{n}$ and $\kwann{\bfR'}{n}$ transform into each other under translation by the Bravais lattice vector $\bfR-\bfR'$\cite{Blount1962}.

For an isolated set of $J$ bands describing,
\eg{}, the valence bands of a semiconductor, the most general phase choice for a Wannier transform can be written as
\begin{equation}
\kwann{\bfR}{n} =\frac{V}{(2\pi)^3} \intBZ{\diffk}\left[\;\sum_{m=1}^{J} \kpsi{m}{\bfk}\umn\;\right]e^{-i\bfk\cdot\bfR},
\label{eq6:wannier-general} 
\end{equation}
where \matUk is a unitary matrix that, at each wave vector \bfk, mixes Bloch states belonging to different bands, giving as a result a set of $J$ composite WFs.
The localisation of the WFs may be improved by choosing the unitary matrices \matUk such that $|\widetilde{\psi}_{n\bfk}\rangle=\sum_{m}\kpsi{m}{\bfk}\umn$ in Eq.~(\ref{eq6:wannier-general}) is as smooth as possible, \ie{}, analytic with respect to $\bfk$ (see, e.g., Duffin~\cite{duffin1953}). Different approaches have been put forward\cite{Stephan_PRB_62,Ku_PRL_89,Lu_PRB_70,Qian_PRB_78,Andersen_PRB_62} to generate well-localised WFs. In the Marzari--Vanderbilt (MV) approach\cite{MV_PRB56} \matUk is chosen to minimise the sum of the quadratic spreads of the WFs, given by
\begin{equation}
\Omega =\sum_{n=1}^{J}\left[\Braket{(\bfr - \averagern)^2}_n \right] =\sum_{n=1}^{J} \left[\braket{r^2}_n-\averagern^2\right],
\label{eq:wannier-omega}
\end{equation}
where $\braket{\cdot}_n \equiv \elm{\wann{n}{\bfO}}{\cdot}{\wann{n}{\bfO}}$ and $\averagern = \braket{\bfr}_n = \elm{\wann{n}{\bfO}}{\bfr}{\wann{n}{\bfO}}$ is the centre of the $n$-th Wannier function. The resulting WFs are known as maximally-localised Wannier functions (MLWFs), and are the solid-state equivalent of the Foster-Boys molecular orbitals\cite{Boys_RMP32,Foster_Boys_a_RMP32,Foster_Boys_b_RMP32} in quantum chemistry.

The total quadratic spread $\Omega$ may be separated into two positive-definite terms: $\Omega =\omi + \omt$,
where
\begin{equation}
\omi =\sum_n  \left[\braket{r^2}_n - \sum_{m\bfR}\modulussqd{\elm{\wann{m}{\bfR}}{\bfr}{\wann{n}{\bfO}}}\right]
\label{eq:omi}
\end{equation}
and
\begin{equation}
\omt = \sum_n\sum_{m\bfR\neq n\bfO} \modulussqd{\elm{\wann{\bfO}{ n}}{\bfr}{\wann{\bfR}{m}}} 
\label{eq:omt}.
\end{equation}  
It can be shown that\cite{MV_PRB56,MMYSV_RMP84} $\omi$ is gauge invariant, whereas $\omt$ depends on the particular choice of the gauge (i.e., on the choice of $\matUk$). 
For an isolated group of bands, therefore, $\omi$ is evaluated once and for all in the initial gauge and minimising the total spread $\Omega$ is equivalent to minimising only the gauge-dependent part $\omt$.

For crystalline solids with translational symmetry, it is natural to work in reciprocal space, henceforth referred as $k$-space. Applying Blount's identities \cite{Blount1962} for the representation of the position operator $\bfr$ and $r^2$ in $k$-space and discretising in $\bfk$ (on a uniform grid) gives\cite{MV_PRB56} 
\begin{equation}
\omi = \frac{1}{N_{\bfk}} \sum_{\bfk,\bfb}w_b\sum_{m=1}^J\left[1-\sum_{n=1}^J \modulussqd{M_{mn}^{(\bfk,\bfb)}}\right] ,
\label{eq:omi_rs1} 
\end{equation}  
and
\begin{align}
\omt & = \frac{1}{N_{\bfk}} \sum_{\bfk,\bfb} w_b \Bigg[\sum_{n=1}^J \left(- \mathrm{Im} \ln M_{nn}^{(\bfk,\bfb)} - \bfb\cdot\averagern\right)^2 \notag \\ 
 &\phantom{=} + \sum_{m\neq n} \modulussqd{M_{mn}^{(\bfk,\bfb)}}\Bigg]
\label{eq:omt_rs},
\end{align}
where the vectors $\{\bfb\}$ connect a BZ mesh point $\bfk$ to its nearest neighbours $\bfk + \bfb$, the associated weights $w_b$ come from the finite difference representation of the gradient operator in $k$-space (a result of the change of representation $\bfr \rightarrow i/\hbar\nabla_\bfk$), and $\matMbk$ is given by 
\begin{equation}
M_{mn}^{(\bfk,\bfb)} \;=\; \inprod{u_{m,\bfk}}{u_{n,\bfk+\bfb}}.
\label{eq:wannier-M}
\end{equation}
Since the gradient of $\Omega$ with respect to the $U_{mn}^{(\bfk,\bfb)}$ degrees of freedom can be expressed analytically as function of the $M_{mn}^{(\bfk,\bfb)}$, the minimisation of the spread functional may be obtained, for instance, by steepest-descent or conjugate-gradient methods (see Refs.~[\onlinecite{MV_PRB56,MMYSV_RMP84}]).

Interestingly, even though the global minimisation of $\Omega$ fixes the gauge, a certain degree of non-uniqueness may remain for instance if the minimum is very shallow or flat as in the case of LiCl\cite{MV_PRB56}. This results in different configurations to be degenerate and therefore different solutions (usually related by a global rotation of the MLWFs) can be obtained depending on the initial guess. Moreover, MLWFs are only defined modulo a lattice vector by definition.

In many applications, the group of bands of interest are ``entangled'', i.e., are not separated by an energy gap from other bands throughout the whole Brillouin zone.

Souza, Marzari and Vanderbilt\cite{SMV_PRB65} (SMV) proposed a ``disentanglement'' strategy that involves two steps. 
In the first step, one defines an energy window that encompasses the states of interest and which contains $J\tinysup{win}_\bfk$ bands at each $\bfk$. This defines a local Hilbert space $\mathcal{F}(\bfk)$ at each $k$-point, which is spanned by the $J\tinysup{win}_\bfk$ states.
Then, for a given number $J\leq\min_\bfk J\tinysup{win}_\bfk$ of target Wannier functions, one finds the optimal set of $J$-dimensional subspaces $\{\mathcal{S}(\bfk)\}$, with $\mathcal{S}(\bfk)\subseteq \mathcal{F}(\bfk)$, that have maximum intrinsic smoothness over the BZ, where the intrinsic smoothness of the Hilbert space is measured by $\omi$. 
Heuristically, $\omi$ represents the ``change of character'' of the states across the Brillouin zone. (For a rigorous derivation see Ref.~[\onlinecite{MV_PRB56}].)
The subspaces $\mathcal S(\bfk)$ are defined as the span of $\{\ket{u_{n\bfk}\tinysup{opt}}\}$, which are obtained via a unitary transformation on the $\kunk$ that span $\mathcal F(\bfk)$:
\begin{equation}
    \label{eq:U-dis}
    \ket{u_{n\bfk}\tinysup{opt}} \;=\; \sum_{m=1}^{J\tinysup{win}_{\bfk}}\kumk U_{mn}\tinysup{dis(\bfk)}, \qquad n=1,\ldots,J.
\end{equation}
Note that here the $\mathbf{U}\tinysup{dis(\bfk)}$ are rectangular \mbox{$J\tinysup{win}_{\bfk} \times J$} matrices, and are unitary in the sense that \mbox{$({\mathbf{U}}\tinysup{dis(\bfk)})^\dagger {\mathbf{U}}\tinysup{dis(\bfk)}=\boldsymbol{1}_{J}$} (with $\boldsymbol{1}_J$ being the $J\times J$ identity matrix), ensuring that $\{\ket{u_{n\bfk}\tinysup{opt}}\}$ form an orthonormal set. Maximum intrinsic smoothness is achieved by choosing $\mathbf{U}\tinysup{dis(\bfk)}$ to minimise $\omi$, which, as discussed earlier, is a measure of the ``spillage'' between neighbouring subspaces $\mathcal{S}(\bfk)$\cite{SMV_PRB65}. 

In the second step, having defined a $J$-dimensional subspace $\ket{u_{n\bfk}\tinysup{opt}}$ at each $\bfk$, one proceeds by minimising $\omt$ following the same recipe described in the previous section for the case of an isolated manifold of bands. Further details on the disentanglement procedure can be found in Refs.~[\onlinecite{MMYSV_RMP84}] and [\onlinecite{SMV_PRB65}].

The iterative minimisation of $\omi$ starts with an initial guess for the subspaces $\mathcal{S}(\bfk)$. 
However, the spread functional is non-convex and the minimisation may get trapped in a local minimum, often resulting in complex-valued WFs\cite{MV_PRB56} (in the absence of spin-orbit coupling, the WFs at the global spread minimum are expected to be real\cite{Panati2013}). 
For gradient-based minimisation methods, thus, the ability to reach the global minimum strongly depends on 
the choice of an appropriate starting point, sufficiently close to the final solution.
To this aim, if one has a chemical intuition of the target $J$ Wannier functions, an initial guess of $J$ trial localised functions $g_n(\bfr)$ can be defined. 
These are then projected at every $\bfk$ onto the 
$J\tinysup{win}_{\bfk}$ Bloch states inside the target energy window (for isolated bands, $J\tinysup{win}_{\bfk}=J,\quad\forall~\bfk$), yielding:
\begin{equation}
\ket{\phi_{n\bfk}}=
\sum_m^{J\tinysup{win}_{\bfk}}
\ket{\psi_{m\bfk}}\inprod{\psi_{m\bfk}}{g_n}\equiv\sum_m^{J\tinysup{win}_{\bfk}}\ket{\psi_{m\bfk}}A^{(\bfk)}_{mn}\label{eq:wannier-Anm},
\end{equation}
where, at every $\bfk$, $A^{(\bfk)}_{mn} = \inprod{\psi_{m\bfk}}{g_n}$ is a $J\times J$ square matrix in the case of an isolated manifold of bands and a $J\tinysup{win}_{\bfk}\times J$ rectangular matrix in the case of entangled bands. The initial unitary matrix $\mathbf{U}\tinysup{dis(\bfk)}$ can then be obtained by orthonormalising the projected guess orbitals $\ket{\phi_{n\bfk}}$ through a L\"owdin orthogonalisation of $\mathbf{A}^{(\bfk)}$:
\begin{equation}
    \label{eq:lowdin}
\mathbf{U}\tinysup{dis(\bfk)} = \mathbf{A}^{(\bfk)}\left({\mathbf{A}^{(\bfk)}}^\dag \mathbf{A}^{(\bfk)}\right)^{-1/2}.
\end{equation}
One possible choice, for instance, is to start from the Bloch states themselves as the projection functions ($g_n(\bfr)=\psi_{n\bfk}(\bfr)$), so that the elements of $A^{(\bfk)}$ are the (random) phases of the Bloch states that are computed by the {\it ab initio} code. In the case of isolated bands, even a poor initial choice such as this is often sufficient to reach the global minimum of the spread functional (with enough iterations of the minimisation algorithm). Conversely, in the case of entangled bands, the two-step ``disentanglement'' procedure is usually unable to reach the global minimum of the spread functional unless the initial trial orbitals are already quite close to the final solution.

This strong dependence of the SMV minimisation algorithm on the initial trial functions, and hence on the user's intuition and intervention, has been the main obstruction in the development of fully-automated workflows for generating \MLWFs{} for high-throughput applications.

\section*{Results and Discussions}\label{sec2:results}
\subsection*{The SCDM algorithm and its physical interpretation}\label{subsec2.2:scdm}
\label{sec:SCDM}
An alternative method to the SMV approach described in the \nameref{sec1:intro} has recently been proposed by Damle, Lin and Ying\cite{DL_2015_SCDM,DL_2018_SIAM} in the form of the aforementioned selected columns of the density matrix (SCDM) algorithm. The method uses a QR factorisation with column pivoting (QRCP)\cite{golub1996matrix} of the single-particle density matrix (DM),
\begin{equation}
	P_\bfk=\sum_{n=1}^{J}\ket{\psi_{n\bfk}}\bra{\psi_{n\bfk}},
\label{eqn:P-definition}	
\end{equation}
to fix the gauge freedom in a single step, without the need for an iterative minimisation algorithm. 
In this section, we outline the core concepts of the SCDM method, focusing mainly on the aspects needed to provide a physical interpretation and facilitate its understanding. We refer to the original publications\cite{DL_2015_SCDM,DL_2018_SIAM} for additional details.

For clarity, we start by considering a system sampled at a single $k$-point, e.g. $\Gamma$, and so we drop the index $\bfk$ from the DM and other quantities; the extension to multiple $k$-points is given in the next subsection.
We start by considering systems with a finite band-gap between the $J$ valence bands and the conduction bands, \eg{}, insulators and semiconductors.

Let us first recall that $P=\sum_{n=1}^{J}\ket{\psi_{n}}\bra{\psi_{n}}$ is gauge-invariant and it is a projector on the space $\mathcal S$ spanned by the $J$ valence wavefunctions $\{\ket{\psi_{n}}\}$.
Moreover, in the insulating case, the real-space representation \mbox{$P(\bfr,\bfrp)\equiv \elm{\bfr}{P}{\bfrp}$} of the DM decays exponentially with the distance between two points $\bfr$ and $\bfrp$: \mbox{$P(\bfr,\bfrp) \sim e^{-\gamma\modulus{\bfr-\bfrp}}$}. This is the well-known near-sightedness principle\cite{desCloizeaux_PR135,Prodan_PNAS_2005,Benzi_SIAM_2013}. 
In particular, this means that for a given fixed $\bfr' = \bfr_0$, the function
\begin{equation}
	\varphi_{\bfr_0}(\bfr)\equiv P(\bfr, \bfr'=\bfr_0) = \int \fulld{\bfrp} P(\bfr,\bfrp)\delta(\bfrp-\bfr_0)
	\label{eqn:delta-function-continuum}
\end{equation}
	represents the projection on the subspace $\mathcal S$ of a delta function centred at $\bfr_0$,
and that this projection is an exponentially-localised orbital.

To understand the numerical implementation of the method, we consider from now on the real-space discretised version of the DM.
The $J$ valence wavefunctions (or, in the case of periodic systems, the periodic part $u_{n\bfk}(\bfr)$ of the $J$ valence Bloch states) can be stored on a grid of $n_G$ points in real space $\bfr_1, \bfr_2, \ldots, \bfr_{n_G}$.
We can then define the following $n_G\times J$ matrix $\Psi$ that contains the values of the $J$ wavefunctions on the grid points:
\begin{equation}
{\Psi} = \begin{pmatrix} \psi_{1}(\bfr_1) & \dots & \psi_{J}(\bfr_1) \\ 
 \vdots & \ddots & \vdots \\
 \psi_{1}(\bfr_{n_G}) & \dots & \psi_{J}(\bfr_{n_G}) \end{pmatrix}.\label{eq:psi_on_grid}
\end{equation}

With this definition, the orthonormality condition is written as $\Psi^\dagger\Psi=1_J$, while the density matrix (which in discretised form is an $n_G\times n_G$ matrix) can be written as $ P = \Psi{\Psi}^\dag$, i.e., $P_{ij}=\sum_{n=1}^J \psi_n(\bfr_i)\psi_n^*(\bfr_j)$. 

We can now interpret the $j$-th column $\mathcal C^j$ of the DM, $\mathcal C^j_i\equiv P_{ij}$, as the projection on the valence subspace $\mathcal S$ of a test orbital $\phi_j$ that is zero everywhere except at the $j$-th grid position (i.e., at position $\bfr_j$).
This statement is the discretised version of the projection of a delta function in Eq.~\eqref{eqn:delta-function-continuum}, i.e., apart from normalisation, $\phi_j$ is the discretised version of $\delta(\bfr - \bfr_j)$. Therefore, thanks to the 
near-sightedness principle, the orbitals represented by the columns of the DM are localised. 

This statement is at the core of the SCDM method. 
In fact, when searching for Wannier functions, we are looking for a complete and orthogonal basis set of $J$ localised functions that span the subspace $\mathcal S$.
In our case, the set of all columns $\mathcal C^j$ clearly spans the whole subspace $\mathcal S$ (since the $P$ operator is the projector on $\mathcal S$). However, in essentially all practical situations, $J\ll n_G$ and the set of all these $n_G$ orbitals is redundant.
In addition, these orbitals are not orthogonal---intuitively, projecting on delta functions centred at two neighbouring points will typically result in a large overlap between the projected orbitals---and not normalised (e.g., in the limiting case of a delta function centred at a position in space where there is no charge density, the resulting projection will have zero norm).
Selecting any set of $J$ linearly-independent columns would form a basis for $\mathcal{S}$, and an initial guess for the Wannier functions could be obtained by orthonormalising these $J$ columns, \eg{}, with a L\"owdin symmetric orthogonalisation. However, if these $J$ columns are not already almost orthogonal, the orthogonalisation will be numerically unstable and, most importantly, will mix them and thereby degrade their localisation.
Therefore, the goal of the SCDM method is to select the ``most representative'' $J$ columns, \ie{}, the columns that possess the largest norm and that are as orthogonal to each other as possible, i.e. the most ``well-conditioned subset'', so that the L\"owdin orthogonalisation will mix these orbitals as little as possible (L\"owdin orthogonalisation minimises the squared difference between the original and orthogonalised functions\cite{Carlson_PR105}).
Equivalently, as every column is the projection of a delta-like test orbital centred at $\bfr_j$, we can say that the SCDM algorithm selects $J$ points, from among the original $n_G$ grid points, that define the ``most representative'' localised projected orbitals.

To achieve this goal, SCDM uses the standard linear algebra QRCP method\cite{golub1996matrix}, which factorises a matrix $P$ as $ P\Pi = QR$, where $Q$ is a matrix with orthonormal columns, $R$ is a upper-triangular matrix, and $ \Pi$ is a permutation matrix that swaps the columns of $P$ so that the diagonal elements of $R$ are in order of decreasing magnitude $|R_{11}| \geq |R_{22}| \geq \cdots \geq | R_{n_Gn_G}|$ (see Methods section of the Supplementary Material for more details).
The relevant output of the algorithm is the $\Pi$ permutation matrix, or more specifically the indexes of the first $J$ columns chosen by the algorithm: these are the ``most representative'' columns discussed above and, after orthonormalisation, they provide the best guess for the localised Wannier functions of the system. With a slight abuse of notation, in the following we will use the symbol $\Pi$ also to identify the vector of indexes of the permutation matrix, such that $\Pi(i)=j$ has the following meaning: $\Pi_{ij}=1$, 
and all of the other elements in the $j$-th column are equal to zero. 

QRCP (a greedy algorithm) selects columns as follows: since $R$ is triangular (and $Q$ has orthonormal columns), the norm of the first selected column $\mathcal C^{\Pi(1)}$ of $P$ is $|R_{11}|^2$ and must be the largest possible, therefore the algorithm will choose the column with the largest norm. The second column $\mathcal C^{\Pi(2)}$ is chosen to maximise $|R_{22}|^2$ that, due to the properties of $Q$ and $R$, is the component of $\mathcal C^{\Pi(2)}$ orthogonal to $\mathcal C^{\Pi(1)}$, as shown in the Methods section of the Supplementary Material. So, the QRCP algorithm will select as the second vector the one with the largest orthogonal component to the first, and in general will select the $k$-th vector as the one with the largest orthogonal component to the subspace spanned by the previous $(k-1)$ columns (to be more precise the actual selection process is a heuristic for trying to keep principal sub-matrices of $R$ as well-conditioned as possible). It is worth mentioning that this approach is related to the Cholesky orbitals approach of Aquilante \etal{} \cite{Aquilante_2006}, that applies to finite (non-periodic) systems and for a different basis set (a basis of atomic orbitals rather than a real-space grid discretisation). In particular, the Cholesky algorithm used in Ref.~[\onlinecite{Aquilante_2006}] is a refined version of the original Cholesky decomposition specifically adapted for positive semi-definite matrices, \ie{}, Cholesky decomposition with full column pivoting (CholCP)
\mbox{$\widetilde{\Pi}^T P \widetilde{\Pi} = L^\dag L$}, where $L$ is an upper triangular matrix and $\widetilde{\Pi}$ is a permutation matrix. In the Methods section of the Supplementary Material we demonstrate that the selection of the columns in CholCP is the same as in QRCP, at least for the first $J=\mathrm{rank}(P)$ columns, \ie, $(P\Pi)_{:,1:J} = (P\tilde{\Pi})_{:,1:J}$. This is due to well-known connections between QR factorizations and Cholesky factorizations\cite{golub1996matrix}. Finally, the two methods use undoubtedly related ideas but they are not direct analogues since there are multiple ``variants'' of SCDM when using localised orbitals.

For an effective practical implementation of the method, a final step is required. In fact, the $P$ matrix can be extremely large, since $n_G$ can be of the order of 100\,000 or more (while $J$ is often of the order of 10--100).
Therefore, applying the QRCP algorithm directly to $P$ is impractical, both for the memory required to store it ($\mathcal{O}(n_G^2)$), and for the time needed to compute the result ($\mathcal{O}(J\times n_G^2)$). Instead, using the fact that $P={\Psi}{\Psi}^\dagger$ and that the original columns of ${\Psi}$ are orthonormal, one can prove (see Methods section of the Supplementary Material) that the same permutation matrix $\Pi$ can be obtained applying the QRCP algorithm directly to the much smaller matrix $\Psi^\dagger$ (of size $J\times n_G$), with a computational cost that scales as $\mathcal{O}(J^2\times n_G)$.
Moreover, the matrix obtained from the first $J$ columns of $({\Psi}^\dagger \Pi)$ may be used as the $A_{mn}$ projection matrix of Eq.~\eqref{eq:wannier-Anm} as a starting point for the usual Wannierisation procedure in order to obtain MLWFs.

Finally, it is worth noting the connection with the ``canonical'' approach of user-defined initial guesses (e.g., atomic-like orbitals at specified centres): the SCDM method may be thought of as using as initial guesses a set of extremely localised $s$-like ``orbitals'' (actually, $\delta$ functions), whose centres (located at the points of the real-space grid) are optimally chosen by the SCDM algorithm via the QRCP factorisation.

\subsection*{SCDM for periodic systems: SCDM-k}\label{sec:scdm-k}
We now extend the discussion to the case of $k$-point sampling with more than one $k$-point (i.e., not only at $\Gamma$), still considering an isolated manifold (e.g., the valence bands). The DM $P_\bfk=\sum_{n}\ket{\psi_{n\bfk}}\bra{\psi_{n\bfk}}$ is an analytic function of $\bfk$\cite{Nenciu_RMP_63,Panati2013}, and it is also proven that WFs with an exponential decay exist\cite{Brouder_PRL_98}; numerical studies for the specific case of \MLWFs{} have confirmed this claim for several materials\cite{Brouder_PRL_98,He_PRL_86}, and recently there has been a formal proof for 2D and 3D \mbox{time-reversal-invariant} insulators\cite{Panati2013}.
The SCDM method has been extended also to the case of $k$-sampling\cite{DL_2018_SIAM} and named in this case ``SCDM-$k$''.
In summary, the goal is now to select a common set of columns for all the $k$-dependent density matrices $P_{\bfk}$. Ref.~[\onlinecite{DL_2018_SIAM}] discusses extensively how the method can be extended to a $k$-point sampling with more than one $k$ point and it shows detailed results of the convergence as a function of the number of $k$ points used in the column-selection algorithm. The final conclusion of the authors is that it is typically sufficient to select the columns using a single ``anchor'' $k$ point (typically chosen to be $\Gamma$), i.e., it is sufficient to compute the permutation matrix $\Pi$ using a QRCP on $P_{\bfk=\Gamma}$ only. Then, this selection of columns can be used for all other $k$-points. 
\subsection*{Extension to entangled bands}\label{sssec:scdm_entangled}
Finally, the extension to the entangled case (e.g., for metals or when considering also the conduction bands of insulators and semiconductors) has been proposed in Ref.~[\onlinecite{DL_2018_SIAM}].
In this case, a so-called quasi-density matrix is defined,
\begin{equation}
P_{\bfk} = \sum_n \ket{\psi_{n\bfk}}f(\epsilon_{n\bfk})\bra{\psi_{n\bfk}},\label{eq:gen_Pk}
\end{equation}
where $f(\epsilon_{n\bfk})$ is an occupancy function. The isolated-bands case can be recovered by setting $f(\epsilon_{n\bfk})=1$ for energy values $\epsilon_{n\bfk}$ within the energy range of the isolated bands, and zero elsewhere. 
For the typical cases of interest of this work (metals, and valence bands and low-energy conduction bands in semiconductors and insulators), one needs bands up to a given energy (typically slightly above the Fermi energy). Then, as suggested in Ref.~[\onlinecite{DL_2018_SIAM}], $f(\epsilon)$ can be chosen as the complementary error function:
\begin{equation}
	f(\epsilon)=\frac{1}{2}\mathrm{erfc}\left(\frac{\epsilon - \mu}{\sigma}\right).\label{eq:erfc-smearing}
\end{equation}	
This function depends on two free parameters $\mu$ and $\sigma$, whose choice is critical to tune the algorithm and obtain a set of Wannier functions that correctly interpolate the low-energy electronic bands of a given material. In the~\nameref{sec5:entangled} section we describe our protocol to choose the values of $\mu$ and $\sigma$ based on the electronic structure of the material, allowing us to implement a fully automated workflow to construct its Wannier functions via the SCDM method.

The algorithm then proceeds as in the case for isolated bands, computing the QRCP factorisation on the quasi-density-matrix or, in practice, on the matrix ${F}_{\bfk}\Psi_{\bfk}^\dagger$ at the $\bfk=\Gamma$ anchor point, with ${F}_{\bfk}$ a diagonal matrix with matrix elements $\{f(\epsilon_{1,\bfk}), \ldots, f(\epsilon_{J\tinysup{win}_\bfk,\bfk})\}$. This approach, therefore, constitutes an alternative to the SMV disentanglement procedure described in the~\nameref{sec1:intro} section: matrices obtained from the first $J$ selected columns of ${F}_{\bfk}\Psi_{\bfk}^\dagger$ at each $\bfk$ form the projection matrices $\mathbf{A}^{(\bfk)}$, and the $\mathbf{U}\tinysup{dis(\bfk)}$ matrices of Eq.~\eqref{eq:U-dis} are obtained using the L\"owdin transformation of Eq.~\eqref{eq:lowdin}. 

\subsection*{SCDM and \MLWFs{}}
The SCDM algorithm is able to robustly generate well-localised functions that are used to generate Wannier functions without the need for an initial guess. Whilst this makes the algorithm well-suited for direct integration within HT frameworks, the selection of the columns cannot be controlled by external parameters (at least for isolated bands), and therefore it is not possible to enforce constraints that might be desirable, such as point symmetries. On the contrary, when explicitly specifying atomic-like initial projections, these (if appopriately chosen) provide at least some degree of chemical and symmetry information. In the~\nameref{sec3:scdm_vs_mlwfs} section we discuss how this affects the WFs obtained by the algorithm.
Our aim is to leverage on the ability of SCDM to automatically generate a good set of localised functions, and to use these to seed the MV algorithm for the minimisation of the total spread functional, which will give in turn an automated protocol to generate \MLWFs{}. Being able to automatically generate \MLWFs{} will also allow users to seamlessly exploit the set of computational tools that have been developed in recent years for \MLWFs{} and implemented in various codes, such as \Wannier.
 In practice, this entails employing the SCDM algorithm to compute the $\mathbf{A}^{(\bfk)}$ matrices of Eq.~\ref{eq:wannier-Anm} as follows:
\begin{equation}
A_{mn}^{(\bfk)} = f(\varepsilon_{m\bfk})\psi^*_{m\bfk}(\bfr_n),
\end{equation}
where the $J$ points $\bfr_n$ are obtained from the first $J$ columns of the permutation matrix $\Pi$, computed at $\Gamma$, \ie{}, $\mathbf{A}^{(\bfk)}=F_\mathbf{k}\Psi^\dag_\bfk\Pi_\Gamma(J)$, with $\Pi_\Gamma(J)$ representing the reduced matrix formed by the first $J$ columns of $\Pi_\Gamma$.

\subsection*{SCDM and ``disentanglement''}
It is worth noting that the SCDM method can be also combined with the SMV disentanglement procedure, as a means of seeding the initial subspace projection. However, this introduces two additional parameters associated with the SMV approach, namely $\varepsilon\tinysub{outer}$, and $\varepsilon\tinysub{inner}$, giving a total of four  parameters (together with $\mu$ and $\sigma$). $\varepsilon\tinysub{outer}$ defines the upper limit of the so-called ``outer'' energy window discussed in the~\nameref{sec1:intro} section, and
$\varepsilon\tinysub{inner}$ defines the upper limit of a smaller energy window contained within the outer energy window. This inner window is used to ``freeze'' the Bloch states within during the minimisation of $\omi$, such that they are fully preserved within the selected subspaces $\{\mathcal{S}(\bfk)\}$
(see Ref.~[\onlinecite{SMV_PRB65}] for a comprehensive description of the outer and inner energy windows). Each additional parameter makes it increasingly difficult to find a robust and automated protocol for obtaining \MLWFs. Consequently, when combining SCDM with SMV disentanglement, an optimal selection of all the parameters can be achieved only in an {\it ad hoc}, non-automatic fashion (hence only for few materials). As shown in the~\nameref{sec:SCDM} section, SCDM employs a generalised form of the density matrix Eq.~\eqref{eq:gen_Pk}, which implicitly defines an energy window via the function $f(\varepsilon)$ and selects a smooth manifold by construction. Intuitively, this suggests that SCDM can be used \textit{in lieu} of the SMV disentanglement procedure. In general, we have found that for the sole purpose of interpolating the energy bands up to a given energy, performing SMV disentanglement step on top of SCDM has at best a marginal improvement on the quality of the interpolation (see~\nameref{sec5:entangled}), and in some cases can even be detrimental due to the case-by-case sensitivity on the choice of energy windows. 
For this reason, in the~\nameref{sec5:entangled} section we focus exclusively on a protocol for the automatic selection of the free parameters in SCDM, \ie{}, $\mu$ and $\sigma$, without considering any additional SMV disentanglement.

\subsection*{SCDM vs MLWFs in well-known materials}\label{sec3:scdm_vs_mlwfs}
As a precursor to the fully-automated high-throughput study on a set of 200 materials that focuses on automatic Wannierisation and band interpolation from SCDM projections and which will be presented in the~\nameref{sec5:entangled} section, in this section we consider in greater depth and detail the performance of the SCDM method on a small set of simple systems with well-known Wannier representations of the electronic structure. Specifically, we compare quadratic spreads, centres and symmetries of the WFs computed from the SCDM gauge (as described in the~\nameref{subsec2.2:scdm} section) with the ones computed from carefully chosen initial projections. 
Comparative studies between SCDM localised functions and \MLWFs{} on well-known materials have recently appeared in the literature\cite{DL_2018_SIAM,Damle_SIAM_2019}. However, here we expand on different aspects, focusing in particular on the combination of the SCDM and the MV approaches (SCDM+\MLWFs{}), to better assess its range of applicability, for instance for beyond-DFT methods, \eg{}, {\it ab initio} tight-binding\cite{Horsfield_1999,Kaxiras_PRB92}, DFT+U\cite{Anisimov_1997,Schnell_PRB65,Novoselov_2015} and DMFT\cite{Georges_RMP68,Georges_PRB74}, where the symmetries of the Wannier functions are important.

All DFT calculations have been carried out with \QE{}, using the PBE exchange-correlation functional and Vanderbilt ultrasoft pseudopotentials\cite{US_Vanderbilt}. \MLWFs{} are generated from Bloch states calculated on a $10\times10\times10$ Monkhorst-Pack grid of $k$-points. The SCDM method has been implemented in the \pwtowannier{} code, which interfaces \QE{} with the \Wannier{} code\cite{MOSTOFI20142309,Mostofi_CPC}, as explained in~\nameref{sec6:implementation}. \Wannier{} is used throughout this work to generate the WFs on a real-space grid and to perform the interpolation of band structures in reciprocal space.

We consider four different schemes for generating Wannier functions: (1) Full minimisation of $\Omega$ using the SMV disentanglement algorithm to minimise $\omi$ and the MV algorithm to minimise $\omt$ ({\tt DIS+MLWF}); (2) Minimisation of $\omi$ only, using the SMV algorithm ({\tt DIS}); (3) Minimisation of $\omt$ only, using the MV algorithm ({\tt MLWF}); and (4) No minimisation of $\Omega$ ({\tt proj-ONLY}). In each case, the initial $J$-dimensional subspace at each $\bfk$ is determined in one of two ways, either by the SCDM method or by projection onto specific atomic-like localised orbitals (Eq.~(\ref{eq:wannier-Anm})). 

\newcolumntype{A}{>{\centering\arraybackslash}m{6em}}
\newcolumntype{C}{>{\centering\arraybackslash}m{6em}}

\begin{figure*}
\begin{adjustbox}{angle=0}
\begin{tabular}{@{}l*{4}{C}@{}}
\toprule[0pt]
 & {\tt DIS+MLWF} &  {\tt DIS} & {\tt MLWF} & {\tt proj-ONLY} \\
 \midrule[1pt]
 \makecell[c]{$sp^3$ projections \\ back-bonding case} & {\includegraphics[width=0.15\columnwidth,trim={60px 0px 60px 0px},clip]{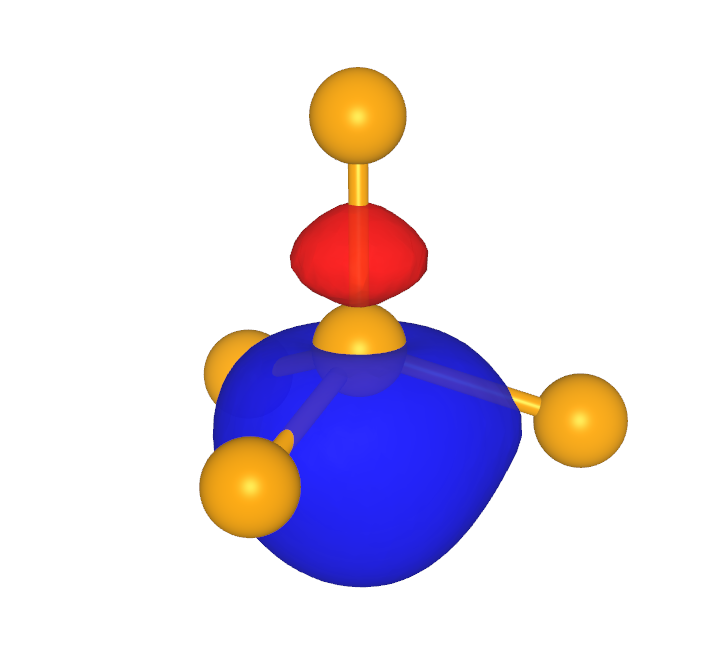}} & {\includegraphics[width=0.15\columnwidth,trim={60px 0px 60px 50px},clip]{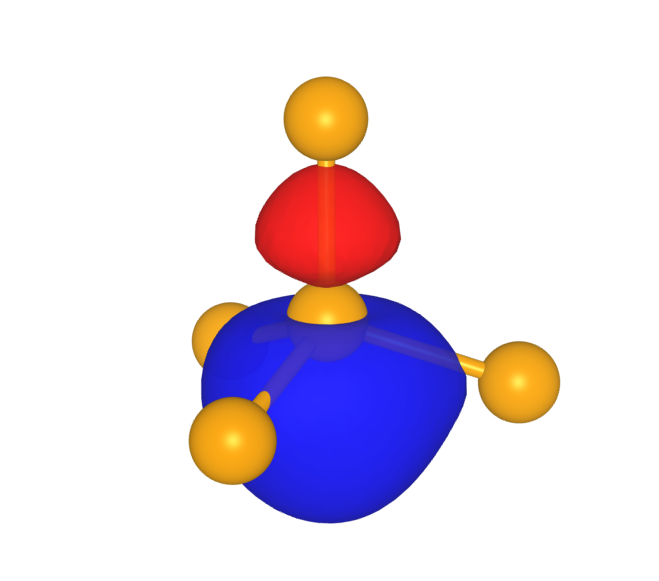}} &  {\includegraphics[width=0.15\columnwidth,trim={60px 0px 60px 50px},clip]{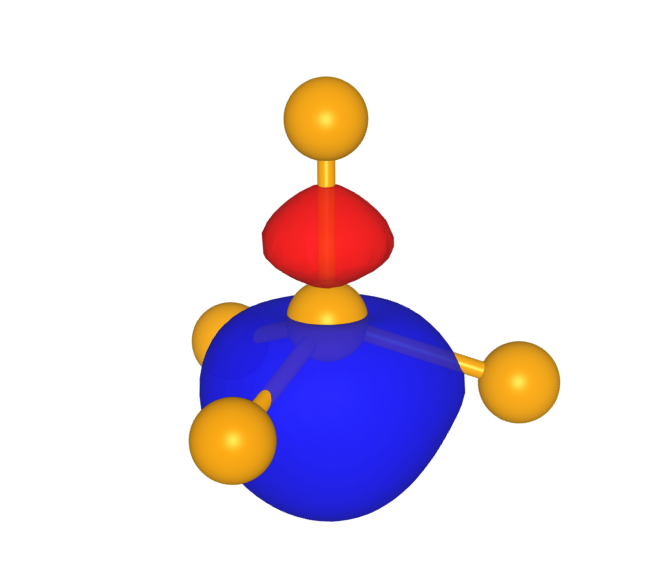}} & {\includegraphics[width=0.15\columnwidth,trim={60px 0px 60px 50px},clip]{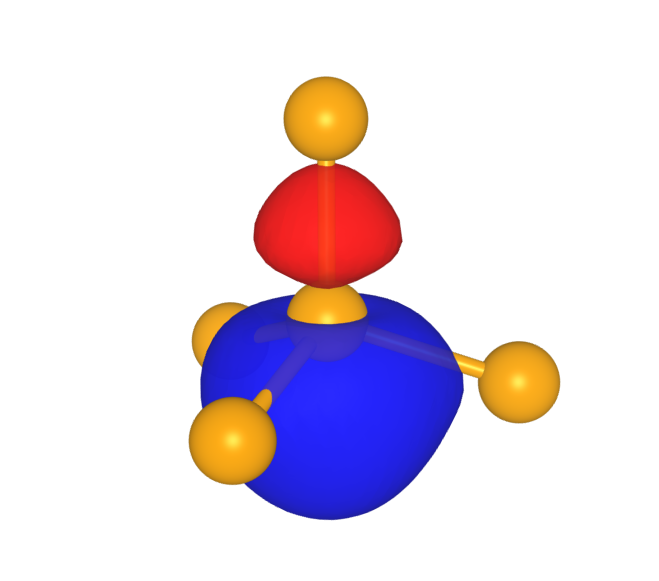}}\\
 & 2.93 & 3.09 & 4.23 & 4.37  \\
 \makecell[c]{$sp^3$ projections \\ front-bonding case} & {\includegraphics[width=0.15\columnwidth,trim={60px 0px 60px 0px},clip]{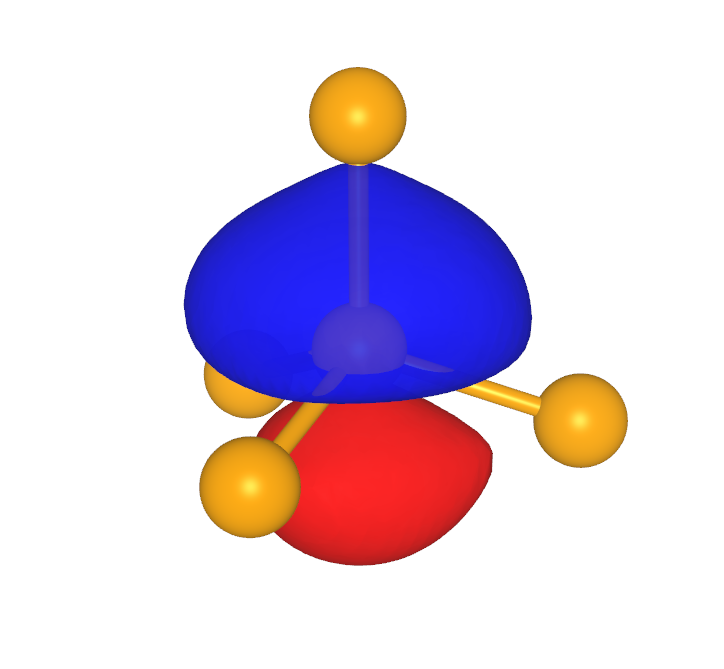}} &  {\includegraphics[width=0.15\columnwidth,trim={60px 0px 60px 50px},clip]{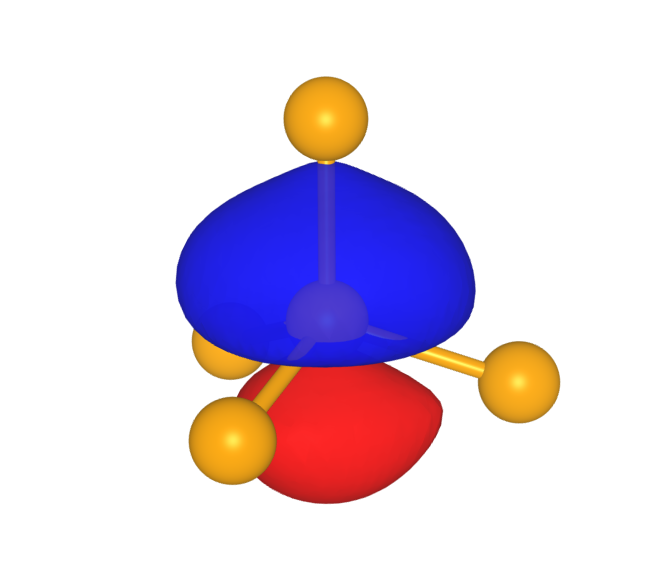}} & {\includegraphics[width=0.15\columnwidth,trim={60px 0px 60px 50px},clip]{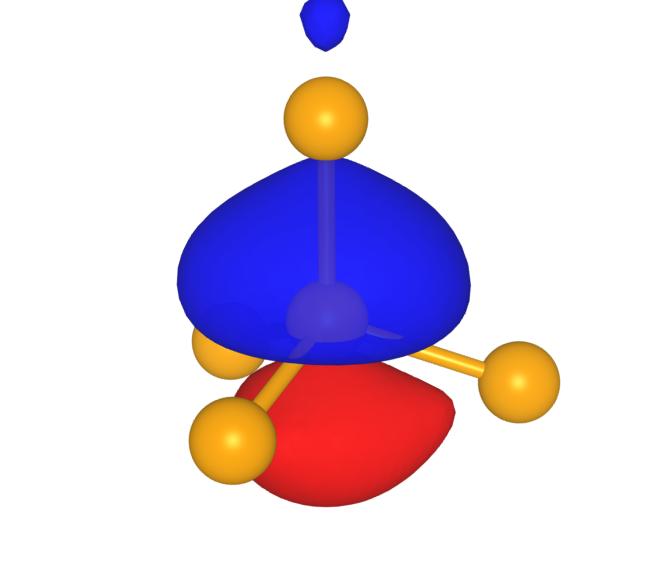}} &  {\includegraphics[width=0.15\columnwidth,trim={60px 0px 60px 50px},clip]{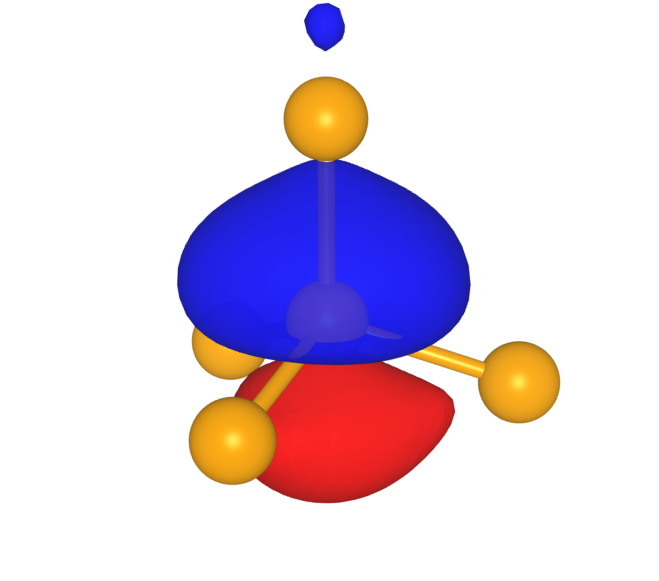}}\\
 & 3.36 & 3.46 & 4.57 & 4.66\\
 \midrule[1pt] \\
 \multirow{4}{*}{SCDM projections} & \multirow{4}{*}{
    \begin{minipage}{0.15\columnwidth}\includegraphics[width=\linewidth,trim={20px 0px 20px 0px},clip]{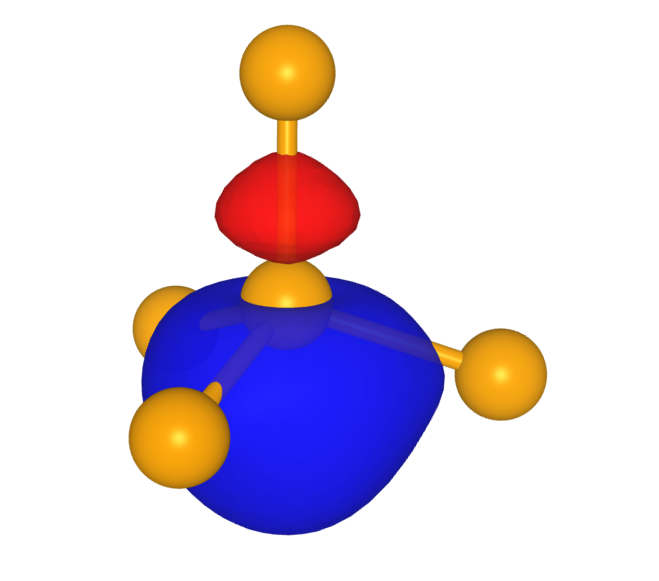}\\2.93\end{minipage}} 
        & {\includegraphics[width=0.15\columnwidth,trim={20px 0px 20px 0px},clip]{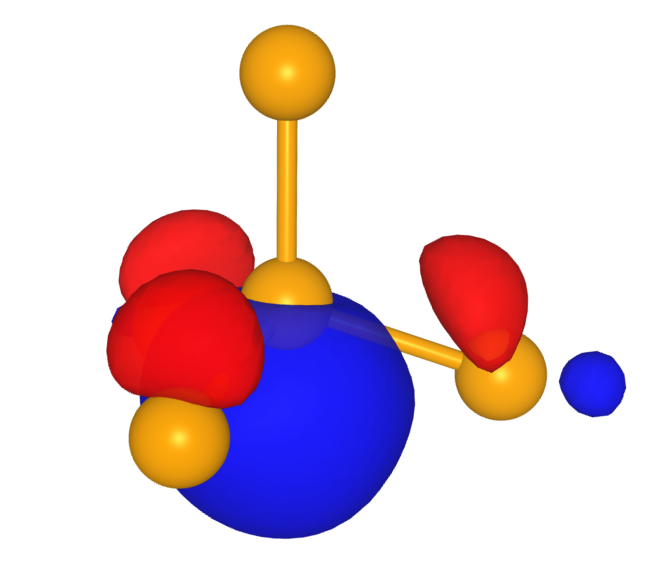}} &  {\includegraphics[width=0.15\columnwidth,trim={20px 0px 20px 0px},clip]{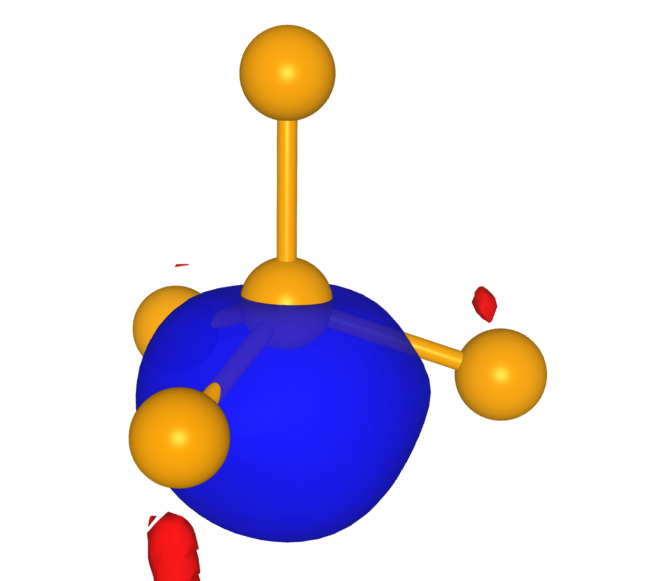}} & {\includegraphics[width=0.15\columnwidth,trim={20px 0px 20px 0px},clip]{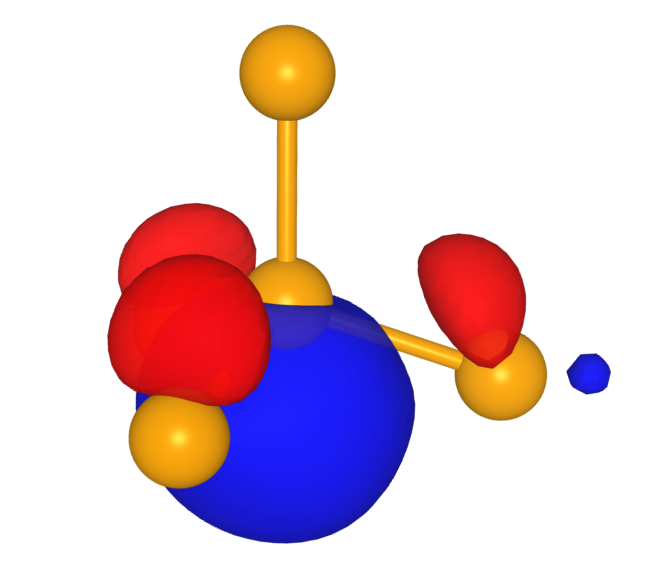}}\\
 &  & 7.66 & 5.67 & 8.25 \\
 &  & {\includegraphics[width=0.15\columnwidth,trim={20px 0px 20px 0px},clip]{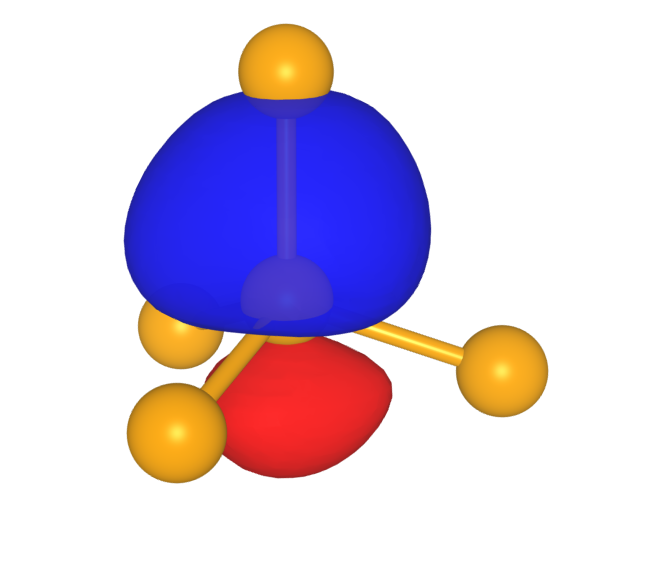}} & {\includegraphics[width=0.15\columnwidth,trim={20px 0px 20px 0px},clip]{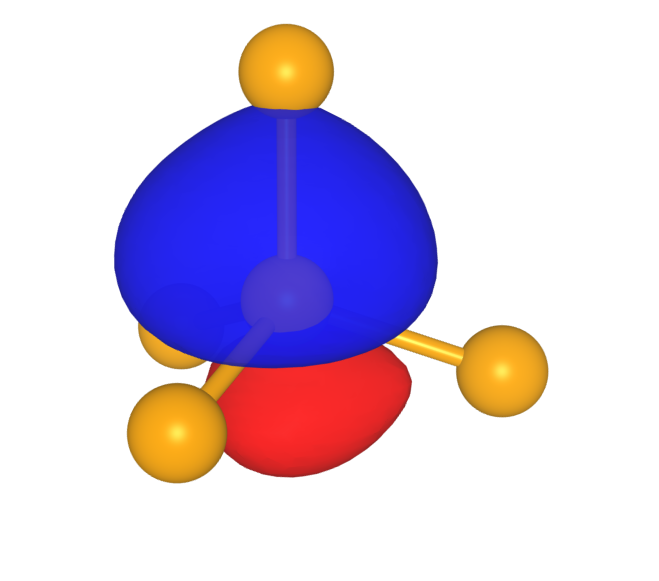}} & {\includegraphics[width=0.15\columnwidth,trim={20px 0px 20px 0px},clip]{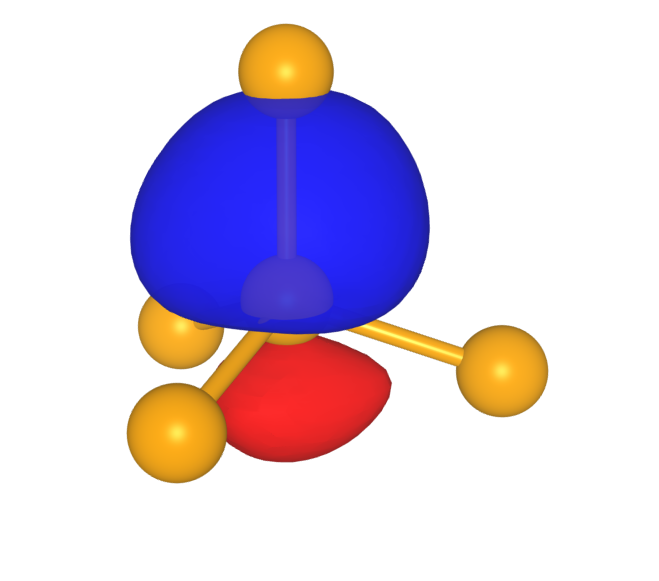}} \\   
 &  & 5.35 & 5.58 & 5.69 \\
\bottomrule[1pt]
\end{tabular}
\end{adjustbox}
\caption{\titlefig{Wannier functions obtained by wannierising the four valence bands plus the four low-lying conduction bands in silicon}
First row: the initial subspace is defined by projecting the Bloch states $\psink{n}$ on eight appropriately oriented $sp^3$-type orbitals giving {\it back-bonding} (BB) \MLWFs{} in all cases.  Second row: as above but with different orientations for the $sp^3$-type orbitals, resulting in {\it front-bonding} (FB) \MLWFs{} in all cases. Third row: the initial subspace is obtained from the SCDM method. Here, the eight $sp^3$-type WFs are in the BB configuration only when a full minimisation is performed. In all other cases a mixture of configurations is obtained instead. The values below each WF isosurface (isovalue=$\pm 0.45$ \AA$^{-3/2}$) is the value of the individual spread in\angstrom$^2$.}
\label{fig3.1}
\end{figure*}

We start by studying the Wannierisation of a manifold of bands consisting of the four valence bands plus the four low-lying conduction bands in silicon, the latter being entangled with bands at higher energies. 
For the SCDM method, we use $\sigma = 2$~eV and $\mu = 10$~eV. This choice is equivalent to that of Ref.~[\onlinecite{DL_2018_SIAM}], taking into account a shift in the absolute energy scale, which shifts the value of $\mu$. The outer and inner energy windows (described in the~\nameref{sec1:intro}), obtained through convergence tests, are set to $\varepsilon\tinysub{outer} = 17.0$~eV and $\varepsilon\tinysub{inner}=6.5$~eV.

When using initial projections onto atomic-like orbitals, we find that the spread functional $\Omega$ has three minima that are very close to each other and each of which gives eight real \MLWFs. The global minimum corresponds to four $sp^3$-type \MLWFs{} per Si atom in the two-atom unit cell, oriented in a {\it back-bonding} (BB) configuration, \ie{}, with the major lobes of the $sp^3$-type \MLWFs{} pointing towards the tetrahedral interstitial sites. A representative example of one such BB MLWF is shown in the isosurface plots in the first row of Fig.~\ref{fig3.1}.
Intuitively, from an atomic orbital perspective, one might instead expect the $sp^3$-type \MLWFs{} to be in a front-bonding (FB) configuration, \ie{}, with the major lobes pointing towards the vertices of the tetrahedra centred on the two non-equivalent Si atoms, 
as shown in the isosurface plots in the second row of Fig.~\ref{fig3.1}. However, this FB configuration corresponds to a slightly larger value of the total spread $\Omega$ and, therefore, constitutes a local minimum of the spread. 
A third (intermediate) local minimum gives four $sp^3$-type \MLWFs{} that are in the BB configuration on one Si atom in the unit cell and four $sp^3$-type in the FB configuration on the other Si atom. At variance with what is stated in Ref.~[\onlinecite{Damle_SIAM_2019}], all these cases can be found by specifying as initial projections four appropriately oriented $sp^3$-type orbitals on each Si atom in the unit cell. For the BB configuration: four $sp^3$-type orbitals centred on the Si atom at $(0.0, 0.0, 0.0)$ (\texttt{Si1}), and four rotated $sp^3$-type orbitals centred on the other Si atom (\texttt{Si2}) at (-\nicefrac{1}{4},\nicefrac{3}{4},-\nicefrac{1}{4}) in fractional coordinates with respect to the lattice vectors $\mathbf{a}_1=(-5.10,0.00,5.10)$, $\mathbf{a}_2=(0.00,5.10,5.10)$ and $\mathbf{a}_3=(-5.10,5.10,0.00)$ (in $a_0$). In the \Wannier{} code this can be specified in the projection block of the input file as: {\tt Si1:sp3:z=0,0,-1:x=0,1,0; Si2:sp3}.
For the FB configuration: same as above but with the labels 1 and 2 on the Si atoms interchanged.

With these initial projections, the four different minimisation options described earlier give the same qualitative results. Going from the {\tt DIS+MLWF} case to {\tt DIS} to {\tt MLWF} to {\tt proj-ONLY}, the spreads of the \MLWFs{} increase, as expected, but the FB/BB character is consistently present (see the top two rows of Fig.~\ref{fig3.1}, the spread of the individual \MLWFs{} (in units of\angstrom$^2$) is reported underneath each isosurface plot). Performing the SMV disentanglement step results in a reduction of $\omi$ from $26.54$\angstrom$^2$ to $20.06$\angstrom$^2$ in both the FB and BB cases, showing that the initial and final selected subspaces from the two different choices of projection have the same intrinsic smoothness.

Instead, starting from SCDM to define the initial subspace, we obtain different qualitative results for the four different minimisation schemes. Wannier functions in the BB configuration are found when a full minimisation is performed (i.e., SCDM followed by SMV and MV minimisation). A representative example of one such WF is shown in the third row and first column of Fig.~\ref{fig3.1}. SCDM selects a less smooth initial subspace ($\omi=27.54$\angstrom$^2$) than specifying atomic orbital initial projections ($26.54$\angstrom$^2$), but the final spreads are the same as in the equivalent BB case with atomic orbital initial projections. We also observed that in the case of SCDM, the minimisation of both $\omi$ and $\omt$ required more iterations to achieve the same level of convergence, perhaps reflecting the fact that the initial subspace is less smooth.
When using the other minimisation schemes, we find functions of both FB and BB character, all with slightly different individual spreads. Representative isosurfaces are shown in the last three columns of the row labelled ``SCDM'' in Fig.~\ref{fig3.1}. It is clear that the tetrahedral site symmetry is not preserved in the resulting WFs. Moreover, there is no clear pattern in the individual spreads going from the {\tt DIS} case to the {\tt proj-ONLY} case.

When looking at the interpolated band structure, however, a different picture 
emerges. In the case of choosing atomic orbital projections, the interpolation is very poor if no SMV disentanglement step is included in the minimisation. This shows the importance of disentangling the correct manifold and it is in agreement with what has been previously reported in the literature\cite{MMYSV_RMP84}. On the other hand, in the case of an SCDM-generated initial subspace, the interpolation is only marginally affected by the minimisation scheme employed (see Fig.~S1 in Supplementary Note 1).

To summarise, in silicon SCDM performs very well when combined with full spread minimisation, both in terms of the symmetries of the WFs and band interpolation (see Fig.~S1). 
When SCDM is used in isolation, the individual spreads of the resulting WFs are larger than WFs generated from user-defined atomic orbital projections; the quality of band structure interpolation, however, is almost independent of whether or not subsequent spread minimisation is carried out. 

\begin{figure}[tbp]
 \begin{tabular}{@{}cc@{}}
 \toprule[0pt]
  Projections ($s,d$) & SCDM  \\
 \midrule[1pt]
 {\includegraphics[width=0.25\columnwidth,trim={0px 0px 0px 0px},clip]{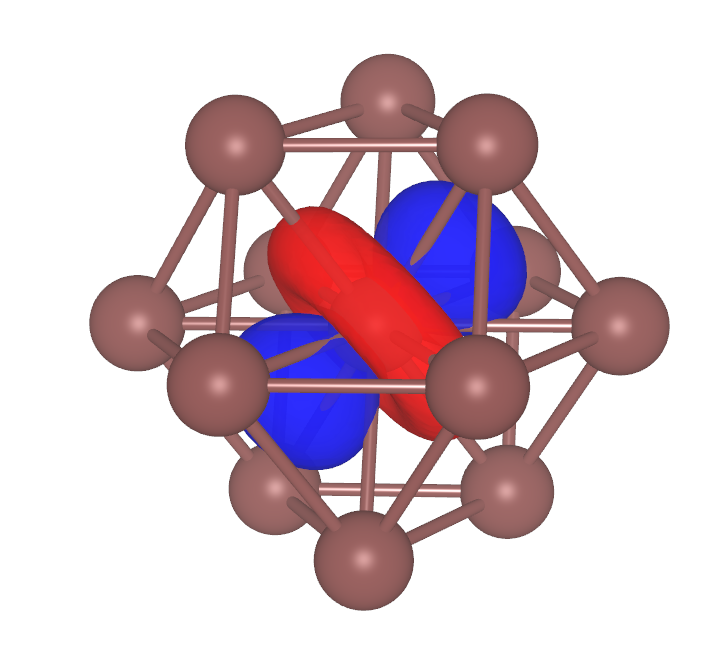}} & 
 {\includegraphics[width=0.25\columnwidth,trim={0px 0px 0px 0px},clip]{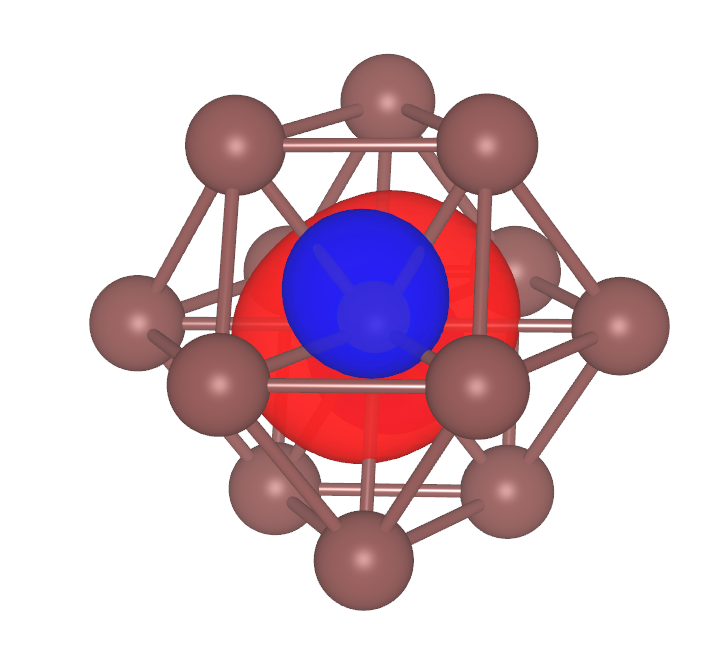}} \\
 (a) $t_{2_g}$ (0.404) & (d) $t_{2_g}/e_g$ (0.389) \\ 
 {\includegraphics[width=0.25\columnwidth,trim={0px 0px 0px 0px},clip]{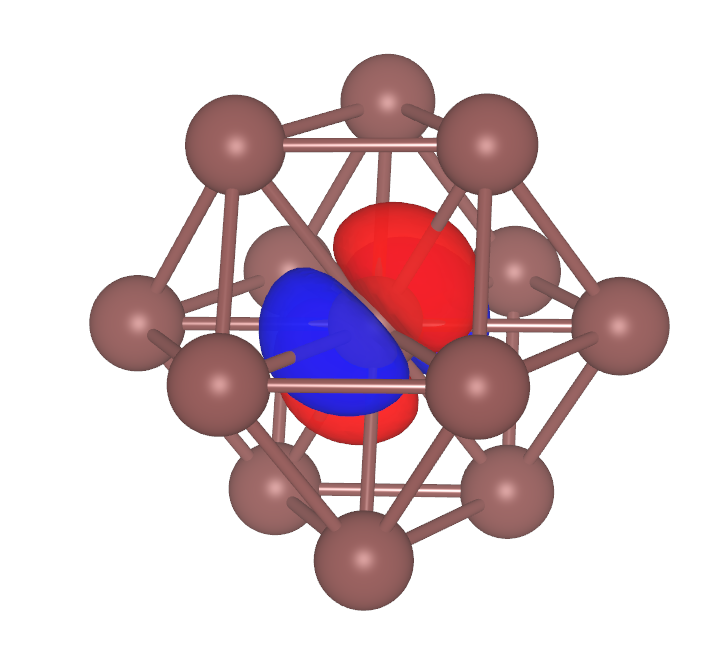}} & 
 {\includegraphics[width=0.25\columnwidth,trim={0px 0px 0px 0px},clip]{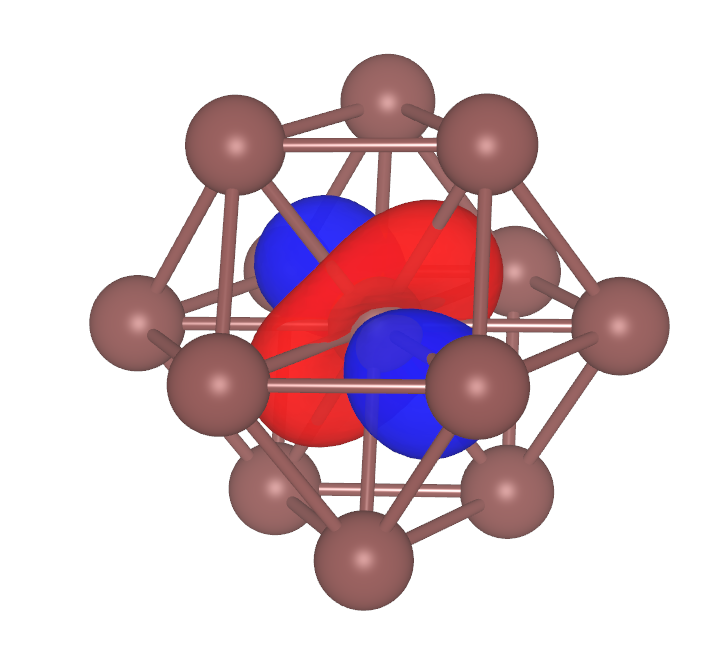}} \\
 (b) $e_g$ (0.377) & (e) $t_{2_g}/e_g$ (0.389) \\ 
 {\includegraphics[width=0.25\columnwidth,trim={0px 0px 0px 0px},clip]{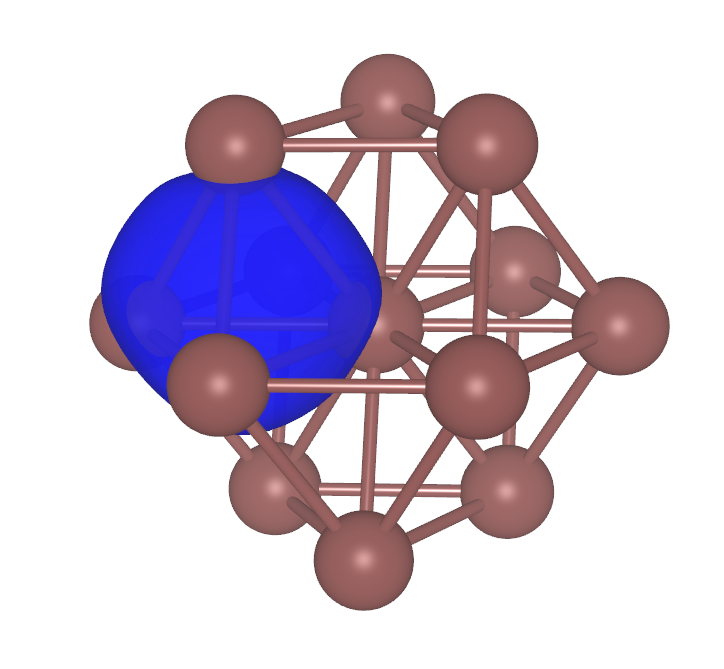}} & 
 {\includegraphics[width=0.25\columnwidth,trim={0px 0px 0px 0px},clip]{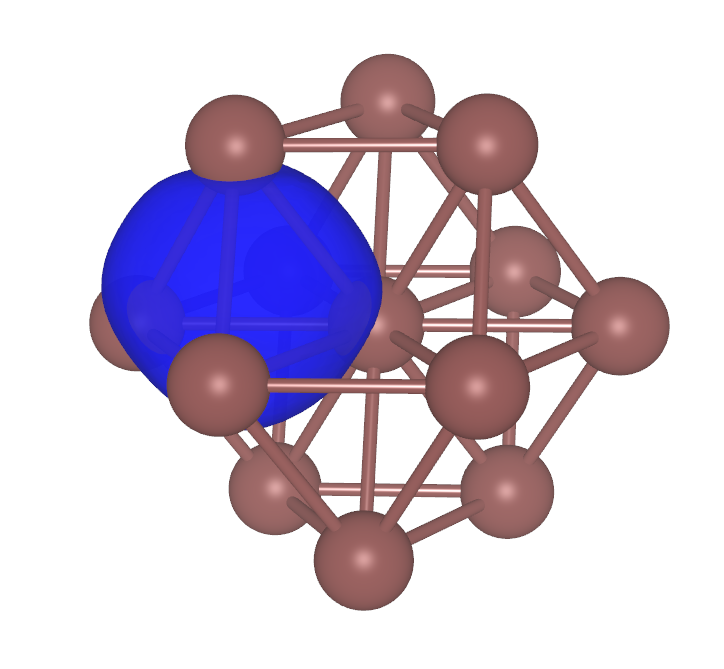}} \\
 (c) $a_1$ (2.11) & (f) $a_1$ (2.11) \\
  \bottomrule[1pt]
 \end{tabular}
\caption{\titlefig{\MLWFs{} obtained by wannierising the $s$-$d$ complex in copper} First column: three representative \MLWFs{} obtained from using atomic orbital projections to define the initial subspace (see main text for description). 
Panel (a) shows one of the three \MLWFs{} with $t_{2_g}$ character; panel (b) shows one of the two \MLWFs{} with $e_g$ character; panel (c) shows one of the two broad $s$-like orbitals centred on a tetrahedral-interstitial site. Second column: three representative \MLWFs{} obtained from using SCDM to define the initial subspace. Panel (d) and (e) show two of the five \MLWFs{} with mixed $t_{2_g}/e_g$ character; panel (f) shows one of the two broad $s$-like orbitals centred on an tetrahedral-interstitial site. Below each function its individual spread in\angstrom$^2$ is reported. Isosurfaces are plotted with an isovalue of $\pm 0.45$ \AA$^{-3/2}$.}
\label{fig3.2}
\end{figure}

Copper presents a paradigmatic case of a noble metal where a set of bands (e.g., of $d$-orbital character) cross and mix in a narrow energy window around the Fermi energy with a set of broad, nearly-free-electron bands. In this case, the SMV algorithm turns out to be very sensitive to the choice of the initial gauge and a good Wannier representation of the band structure can be achieved only by a careful choice of both initial projections and energy windows. Consequently, the possibility of bypassing these user-intensive steps makes the SCDM an attractive approach. This is particularly important for methodologies such as \abinitio{} tight binding\cite{Kaxiras_PRB92}, DFT+U\cite{Schnell_PRB65} and DMFT\cite{Georges_PRB74}, which deal with strong correlation in a local subspace, \eg{}, the subspace spanned by $d$ orbitals (for transition metals or transition-metal oxides) or $f$ orbitals (for rare-earth or actinide intermetallics).
For copper, as suggested by Souza~\etal{}\cite{SMV_PRB65}, in order to generate a faithful representation of the band structure around the Fermi level, we work with a manifold of dimension $J=7$, which contains one more function than the conventional minimal basis usually employed in tight-binding models. For this system, we focus only on the full minimisation scheme ({\tt DIS+MLWF}), as it is the most representative when comparing the symmetries of the WFs, as shown in the previous section.
For the disentanglement step we set $\varepsilon\tinysub{outer} = 38.0$~eV and $\varepsilon\tinysub{inner}=19.0$~eV. For SCDM, we set $\mu=11.40$~eV and $\sigma=2.0$~eV. The Fermi energy in our calculation is at $12.18$~eV. 
As shown in Ref.~[\onlinecite{SMV_PRB65}], appropriately selected initial projections are five $d$-type orbitals centred on the Cu atom and two $s$-type orbitals, each centred on one of the two tetrahedral interstitial sites. The resulting seven \MLWFs{} respect the symmetries one would expect from group theory. In fact, the five $d$-like functions give a representation of dimension 3+2 of the $O_h$ point group (which is isomorphic to the site-symmetry group of the origin), with the usual $t_{2g}$ and $e_g$ character (see Fig.~\ref{fig3.2}(a) and Fig.~\ref{fig3.2}(b)). The two $s$-like functions give each a one-dimensional representation ($a_1$) of $T_d$ (which is the site-symmetry group of the tetrahedral interstitial sites), as shown in Fig.~\ref{fig3.2}(c).

When using SCDM projections, the symmetries of the $d$-type \MLWFs{} are not fully recovered. This can clearly be seen in Figs.~\ref{fig3.2}(d) and \ref{fig3.2}(e), where the $d$-type functions show mixed $t_{2g}/e_g$ character (this is a feature of all five $d$-type functions).

\subsection*{Isolated bands}\label{sec4:isolated}
Until here, we have looked into the details of the Wannier functions that can be obtained from SCDM projections, by focusing on the paradigmatic examples of silicon and copper (see \nameref{sec3:scdm_vs_mlwfs}). We focussed on comparing Wannier functions as obtained by adopting different initial projections, given that good atomic-like projections can often be easily identified through chemical intuition. Now we take a complementary perspective, by considering any given crystal structure, where we face the problem of finding good initial projections without any prior chemical knowledge of the system. This is particularly relevant for high-throughput studies, where crystal-structure databases are systematically screened with first-principles simulations. In order to produce high-throughput Wannier functions, it is fundamental to provide an algorithm that does not require human interaction in the choice of the initial projections. In addition, such an algorithm must be able to use only information that is either contained in the crystal structure and the pseudopotential, or that can be computed by a simple first-principles simulation, such as the projected density of states. To this aim, human-specified atomic-like projections are not suitable, and we propose the SCDM method as the workhorse for the automated choice of the initial projections. 

 In order to ascertain the effectiveness of the SCDM method in generating well-localised Wannier functions in an automated way,
we start by testing the algorithm for isolated manifolds. We compare Wannier interpolations and direct DFT calculations for the band structure of the valence bands of a set of 81 insulating bulk crystalline materials spanning a wide range of chemical and structural space, for the full list the Reader is referred to Ref.~[\onlinecite{MaterialsCloudArchiveEntry}]. 
We quantify the differences between two band structures by introducing a simple metric that is inspired by the so-called ``bands distance'' introduced in Ref.~[\onlinecite{sssp_paper}].
  Here we define the distance between DFT and Wannier-interpolated bands as:
\begin{equation}
    \label{eta_def}
    \eta = \sqrt{\sum_{n\mathbf{k}} \left(\varepsilon_{n\mathbf{k}}\tinysup{DFT}-\varepsilon_{n\mathbf{k}}\tinysup{Wan}  \right)^2} ,
\end{equation}
where $\varepsilon_{n\mathbf{k}}\tinysup{DFT}$ and $\varepsilon_{n\mathbf{k}}\tinysup{Wan}$ are respectively the DFT and Wannier-interpolated band structures, and the summation runs over the occupied bands only. Later in the~\nameref{sec5:entangled} section, we will introduce a finite smearing to deal with conduction-band states and metallic systems.
As in Ref.~[\onlinecite{sssp_paper}], to take into account the possibility that significant differences between band structures may occur only in sub-regions of the Brillouin zone or in small energy ranges, we also compute
\begin{equation}
\eta^{\max} = \max_{n\mathbf{k}} \left(\left|\varepsilon_{n\mathbf{k}}\tinysup{DFT} - \varepsilon_{n\mathbf{k}}\tinysup{Wan}\right|\right)
\end{equation}
where, essentially, we select the point $(n\bfk)$ with the worst interpolation, which is responsible for the largest contribution to $\eta$. We use $\eta$ and $\eta^{\max}$ to assess the effect of iteratively minimising the spread $\omt$ to obtain maximally-localised Wannier functions (``SCDM+MLWF''), compared to the one-shot Wannier orbitals that are obtained by using the SCDM projections only  (``SCDM-only''). We note that in the following MLWF might refer either to a maximally-localised WF or to the maximal localisation procedure itself, the meaning being always clear from the context.

For each of the 81 structures of the benchmark set, we first perform a variable-cell optimisation and we then compute the band structure on a high-symmetry path using DFT. The cell and the path are standardised using {\tt seekpath} according to the prescription of Ref.~[\onlinecite{seekpath_2017}]. The ground-state charge density is obtained using a $k$-point spacing of $0.2$ \angstrom${}^{-1}$ in the irreducible Brillouin zone (unless otherwise stated). Band structures are then calculated using the charge density frozen from the earlier calculation and sampling the high-symmetry path with a spacing of $0.01$ \angstrom$^{-1}$.
Then we compute the WFs and the real-space Hamiltonian with \Wannier{}, starting from a non-self-consistent field (NSCF) DFT calculation performed on a possibly different $k$-point grid on the full BZ and employing the ground-state charge density computed earlier.
At this point, the bands distance is then calculated by diagonalising the Wannier Hamiltonian using the \textsc{TBmodels} code~\cite{tbmodels_prm_2018} on the same $k$-points used in the DFT bands calculation.

\begin{figure}[tbp]
  \centering
  {\includegraphics[width=8cm,trim={15px 0px 0px 0px},clip]{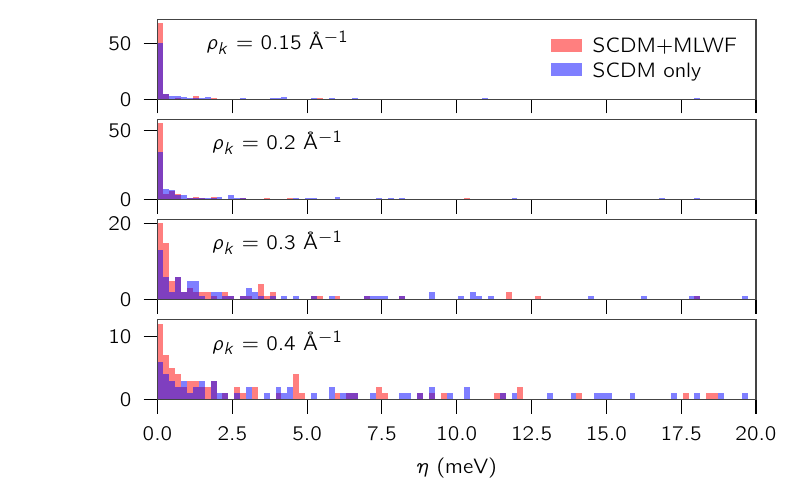}}\\
  {\includegraphics[width=8cm,trim={15px 0px 0px 0px},clip]{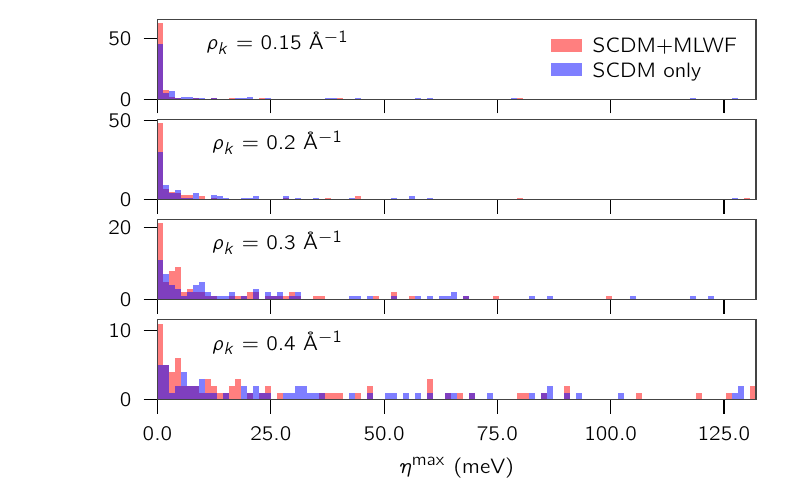}}
  \caption{\titlefig{Average and max band distance $\eta$ using SCDM-only and SCDM+MLWF for the valence bands of 81 insulating materials} Top (bottom) panel: average (max) band distance $\eta$ using SCDM-only (blue) and SCDM+MLWF (red) obtained using four different $k$-point grids with spacing $\rho_k$. 
 The MLWF procedure improves the interpolation accuracy, although SCDM-only Wannier functions perform already remarkably well. The histograms focus on the most relevant interval and few outliers are not shown, in particular at $\rho_k = 0.2$\angstrom$^{-1}$  $98 \%$ (79/81) of the SCDM+MLWF bands and $96\%$ (78/81) of the SCDM-only bands exhibit $\eta<20$ meV, while $98 \%$ (79/81) of the SCDM+MLWF bands and $93\%$ (75/81) of the SCDM-only bands exhibit $\eta^{\max}<130$ meV. }
  \label{fig_histo_ins}
\end{figure}

All DFT calculations are carried out using the \QE{} distribution \cite{giannozzi_qe_2017}, employing  the PBE functional \cite{perdew_pbe_96} and a beta version of the  SSSP v1.0 efficiency pseudopotential library \cite{sssp_paper,gbrv_2014,pslib_2014,oncv_2015,wenz_2014,dojo_2018}, where the norm-conserving ONCV pseudopotentials\cite{oncv_hamann_2013} are recompiled using version 3.3.1 of the code, and the pseudopotentials for Ba and Pb are replaced by \texttt{Ba.pbe-spn-kjpaw\_psl.1.0.0.UPF}	 and \texttt{Pb.pbe-dn-kjpaw\_psl.0.2.2.UPF} of the pslibrary.
In Fig.~\ref{fig_histo_ins} we report histograms of $\eta$ and $\eta^{\max}$ for four different $k$-point densities, namely $\rho_k=$ 0.15, 0.2, 0.3 and 0.4\angstrom$^{-1}$, used in the NSCF step to construct Wannier functions. We stress that for an isolated set of bands, such as for the valence bands of an insulator, the SCDM method involves no free parameters and the only parameter to set is the $k$-point grid spacing $\rho_k$ of a uniform grid that is used to diagonalise the Hamiltonian. Hence it is fundamental to elaborate a strategy for the choice of $\rho_k$, as this finally removes every free parameter from the construction of Wannier functions for isolated bands.

The SCDM method is found to work well for all of the 81 systems studied, with the exception of two that have very poor interpolation.
Notably, these two structures (three if we consider the SCDM-only method) are the ones that exhibit the highest initial spread $\Omega$ per Wannier function. Although a large initial spread does not necessarily imply poor interpolation, it certainly correlates with a potential risk of poor Wannierisation and it could be used as a marker for triggering a check on the quality of bands interpolation within the calculation workflow. We postpone the discussion on the causes of the poor performance of the SCDM method in these systems until the end of this section, where we also provide possible solutions that can be automated.

\begin{figure*}[htbp]
  \centering
  \subfloat[CaO]{\includegraphics[width=0.95\columnwidth]{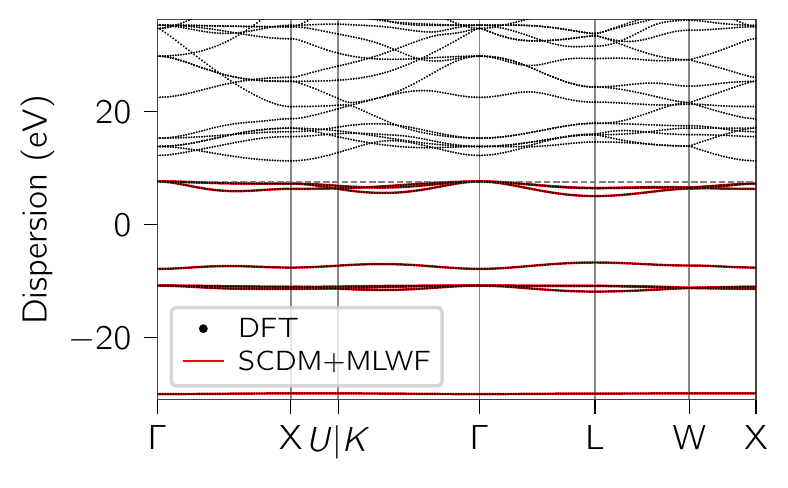}}
  \centering
  \subfloat[C$_3$Mg$_2$ ]{\includegraphics[width=0.95\columnwidth]{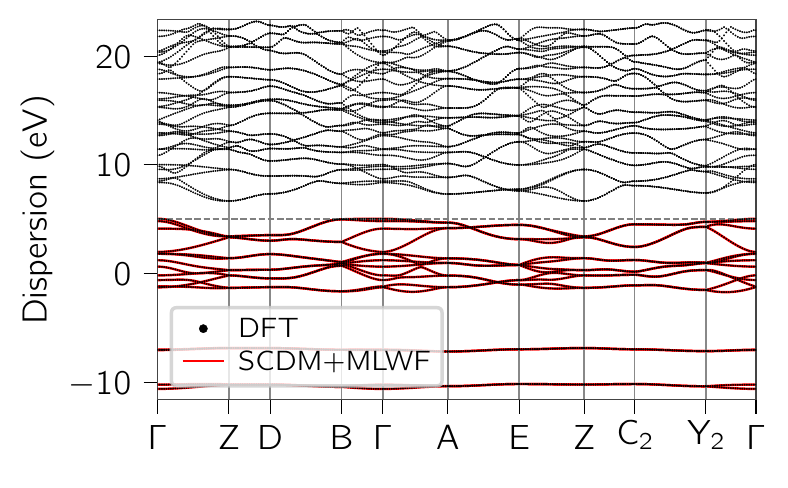}}
  \caption{\titlefig{Comparison between Wannier-interpolated valence bands and the full direct-DFT band structure} Wannier-interpolated (solid red) and full DFT band structure (black dots), using the MLWF procedure on SCDM projections and $\rho_k = 0.2$\angstrom$^{-1}$. The dashed line labels the valence band maximum (VBM). (a) Band structure of CaO ($\eta = 0.06$~meV, $\eta^{\max} = 0.23$~meV, VBM~$=7.52$~eV). (b) Band structure of C$_3$Mg$_2$ ($\eta = 0.4$~meV, $\eta^{\max}= 6.35$~meV, VBM~$=5.0$~eV).}
 \label{fig_comparison}
\end{figure*}

To get a sense of the typical quality of a good SCDM+MLWF interpolation, we report in Fig.~\ref{fig_comparison} the comparison between direct-DFT and SCDM+MLWF interpolated band structures for CaO ($\eta = 0.06$ meV, $\eta^{\rm max} = 0.23$ meV) and C$_3$Mg$_2$ ($\eta = 0.4$ meV, $\eta^{\rm max}= 5.6$ meV) run with a $k$-point spacing $\rho_k=0.2$\angstrom$^{-1}$; the direct and interpolated band structures are essentially indistinguishable (e.g., the largest difference in energy between the bands in the case of CaO is of $\eta^{\rm max}=0.23$~meV).

Fig.~\ref{fig_histo_ins} shows the distribution of $\eta$ and $\eta^{\rm max}$ across the whole set of insulators for the four different $k$-point grids. We find that a grid with spacing $\rho_k = 0.2$ \angstrom$^{-1}$ is typically sufficient to provide accurate interpolated band structures, in particular $96\%$ of the materials (78/81) for SCDM-only and $98\%$  (79/81) for SCDM+MLWF show $\eta < 20$~meV, and $93\%$ (75/81) of the SCDM+MLWF bands and $74 \%$ (60/81) of the SCDM-only bands display $\eta < 2$~meV. As shown in Fig.~\ref{fig_histo_ins}, $\eta^{\max}$ follows a similar trend, with $95\%$ (77/81)  of the SCDM+MLWF bands and $86 \%$  (70/81) of the SCDM-only bands showing an $\eta^{\max} < 50$~meV, and $90\%$ (73/81)  of SCDM+MLWF bands and $77 \%$  (62/81) of the SCDM-only bands showing an $\eta^{\max} < 20$~meV.

Those systems with $\eta > 20$~meV or, in other words, interpolated bands that are significantly less accurate with respect to the majority of the sample, are considered to be outliers. In Table~\ref{tab:outliers_ins}, we report the number of the outliers for the four different $k$-point densities, both in the case of SCDM-only and SCDM+MLWF. Clearly, increasing the $k$-point density produces fewer outliers and, in this respect, the SCDM+MLWF seems to converge slightly faster than SCDM-only, in agreement with the results shown in Fig.~\ref{fig_histo_ins}.
\begin{table}[tbp]
\begin{tabular}{cp{6px}cp{6px}c}
\hline
$\rho_k$ [\simpleangstrom$^{-1}$] && SCDM-only && SCDM+MLWF\\
\hline
0.15 && 3 && 2 \\
0.2 && 3 && 2 \\
0.3 && 6 && 2 \\
0.4 && 16 && 8 \\
\hline
\end{tabular}
\caption{\label{tab:outliers_ins} Number of interpolated bands showing $\eta > 20$~meV, i.e. outliers, with different $k$-point densities $\rho_k$.}
\end{table}

As we will discuss shortly, the superior performance of SCDM+MLWF is linked with the increased localisation associated with the MLWF procedure. As mentioned before, localisation is also related to the poor interpolation of the outliers: at all $k$-point densities, outliers are among the systems with the largest initial spreads. On one hand, a larger initial spread signals a potential problem with the SCDM projections, on the other hand it requires a denser $k$-point grid for convergence (the less localised the Wannier functions are, the more long-range the Wannier Hamiltonian is).

\begin{figure}[tbp]
  \centering
  \includegraphics[width=8cm,trim={30px 30px 30px 30px},clip]{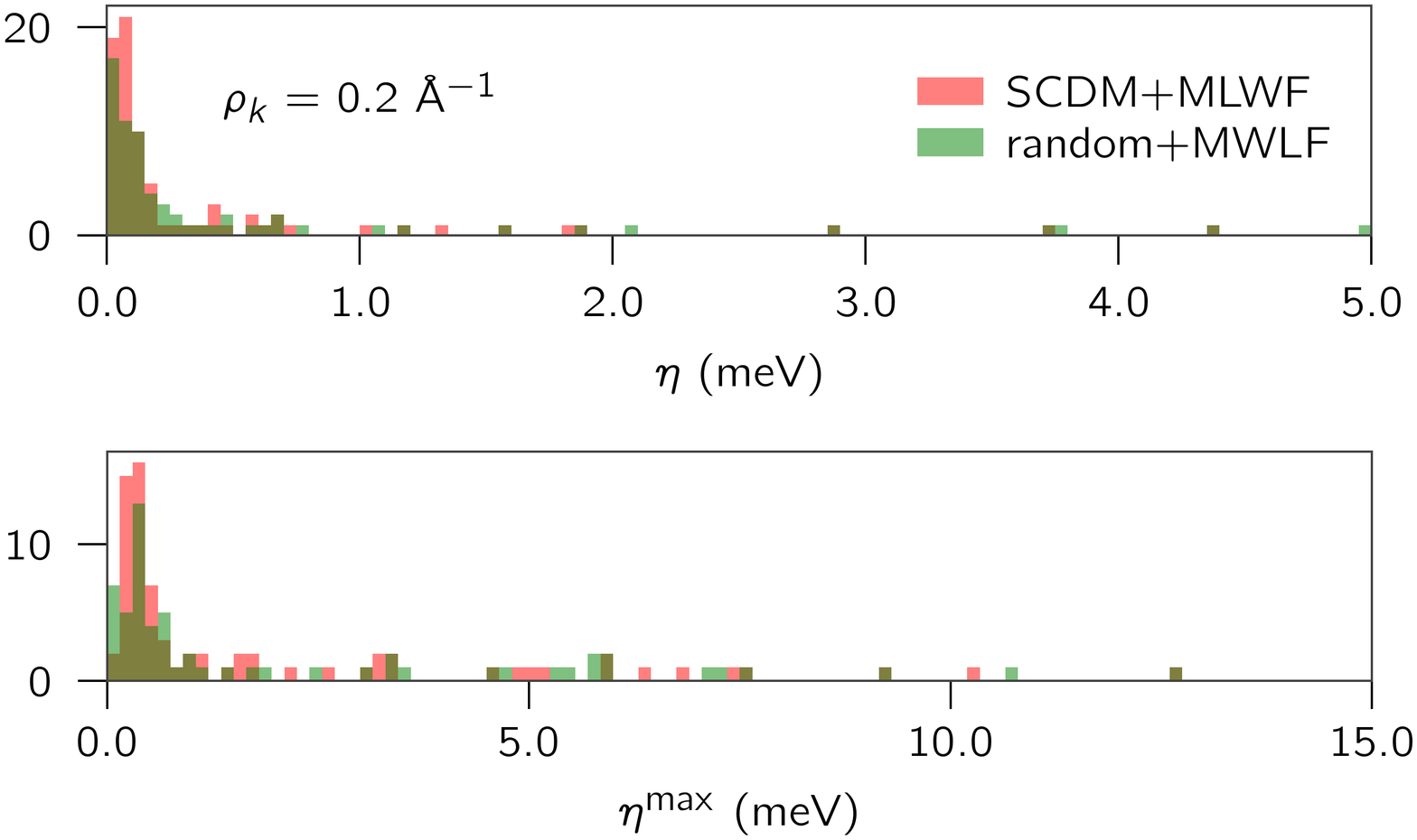}
  \caption{\titlefig{Average and max band distance $\eta$ using random+MLWF and SCDM+MLWF for the valence bands of 81 insulating materials} Top (bottom) panel: average (max) band distance $\eta$ using random+MLWF (green) and SCDM+MLWF (red) obtained using $\rho_k=0.2$\angstrom$^{-1}$. SCDM projections perform better than random projections when used in conjunction with the MLWF procedure. The histograms focus on the most relevant interval and few outliers are not shown, in particular the $96 \%$ (78/81) of the SCDM+MLWF bands and the $83\%$ (67/81) of the random+MLWF bands exhibit an $\eta<5$~meV, while the $90 \%$ (73/81) of the SCDM+MLWF bands and the $74\%$ (60/81) of the random+MLWF bands exhibit an $\eta^{\max}<15$~meV}.
  \label{fig_histo_ins_random}
\end{figure}

Fig.~\ref{fig_histo_ins} also shows that, when considering valence bands only, the MLWF procedure moderately improves the quality of band interpolation with respect to SCDM-only, resulting in narrower $\eta$ and $\eta^{\max}$ distributions, although band interpolation is often already excellent using an SCDM-only approach.
We emphasise, however, that it is known that for the valence bands of gapped systems, a set of randomly-centred Gaussian functions can be often used as starting projections leading to good MLWFs. 
We compare, therefore, the performance of SCDM projections versus randomly-centred Gaussian orbital projections as a starting point for the MLWF procedure (which we refer to as the ``random+MLWF'' scheme), assessing their comparative robustness and accuracy of band interpolation. 
Fig.~\ref{fig_histo_ins_random} reports the distribution of $\eta$ and $\eta^{\max}$ with $k$-point spacing $\rho_k =0.2$\angstrom$^{-1}$. The SCDM projections are found to perform better, leading to narrower distributions: $98\%$ of the materials (79/81) show  $\eta < 20$~meV for SCDM+MLWF against the $89\%$ (72/81) for random+MLWF, and  $93\%$ (75/81) of the SCDM+MLWF bands display $\eta < 2$~meV against $75\%$ (61/81) of random+MLWF bands.  
As shown in Fig.~\ref{fig_histo_ins_random}, $\eta^{\max}$ follows a similar trend, with $95\%$ (77/81)  of the SCDM+MLWF bands and $81 \%$  (66/81) of the random+MLWF bands showing an $\eta^{\max} < 50$~meV, and $90\%$ (73/81)  of SCDM+MLWF bands and $74\%$ (60/81) of the random+MLWF bands showing an $\eta^{\max} < 20$~meV. Therefore, while SCDM is able to provide WFs resulting in a more accurate band interpolation, we emphasise here that for isolated manifolds the minimisation procedure is quite robust also when providing randomly-centred $s$-like Gaussian orbital projections.

We now elaborate on the differences between random and SCDM initial projections. 
First, random projections typically generate a much higher initial spread (7.5\angstrom$^{2}$ per WF) compared to SCDM (1.0\angstrom$^{2}$ per WF).
We find that the MLWF procedure is often sufficient to localise Wannier functions even in the case of large initial spreads: for 63 out of 81 materials the MLWF procedure brings both the random projections and the SCDM projections cases to the same minimum spread value. Notably, it never happens that the spread is similar and the quality of the interpolation is very different, while the opposite happens only in the case of He, a pathological case (1 atom and 2 electrons per cell) where random projections give a poorly localised Wannier function while still being able to provide a very good interpolation.
For 15 materials (16 if we include He), random projections provide a very poor starting point and the MLWF procedure remains trapped in a local minimum with large spread. In these cases, instead, SCDM projections are a good starting point with low spread and the MLWF procedure further reduces it and a higher-quality interpolation is achieved, as demonstrated by the lower $\eta$ values. Finally, there are two materials for which both SCDM-only and SCDM+MLWF do not perform well, 
but where random+MLWF happens to perform better than SCDM+MLWF. 
For one of these cases, Al$_2$Os, we have checked that excluding the semi-core states greatly improves the performance and the quality of the interpolated bands. We believe that the reason lies in the fact that, if semi-core states are present, then there are some projections, centred on the same site, that possess the same symmetry character, \eg{}, $p$-like projections with different principal quantum numbers (for instance $1p$- and $2p$-like). With a relatively low plane-wave energy cutoff, the real-space grid is too coarse and there are not enough degrees of freedom for the column selection in the QRCP step to distinguish or describe sufficiently well these same-symmetry-character states.

In the other case, Se$_2$Sn, there are no semi-core states. Here instead, some SCDM projections show an initial value of $\omd$---the sum of the diagonal elements of $\omt$ in \eqref{eq:omt}---that is not zero or very close to zero ($\omd > 0.5$~\AA$^2$), which could be used as a diagnostic indicator for problematic systems. 
In particular, SCDM+MLWF seems to get trapped in a state in which there are a number of well-localised WFs and two that are diffuse and spread over multiple sites. This set of WF are real with a total spread of $28$~\AA$^2$ and $\omd$ of $2$~\AA$^2$. We found that a possible solution to recover a good interpolation is to add some noise (adding small random numbers to the search direction components, as implemented in \Wannier{}) during the minimisation to help the algorithm escape from the unwanted local minimum.

We propose some technical solutions that could be easily added to a workflow:
\begin{itemize}
\item Automatically detect and exclude semi-core states (if any). This is generally a safe choice as these states are not physically interesting for most applications. Alternatively, one could retain the semi-core states and increase the cutoff energy (or equivalently the density of the real-space grid).
\item If the problem is not in describing semi-core states, then check the value of $\omd$, if it is above a given threshold (e.g., $>1.0$~\AA$^2$) for one or more initial projections, introduce some noise in the minimisation.
\item If none of the above work, switch to random+MLWF projections, which may give a better final result. 
\end{itemize}

\begin{figure}[tbp]
  \centering
  \includegraphics[width=8cm]{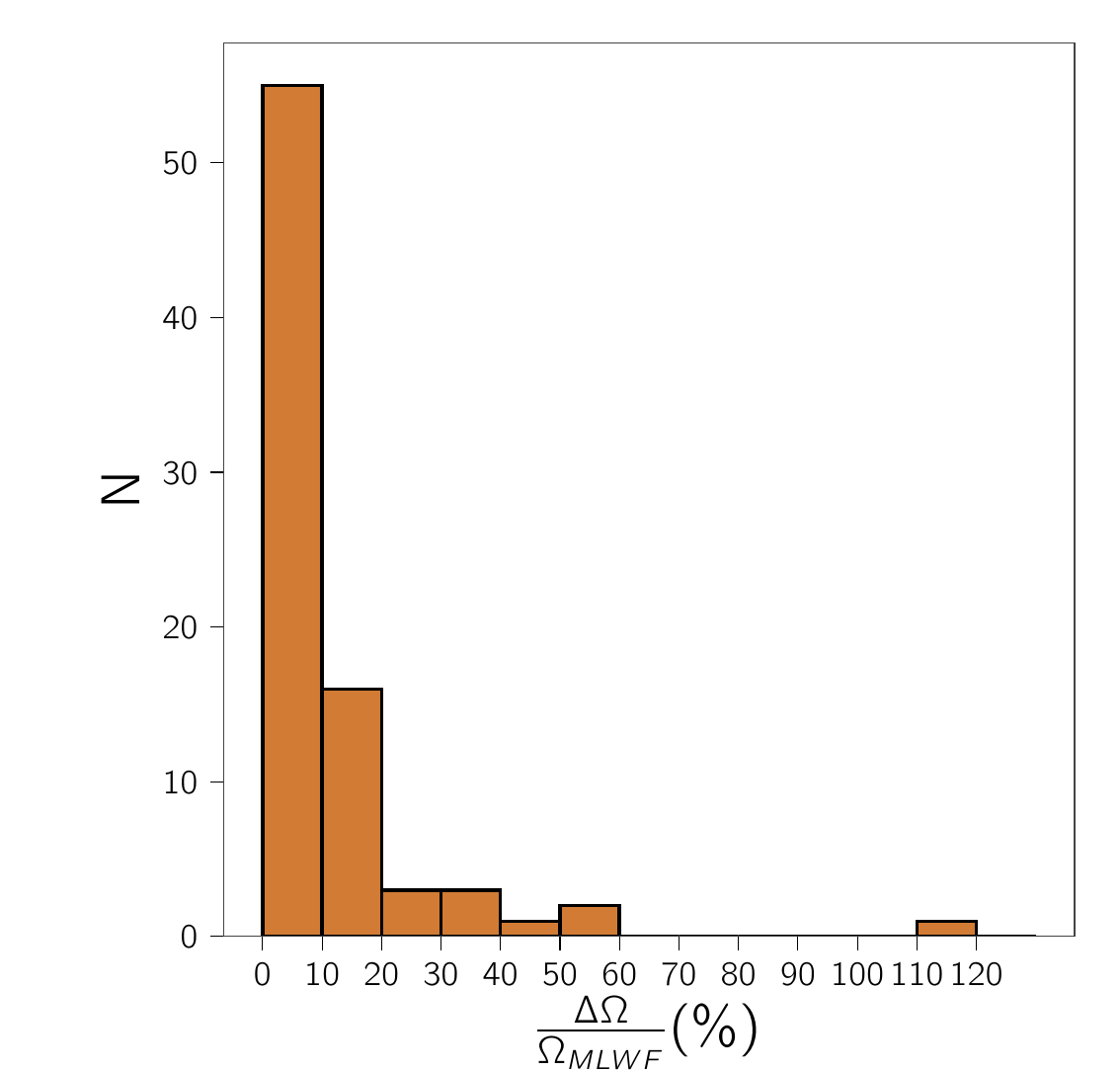}
  \caption{\titlefig{Histogram of the relative variation of the total quadratic spread $\Omega$ before and after the MLWF procedure} The data has been obtained considering the valence bands of our set of 81 insulators, with $\rho_k = 0.2$\angstrom$^{-1}$. The SCDM+MLWF procedure provides Wannier functions that are moderately more localised with respect to SCDM-only, with a relative variation within $10-20 \%$ for most materials.}
  \label{fig_histo_spread_ins}
\end{figure}

To study now more in detail the effect of minimising the spread, we start by comparing the total spread $\Omega$ obtained using SCDM+MLWF and SCDM-only, by computing:
\begin{equation}
\frac{\Delta \Omega }{\Omega^{\rm MLWF}} = \frac{ \Omega^{\rm SCDM} -  \Omega^{\rm MLWF}}{\Omega^{\rm MLWF}}
\end{equation}
where $\Omega^{\rm SCDM}$ and $\Omega^{\rm MLWF}$ are the total spreads obtained with SCDM-only and SCDM+MLWF, respectively.
As reported in Fig.~\ref{fig_histo_spread_ins}, the SCDM-only Wannier functions are already well localised and  $\frac{\Delta \Omega }{\Omega^{\rm MLWF}} $ is less than $10\%$ for $68\%$ (55/81) of systems, and less than $20\%$ for $88\%$ of them (71/81).

\begin{figure}[tbp]
\centering
 \includegraphics[width=8cm]{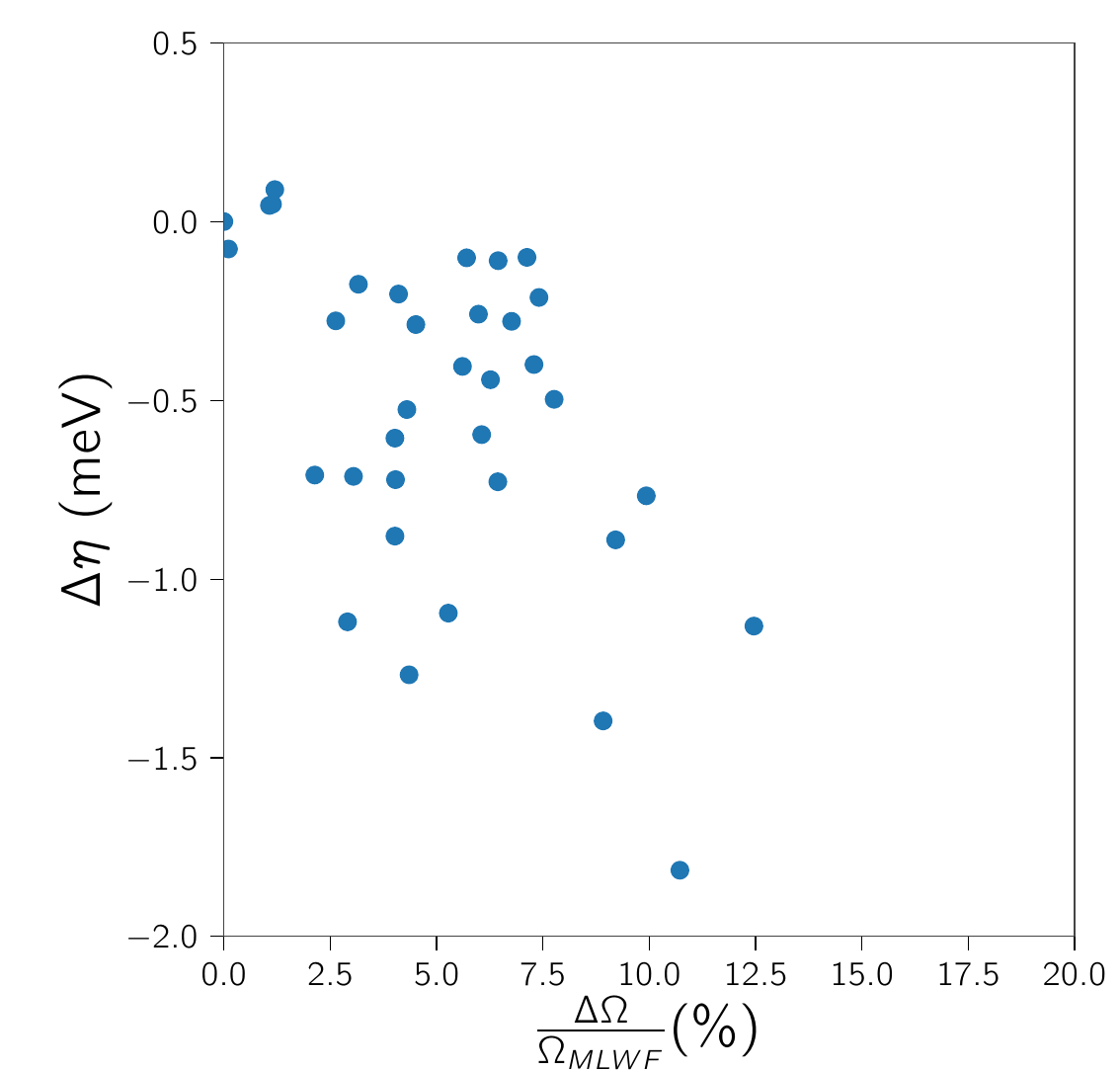}
\caption{\titlefig{$\Delta \eta$  versus  $\Delta \Omega/\Omega^{\rm MLWF}$ scatter plot (valence bands only)} The dataset consists of the 81 insulators described in the main text (only 61 out of 81 visible in the axes range). $\Delta \eta$ and $\Delta \Omega/\Omega^{\rm MLWF}$ represent the quantitative deviation between SCDM+MLWF and SCDM-only in terms of band structures and total spreads respectively. Maximally-localised Wannier functions give comparable and often more accurate interpolated bands. }
 \label{fig_scatter_ins}
\end{figure}

An interesting question is whether the difference in spread due to the MLWF procedure correlates with the difference in the quality of the interpolation. To assess this, we compute the quantity 
\begin{equation}
\Delta \eta = \eta^{\rm MLWF} - \eta^{\rm SCDM},
\end{equation}
where $\eta^{\rm SCDM}$ and $\eta^{\rm MLWF}$ are the band distances obtained with SCDM-only and SCDM+MLWF respectively. Fig.~\ref{fig_scatter_ins} shows a scatter plot of $\Delta \eta$  vs. $\Delta \Omega / \Omega^{\rm MLWF}$ , showing that a reduction in the spread typically implies an improvement in the quality of the interpolation ($\Delta \eta < 0 $).  These findings highlight that SCDM-only Wannier functions are already sufficiently localised and represent well the valence manifold, and the subsequent MLWF procedure (starting from a very good guess) safely refines the initial choice of SCDM, improving the accuracy of the Wannier Hamiltonian by increasing localisation. In general, the greatest benefit from the MLWF procedure is visible in the interpolation of the almost-flat semi-core states. In fact often, when using SCDM-only Wannier functions for the interpolation of these states, the interpolated bands show an oscillatory behaviour, with the maximum absolute difference with respect to the DFT bands of the order of a few meV (comparable to the spread of those bands). From our results, a smoother and more accurate interpolation is usually recovered after a MLWF procedure.

Before discussing the case of entangled bands, we summarise here the main conclusions that can be drawn for isolated bands.
All the results we obtained, displayed in Figs.~\ref{fig_histo_ins}, \ref{fig_comparison}, and \ref{fig_histo_spread_ins}, consistently support the effectiveness of adopting SCDM projections for the Wannier interpolation of the valence bands of insulators. The quality of the interpolation is very high for $98 \%$ of the structures, with only 2 (out of 81) cases showing a poor interpolation. 
Although SCDM-only Wannier functions are shown to provide already accurate band structures, the MLWF procedure appears to improve both the quality of interpolation (lower $\eta$) and localisation (lower spread).
Hence, we suggest the SCDM+MLWF method with $\rho_k = 0.2$\angstrom$^{-1}$ as the standard protocol for producing accurate and efficient Wannier Hamiltonians describing the valence bands of bulk insulating crystals.

\subsection*{Entangled bands}\label{sec5:entangled}
We now consider the case of entangled bands. With the intent of describing a fully automatic protocol,
we limit ourselves to the case of Wannier interpolation of all states up to a given energy (excluding, if appropriate, manifolds of low-lying semicore states that are isolated in energy from the rest of the band structure) and we do not consider the case of computing Wannier functions for a manifold of bands of given symmetry within a narrow energy window (e.g., $d$ states in copper or $t_{2_g}/e_g$ states in a transition-metal oxide, see~\nameref{sec3:scdm_vs_mlwfs}) that is entangled with bands above and below in energy.

In the case of entangled bands, the SCDM method demands the choice of three free parameters: $\mu$ and $\sigma$, as described at the end of~\nameref{subsec2.2:scdm} section, as well as $J$, the target number of Wannier functions.
These parameters play a fundamental role in the selection of the columns of the quasi-DM and hence greatly affect the overall quality of the subspace selection and, consequently, the bands interpolation. 
In particular, since there is no equivalent definition of an inner energy window\cite{SMV_PRB65} in the SCDM method, it is not guaranteed that a subspace that includes the physically-relevant lowest-lying bands will be selected because the greedy QRCP algorithm, owing to an inappropriate choice of $\mu$ and $\sigma$, might favour states that are higher in energy.
It is, therefore, key to the success of the automation process to have a protocol that automatically chooses these parameters in a robust and systematic way. We will now describe such a protocol, and in the~\nameref{sec:entangled-hith-verification} section we show its effectiveness on a large set of chemically diverse materials.

\subsection*{Protocol}\label{sec:mu-sigma-protocol}
To identify appropriate values of $\mu$, $\sigma$ and $J$, we first compute the ``projectability'' $p_{n\bfk}$, which measures how well each Bloch state $\kpsi{n}{\bfk}$ is represented in a Hilbert space $\mathcal{A}$ defined by a given set of localised functions. 
Indeed, in the entangled case, WFs contain contributions from the valence states plus specific conduction states, typically corresponding to the anti-bonding partners of the valence states. The selection of these specific conduction states---out of the very many---can be challenging, because they are not necessarily the lowest energy ones. This idea motivates the use of projectability as a measure to see which conduction states might be more important.

Similarly to Agapito~\etal{}\cite{Agapito_PRB_93}, we choose as our localised functions the set of $N_\mathrm{PAO}$ pseudo-atomic orbitals (PAO)
$\phi_{Ilm}(\bfr)$ 
employed in the generation of the pseudopotentials, where $I$ is an index running over the atoms in the cell and $lm$ define the usual angular momentum quantum numbers. We then construct Bloch sums $\phi_{\mu\bfk}(\bfr)=\frac{1}{N_\mu}\sum_{\bfR}e^{-i\bfk\cdot\bfR}\phi_{\mu}(\bfr-\bfR)$, where $\mu=\{Ilm\}$ and $N_\mu$ is the number of lattice vectors $\bfR$ contained in the Born–von Karman cell
(which is equal to the number of $k$-points sampled in the BZ).
Finally, a Hilbert space $\mathcal{A}^{\bfk}$ at each $k$-point in the BZ is defined as the space spanned by the L\"owdin-orthogonalised functions $\widetilde{\phi}_{\mu\bfk}(\bfr)=\sum_{\nu} ({S^{\bfk}}^{-1/2})_{\mu\nu}\phi_{\nu\bfk}(\bfr)$, with $S^{\bfk}_{\mu\nu} = \braket{\phi_{\mu\bfk}(\bfr)\vert\phi_{\nu\bfk}(\bfr)}$, and $\mathcal{A}$ is given by the direct sum $\mathcal{A}= \bigoplus_{\bfk} \mathcal{A}^\bfk$.

The projectability of each Bloch state onto $\mathcal{A}$ is then defined as 
\begin{equation}
    \label{eq:projectability}
p_{n\bfk} = \sum_{I,l,m}|\braket{\psi_{n\bfk}\vert \phi_{Ilm}^{\bfk}}|^2,
\end{equation}
where $0\leq p_{n\bfk}\leq 1$. 
The projections $\braket{\psi_{n\bfk}\vert \phi_{Ilm}^{\bfk}}$ are computed straightforwardly using the {\tt projwfc.x} code from \QE{}.
In particular, for the pseudopotentials considered in this work, the number of valence electrons and the atomic orbitals included in the pseudopotential files may be found in Table~S1 in Supplementary Note 2.

As the first step of our protocol, we choose $J$ as the total number of projections $N_\mathrm{PAO}$ considered in the sum of Eq.~\eqref{eq:projectability}. Since we aim to interpolate the bands up to a given energy above the Fermi level, fixing $J=N_{\mathrm{PAO}}$ is a conservative choice, as the number of PAOs is usually greater or equal to the number of valence bands plus few conduction bands. 

\begin{figure}[tbp]
    \centering
    \includegraphics[width=8cm]{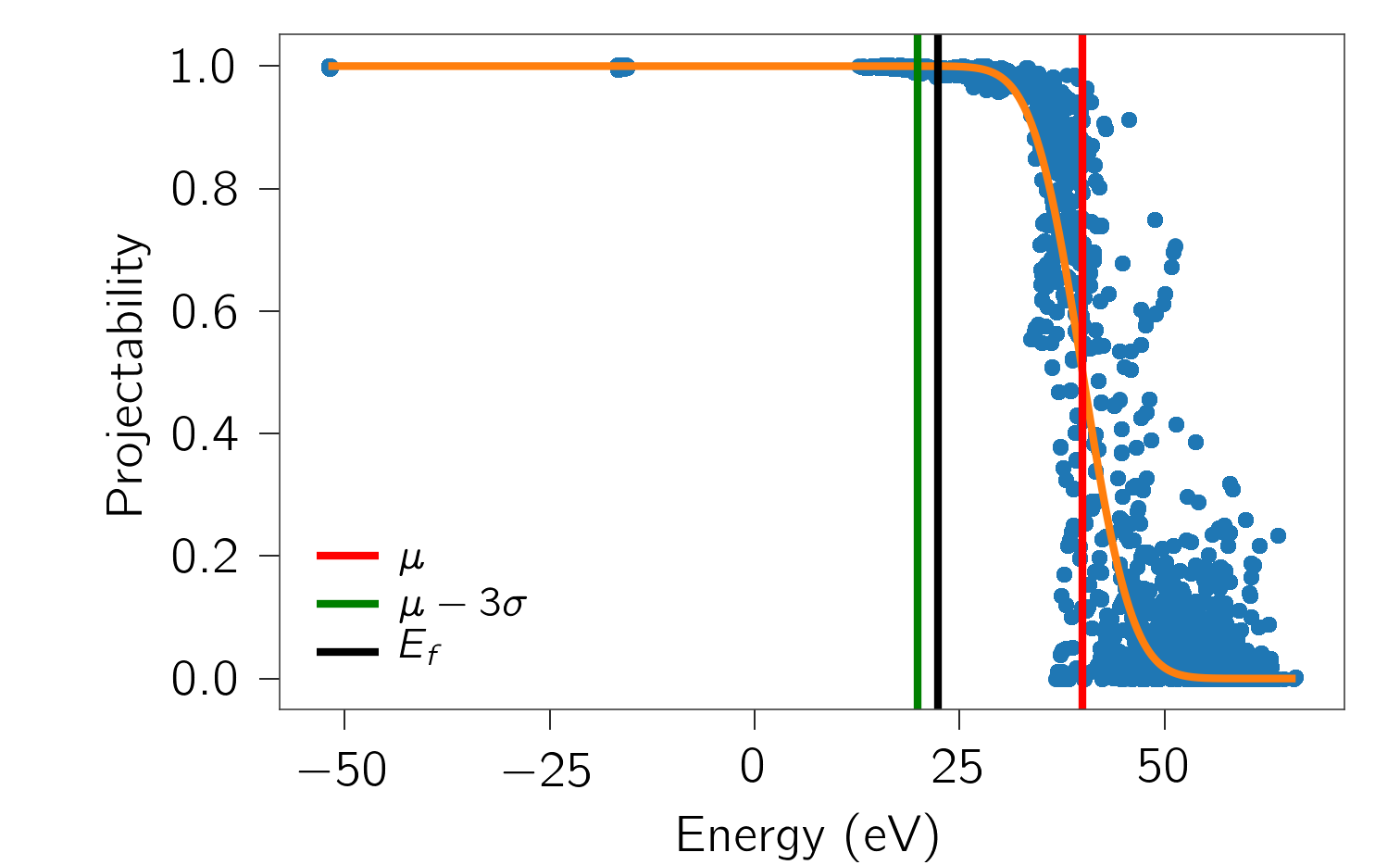}
    \caption{\titlefig{Projectability of the state $\ket{n\bfk}$ as a function of the corresponding energy $\varepsilon_{n\bfk}$ for tungsten} Each blue dot represents the projectability as defined in Eq.~\eqref{eq:projectability}. The yellow line shows the fitted complementary error function. The vertical red line represents the value of $\mu\tinysub{fit}$ while the vertical green line represents the optimal value of $\mu$, \ie{} $\mu\tinysub{opt}=\mu\tinysub{fit}-3\sigma\tinysub{fit}$. The value of the Fermi energy is also shown for reference (black line).}
    \label{fig:W_proj}
\end{figure}

We then use the values of the projectability to inform the choice of $\mu$ and $\sigma$. First, we plot the projectability for all Bloch states as a function of the corresponding band energy $\epsilon_{n\bfk}$, as shown in Fig.~\ref{fig:W_proj} (to illustrate the procedure, we show plots for one prototypical material, namely crystalline tungsten (W), but similar plots and trends also hold for the other materials considered in this work).
The general trend is that $p_{n\bfk}\sim1$ for low-energy states, which are well-represented by the chosen pseudo-atomic orbitals, and $p_{n\bfk}\sim0$ for high-energy states that originate either from free-electron-like states or from localised states with an orbital character that is not included in the set listed in Table~S1 in Supplementary Note 2, e.g., atomic orbitals with principal quantum number $n>3$ (i.e., more than two radial nodes).
We then fit this plot to a complementary error function as in Eq.~\eqref{eq:erfc-smearing}, extracting the two parameters $\mu_\text{fit}$ and $\sigma_\text{fit}$.
The core of our protocol lies on the actual choice of the $\mu$ and $\sigma$ parameters used as input for the SCDM method by setting
\begin{equation}
    \mu = \mu_\text{fit} - 3 \sigma_\text{fit}, \qquad 
    \sigma = \sigma_\text{fit}.
\end{equation}
Let us now motivate this choice. We observe that $\sigma_\text{fit}$ measures the typical energy spread of the bands originating from states within $\mathcal{A}$, and therefore is a good physical guess also for $\sigma$.
The naive choice $\mu = \mu_\text{fit}$, however, produces extremely poor interpolation of the bands for most of the materials that we have tested, see~\nameref{sec:entangled-hith-verification}. The reason is that 
it gives too great a weight in Eq.~\eqref{eq:gen_Pk} to states that have relatively small projectability ($p_{n\bfk} < 1$). 
As a consequence the SCDM algorithm might select columns representing better these states rather than those with projectability close to 1 at low energy, that are essential and physically relevant to include. In these cases, the corresponding band interpolation shows large oscillations and has large errors with respect to the DFT band structure in large portions of the BZ. We need therefore to choose a smaller value $\mu<\mu_\text{fit}$.
On the other hand, however, we note that the weight of states much above $\mu$ becomes numerically zero in Eq.~\eqref{eq:gen_Pk}, i.e., these states become completely unknown to the algorithm. Therefore, by choosing a too low value of $\mu$, i.e., discarding too many relevant states, the SCDM algorithm will fail because it will have to choose $J$ columns within a matrix of smaller rank.

We need, therefore, a general and automatic recipe for choosing an appropriate, intermediate value of $\mu$. Our choice $\mu = \mu_\text{fit} - \kappa \sigma_\text{fit}$ 
is guided by the consideration that states that start to have a significant component of their character outside $\mathcal{A}$ 
should be weighted in SCDM by Eq.~\eqref{eq:erfc-smearing} with a small weight, that is still though not exactly zero, giving the algorithm some freedom to pick up some of their character (for instance, states
at energy $\epsilon\geq\mu_\text{fit}$ have more than 50\% of their character outside $\mathcal{A}$ and are weighted in SCDM with a factor $\leq \frac{1}{2} \text{erfc}(\kappa)$ (e.g., $\kappa=3$ gives $\frac{1}{2}\text{erfc}(3)\approx 10^{-5}$).

\begin{figure*}[tbp]
    \centering
    \subfloat[]{\includegraphics[width=0.85\columnwidth]{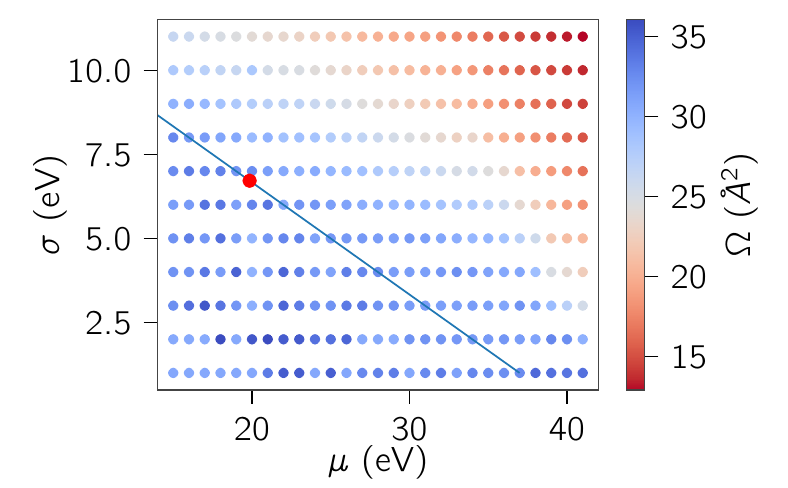}}
    \subfloat[]{\includegraphics[width=0.85\columnwidth]{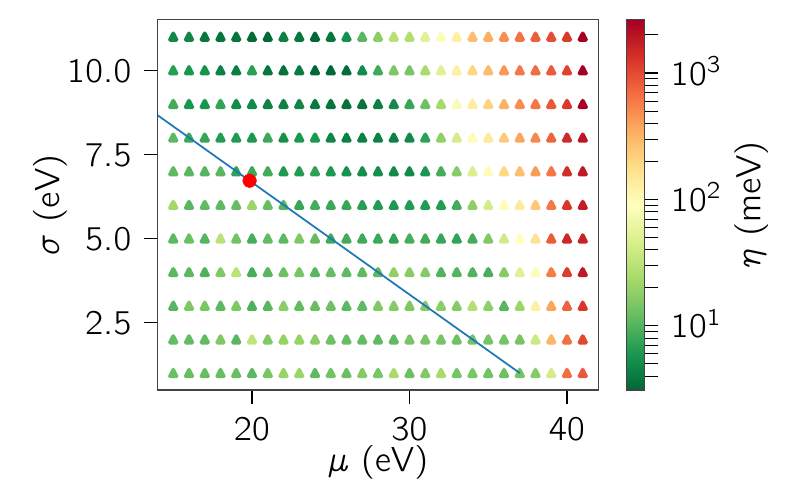}}\\
    \phantom{x}\hspace{1cm}\subfloat[]{\includegraphics[width=0.85\columnwidth]{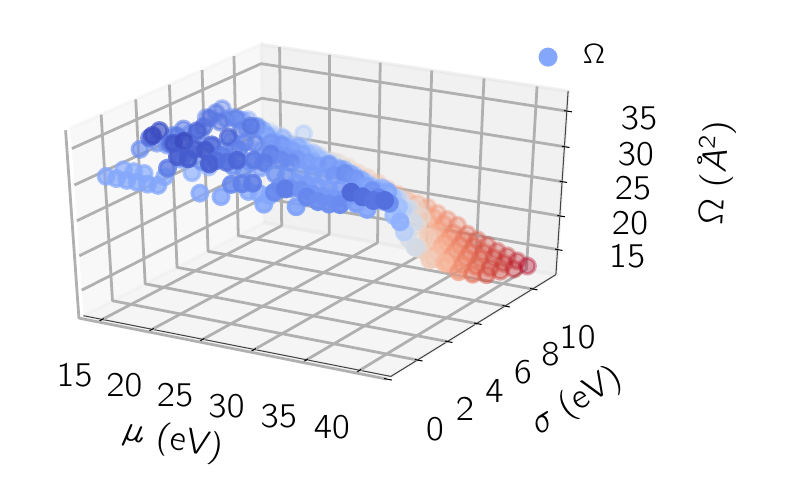}}
    \subfloat[]{\includegraphics[width=0.85\columnwidth]{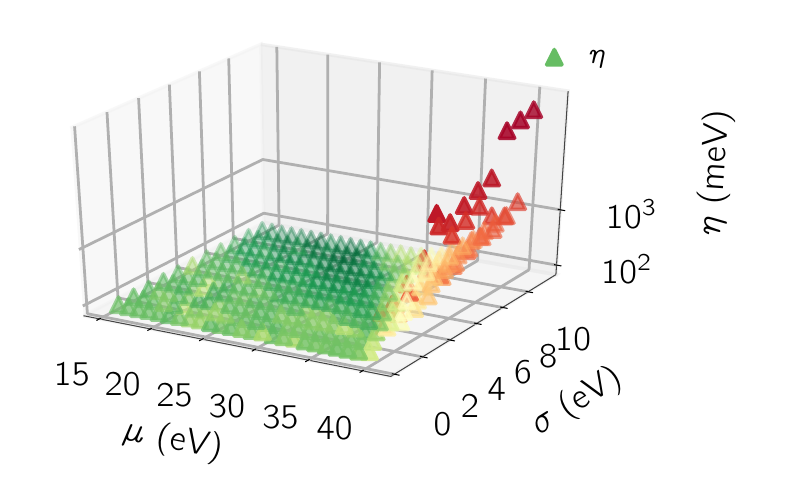}}
\caption{\titlefig{Assessment of the SCDM+MLWF method for tungsten (W) as a function of the SCDM input parameters $\mu$ and $\sigma$} Left panel: bands distance $\eta$. Right panel: total position spread $\Omega$.
    The blue line represent $\mu = \mu_\text{fit} - 3\sigma$ where the red dot corresponds to the choice dictated by our protocol  $\mu = \mu_\text{fit} - 3\sigma_\text{fit}$. The smearing function to compute $\eta$ has smearing $\tau = 0.1$~eV and $\nu$ is set to 1~eV above the Fermi energy.}
    \label{fig:W_all} 
\end{figure*}

In order to explain better our specific choice of $\kappa=3$, we consider again the case of tungsten for the SCDM+MLWF case and we report in Fig.~\ref{fig:W_all} the final total spread $\Omega$ (left-hand side) and the band distance $\eta$ (right-hand side) as a function of a range of values of $\mu$ and $\sigma$. In particular, in the case of entangled bands, we generalise the definition of $\eta$ by introducing a smearing, as we have mentioned in the previous section.
More specifically, we extend the definition of the distance between DFT and Wannier-interpolated bands to:
\begin{equation}
    \label{eta_met}
    \eta = \sqrt{\frac{\sum_{n\mathbf{k}}  \left(\varepsilon_{n\mathbf{k}}\tinysup{DFT}-\varepsilon_{n\mathbf{k}}\tinysup{Wan}  \right)^2\tilde{f}_{n\mathbf{k}}}
                      {\sum_{n\mathbf{k}} \tilde{f}_{n\mathbf{k}} } ,
    }
\end{equation}
where
\begin{equation}
\tilde{f}_{n\mathbf{k}} = \sqrt{f_{n\mathbf{k}}\tinysup{DFT}(\nu,\tau) f_{n\mathbf{k}}\tinysup{Wan}(\nu,\tau)},
\label{eq:fermi_dirac_weight}
\end{equation}
and $f\tinysup{DFT(Wan)}_{n\mathbf{k}}(\nu,\tau)$ is the Fermi-Dirac distribution for the state at energy $\varepsilon\tinysup{DFT(Wan)}_{n\mathbf{k}}$, $\nu$ is a fictitious chemical potential and $\tau$ is a smearing width computed on the direct ($\varepsilon_{n\mathbf{k}}\tinysup{DFT}$) and Wannier-interpolated ($\varepsilon_{n\mathbf{k}}\tinysup{Wan}$)  band structures. 
As in the~\nameref{sec4:isolated} section, we take into account the possibility that significant differences between band structures may occur only in sub-regions of the Brillouin zone or in small energy ranges, so we also compute
\begin{equation}
\eta^{\max} = \max_{n\mathbf{k}} \left(\tilde{f}_{n\mathbf{k}} \left|\varepsilon_{n\mathbf{k}}\tinysup{DFT} - \varepsilon_{n\mathbf{k}}\tinysup{Wan}\right|\right).
\end{equation}

In particular, the value of $\nu$ in $\widetilde{f}_{n\bfk}(\nu,\tau)$ is set to 1~eV above the Fermi energy and the smearing width $\tau$ is $0.1$~eV. In this way, only states up to slightly more than 1~eV above the Fermi level have a weight significantly different from zero when comparing band structures.
In both panels of Fig.~\ref{fig:W_all}, we also show the line representing $\mu=\mu_{\text{fit}}-3\sigma$ to discuss our choice of $\kappa=3$, as well as the point $(\mu_{\text{fit}} - 3\sigma_{\text{fit}}, \sigma_{\text{fit}})$ on this line. 
Our target is to have $\eta$ as small as possible, indicating a good interpolation of the band structure. As visible in Fig.~\ref{fig:W_all}, and as mentioned in the previous two paragraphs, large values of $\mu$ and $\sigma$ degrade significantly the quality of the band interpolation: in this case there are many states at high energy with a non-negligible weight and the QRCP, being a greedy algorithm, might select a subspace that better represents these states rather than the lowest energy states. 
It can also be seen that a larger $\mu$, which results in more states with higher weight, gives the SCDM algorithm more freedom in the choice of the subspace, which in turn results in a lower total spread $\Omega$ (at the expenses of a potentially worse interpolation).

On the other hand, also moving to the region of small $\mu$ and $\sigma$ is detrimental for the quality of the band interpolation (and partially also for the value of $\Omega$). Even if the values of $\eta$ in this region are not so large as in the region of large $\mu$ and $\sigma$, the quality of the interpolation is much less robust and both $\eta$ and $\Omega$ depend strongly on the precise values of the two parameters. In this case, we are discarding relevant states from the initial space used for the column selection of $F_\bfk \boldsymbol{\Psi}_\bfk$, therefore removing important information needed by the method for a good interpolation.

Our choice of $\kappa=3$, thus, together with $\sigma=\sigma_{\text{fit}}$, allows us to locate our choice of $(\mu, \sigma)$ in the intermediate region where $\eta$ is small and both $\eta$ and $\Omega$ are relatively insensitive to small variations of the two parameters. 
Ultimately, this specific choice for $\kappa$ will be justified and validated in our high-throughput study of~\nameref{sec:entangled-hith-verification}, where we show that the automated algorithm resulting from this choice is robust when tested on 200 chemically and structurally different materials, whose full list is available in Ref.~[\onlinecite{MaterialsCloudArchiveEntry}].

We also emphasise here that the choice of $\mu$ and $\sigma$ plays two different roles: the first is to give a relative weight to the states at the anchor point, namely $\Gamma$, that are used for the SCDM column selection; the second is to have a smooth dependence of the subspace as a function of $\bfk$, therefore resulting in a small $\Omega_I$. 

\subsection*{High-throughput verification}\label{sec:entangled-hith-verification}
In this section we present the results of the high-throughput calculations for the general case of 
200 materials that have been chosen to cover a large region of structural (12 different Bravais lattices) and chemical 
(67 different elements) space. The free parameters in the SCDM method have been chosen by the automatic procedure outlined in the previous section. The structure of this section parallels the one for isolated bands; in particular, we make use of the bands distance $\eta$ introduced in Eq.~\eqref{eta_met} to quantitatively assess the Wannier interpolation.
In the case of metals, we also need to appropriately select the value of the fictitious chemical potential $\nu$ and of the smearing width $\tau$ in the distribution $f_{n\bfk}(\nu,\tau)$ of  Eq.~\eqref{eq:fermi_dirac_weight} (the final values used in this work are reported in the previous section), in order for 
$\eta$ and $\eta^{\max}$ to be reliable measures for the interpolation quality of the bands of physical interest. 
Indeed, the Wannier-interpolated bands are not expected to reproduce accurately the dispersion of the DFT bands at high energies; and the energy up to which the Wannier-interpolated bands may be deemed to be accurate depends mainly on the number of target WFs $J$ which, in turn, is determined in our procedure by the number of PAOs in the pseudopotentials.
In most applications, however, the high-energy bands are not of interest;
therefore, $\nu$ and $\tau$ should be chosen so as to define a bands distance that only takes into account the relevant low-energy bands. For most practical applications, this means for states up to a small amount (usually a few eV) above the Fermi energy.

\begin{figure}[tbp]
    \centering
    \includegraphics[width=8cm]{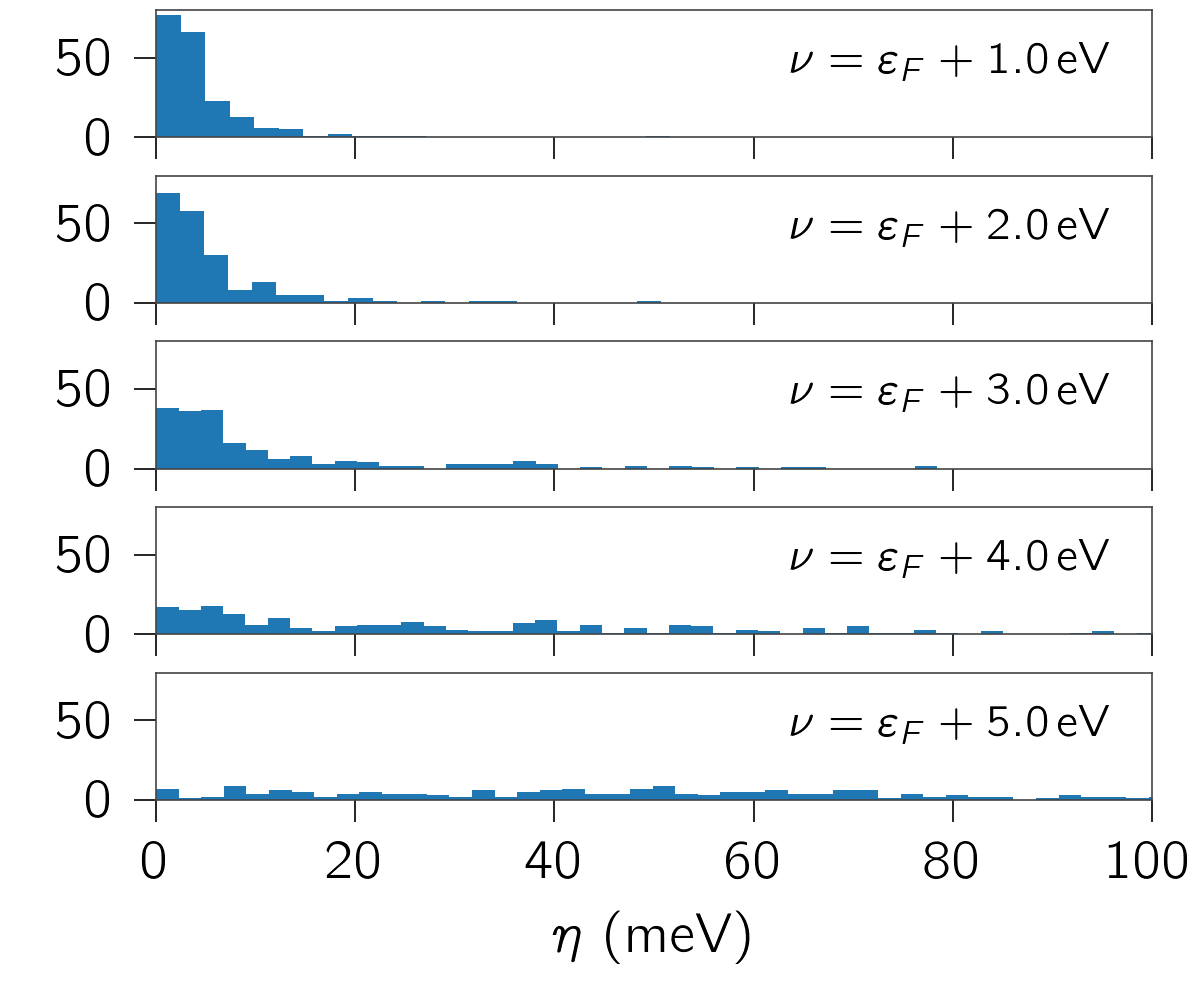}
    \caption{\titlefig{Distribution of the band distance $\eta$ for different values of the fictitious chemical potential $\nu$} The chemical potential is defined as $\nu=\varepsilon_{F} + \Delta$ ($\Delta=1,2,3,4,5$~eV) and the smearing $\tau$ in the Fermi-Dirac distribution is $0.1$~eV. All calculations have been performed with  a $k$-point spacing of $\rho_k = 0.2$ \angstrom$^{-1}$.}
    \label{fig:choice_of_nu}
\end{figure}

To verify up to which energy the interpolation is accurate (for the number of PAOs in the pseudopotentials chosen in this work, see Table~S1 in Supplementary Note 2) we show in Fig.~\ref{fig:choice_of_nu} the distribution of band distances for different values of $\nu=\varepsilon_{F} + \Delta$, with $\Delta=1,2,3,4,5$~eV, and $\tau$ fixed at 0.1~eV in order to have a smooth but sharp-edged Fermi-Dirac distribution. 
When $\nu$ is set at 4~eV or more above the Fermi energy ($\Delta \geq 4$~eV, bottom panels in Fig.~\ref{fig:choice_of_nu}), the distribution is very broad and with a long tail. In this case 
states much above the Fermi energy, where the Wannier interpolation does not reproduce any more the DFT band structure, are given a non-negligible weight $f_{n\bfk}$ which significantly increases the value of the band distance.
The distribution becomes much more narrow and closer to $\eta=0$~eV for $\Delta \leq 3$~eV; in particular, for $\nu=\varepsilon_{F}+1.0$~eV, 98\% of the materials have $\eta < 50$~meV. Since for many applications having a good interpolation up to 1~eV above the Fermi energy is sufficient, in the rest of this work we choose $\nu=\varepsilon_{F}+1.0$~eV (for entangled bands) as a reliable measure of the quality of the interpolation in the energy region of interest.

\begin{figure}[tbp]
    \centering
    \includegraphics[width=8cm]{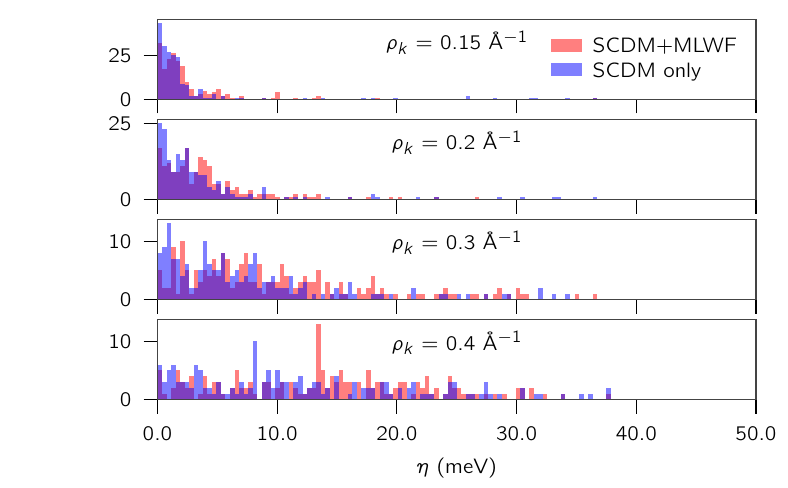}\\
    \includegraphics[width=8cm]{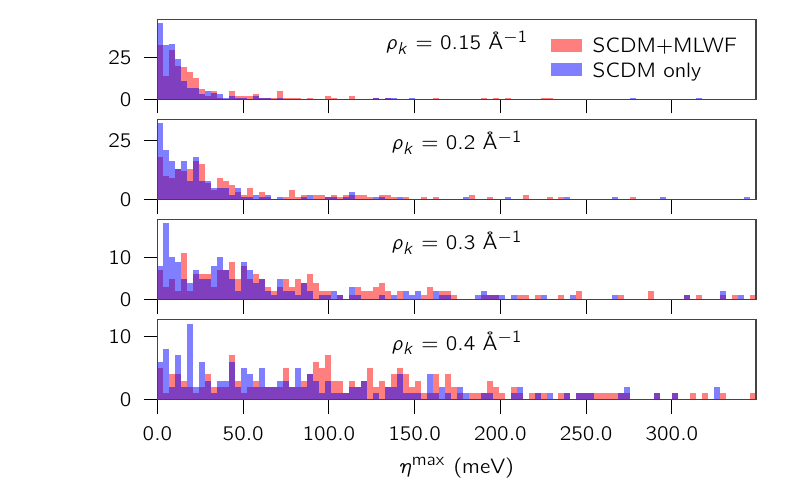}
    \caption{\titlefig{Average and max band distance for the valence and few conduction bands of 200 materials} Top (bottom) panel: histogram of average (max) band distance $\eta$ ($\eta^{\max}$) in meV using SCDM-only (blue) and SCDM+MLWF (red) obtained using four different $\mathbf{k}$-point grids with spacing $\rho_k$. The MLWF procedure slightly worsens the accuracy of the interpolation when compared to SCDM-only Wannier functions. The histograms focus on the most relevant interval and few outliers are not shown, in particular at $\rho_k = 0.2$\angstrom$^{-1}$ $98 \%$ (196/200) of the SCDM+MLWF bands and $99.5\%$ (199/200) of the SCDM-only bands exhibit  $\eta<50$ meV, while $98 \%$ (195/200) of the SCDM+MLWF bands and $94\%$ (188/200) of the SCDM-only bands exhibit $\eta^{\max}<350$~meV.}
    \label{fig_histo_met}
\end{figure}

As in the case of isolated bands, the first step is to study the effect of the $k$-point grid density on the interpolation, to fix the last free parameter in the calculations. 
As shown in Fig.~\ref{fig_histo_met}, a grid with spacing $\rho_k = 0.2$ \angstrom$^{-1}$ is typically sufficient to provide accurate interpolated band structures: in particular, $94\%$ of the materials (187/200) for SCDM-only and $97\%$  (193/200) for SCDM+MLWF show $\eta < 20$~meV, and $72\%$ (144/200) of the SCDM+MLWF bands and $79 \%$ (157/200) of the SCDM-only bands display $\eta < 5$~meV. Moreover, $\eta^{\max}$ follows a similar trend, with $72\%$ (143/200)  of the SCDM+MLWF bands and $82 \%$  (163/200) of the SCDM-only bands showing an $\eta^{\max} < 50$~meV, and $35\%$ (70/200)  of SCDM+MLWF bands and $52 \%$  (104/200) of the SCDM-only bands showing an $\eta^{\max} < 20$~meV, as shown in Fig.~\ref{fig_histo_met}. We therefore set $\rho_k$ to $0.2$\angstrom$^{-1}$ for further analysis in this section. 

\begin{figure*}[htbp]
    \centering
    \subfloat[W]{\includegraphics[width=0.95\columnwidth]{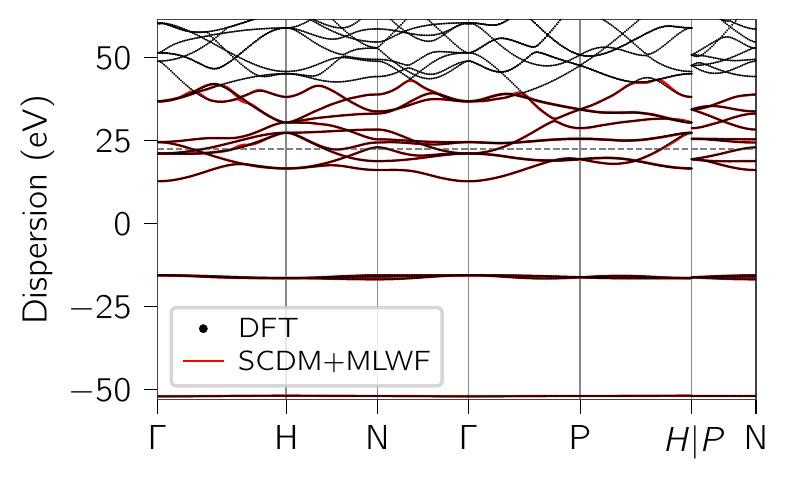}}
    \centering
    \subfloat[C$_3$Mg$_2$ ]{\includegraphics[width=0.95\columnwidth]{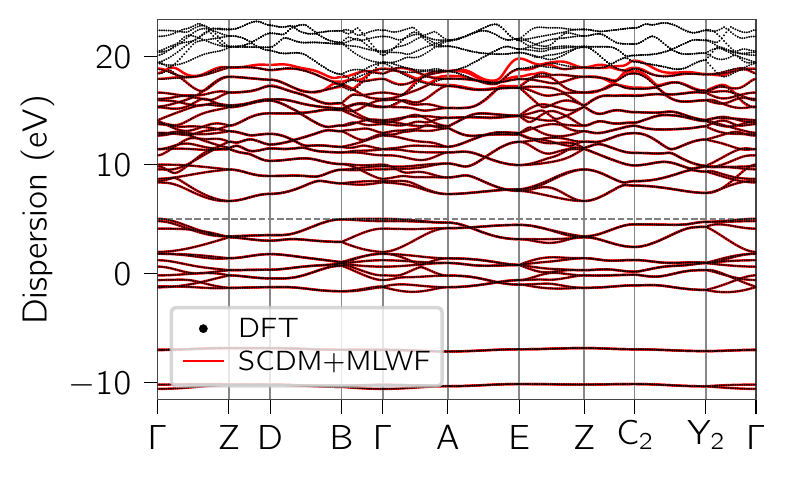}}
    \caption{\titlefig{Comparison between Wannier-interpolated valence bands plus few conduction bands and the full direct-DFT band structure} Wannier-interpolated bands are in solid red and full DFT bands are in solid black. Panel a, $\eta =20$~meV, $\eta^{\max} = 415$~meV, $\mu = 19.85$~eV and $\sigma =6.71$~eV) and C$_3$Mg$_2$ (panel b, $\eta = 2$~meV, $\eta^{\max} = 11$~meV, $\mu = 0.86$~eV and $\sigma = 5.63$~eV) using the MLWF procedure on SCDM projections and $\rho_k = 0.2$\angstrom$^{-1}$. Note that, while we show all Wannier-interpolated bands, the band distance $\eta$ considers only bands up to about $1$ eV above the Fermi level (see text).}
\label{fig_comparison_metals} 
\end{figure*}

\begin{figure}[tbp]
    \centering
    \includegraphics[width=8cm]{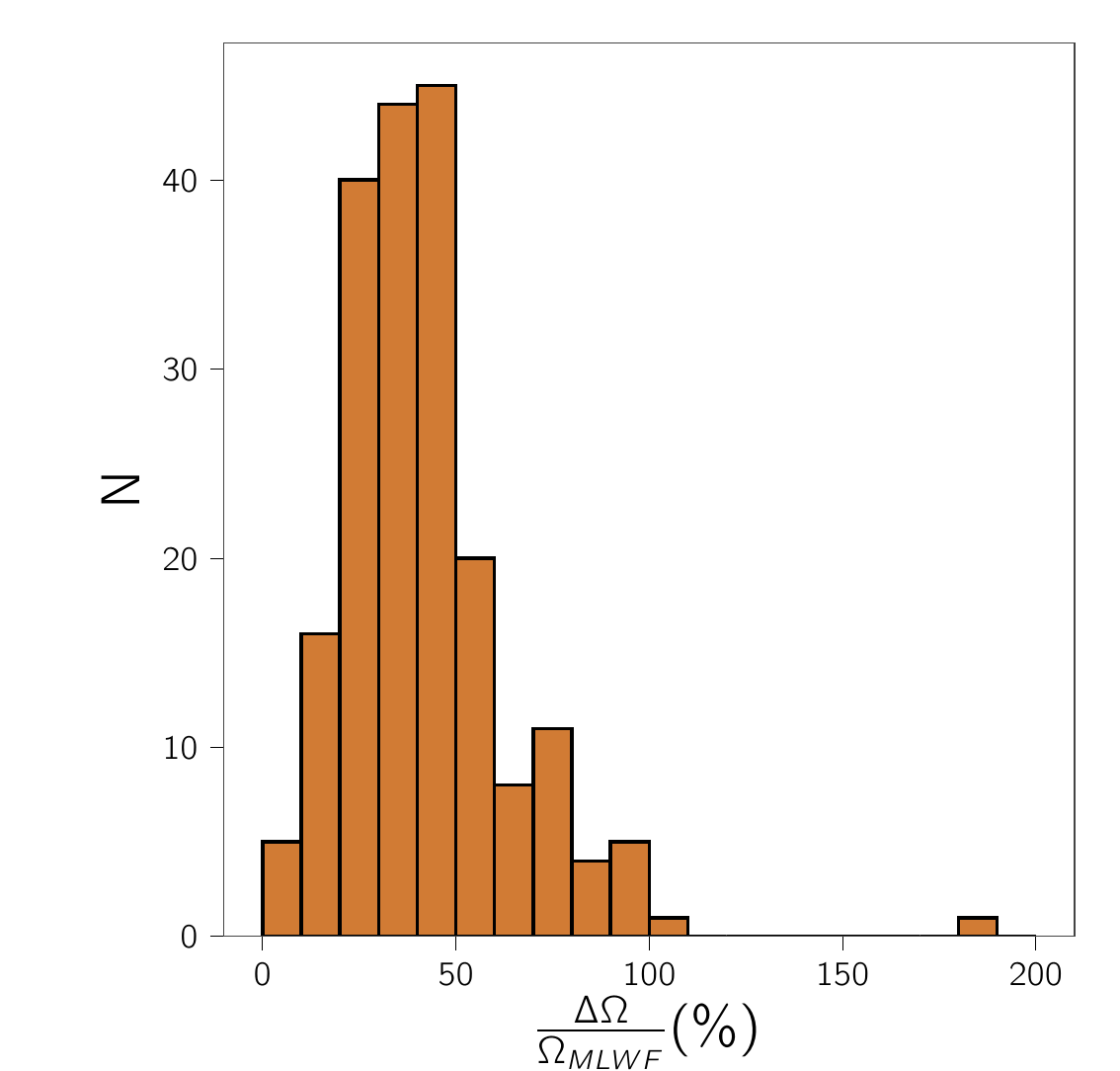}
    \caption{\titlefig{Average and max band distance for the valence and few conduction bands of 200 materials} Histogram of the relative variation of the total quadratic spread $\Omega$ before and after the MLWF procedure for the band structures of our set of 200 materials, obtained for $\rho_k = 0.2$\angstrom$^{-1}$. The SCDM+MLWF procedure provides Wannier functions that are substantially more localised with respect to SCDM-only, with a relative variation between $20-60 \%$ for most materials.}
    \label{fig_histo_spread_met}
\end{figure}

Fig.~\ref{fig_comparison_metals}a shows the Wannier-interpolated bands (red lines) for tungsten (W), a metallic system, and  Fig.~\ref{fig_comparison_metals}b shows the Wannier-interpolated valence bands plus few conduction bands (in red) for the insulator C$_3$Mg$_2$ (and these can be compared with Fig.~\ref{fig_comparison}b for the interpolation of the valence bands only).\\
Unlike the case of isolated bands, for entangled bands the MLWF procedure substantially increases the localisation of the resulting Wannier functions from SCDM projections, giving for instance a $\frac{\Delta \Omega }{\Omega^{\rm MLWF}} $ between $20-60\%$ for $75\%$ (149/200) of materials, with 30 materials showing a $60\%$ or more increase in $\frac{\Delta \Omega }{\Omega^{\rm MLWF}}$, see Fig. \ref{fig_histo_spread_met}. 

\begin{figure}[tbp]
    \centering
    \includegraphics[width=8cm]{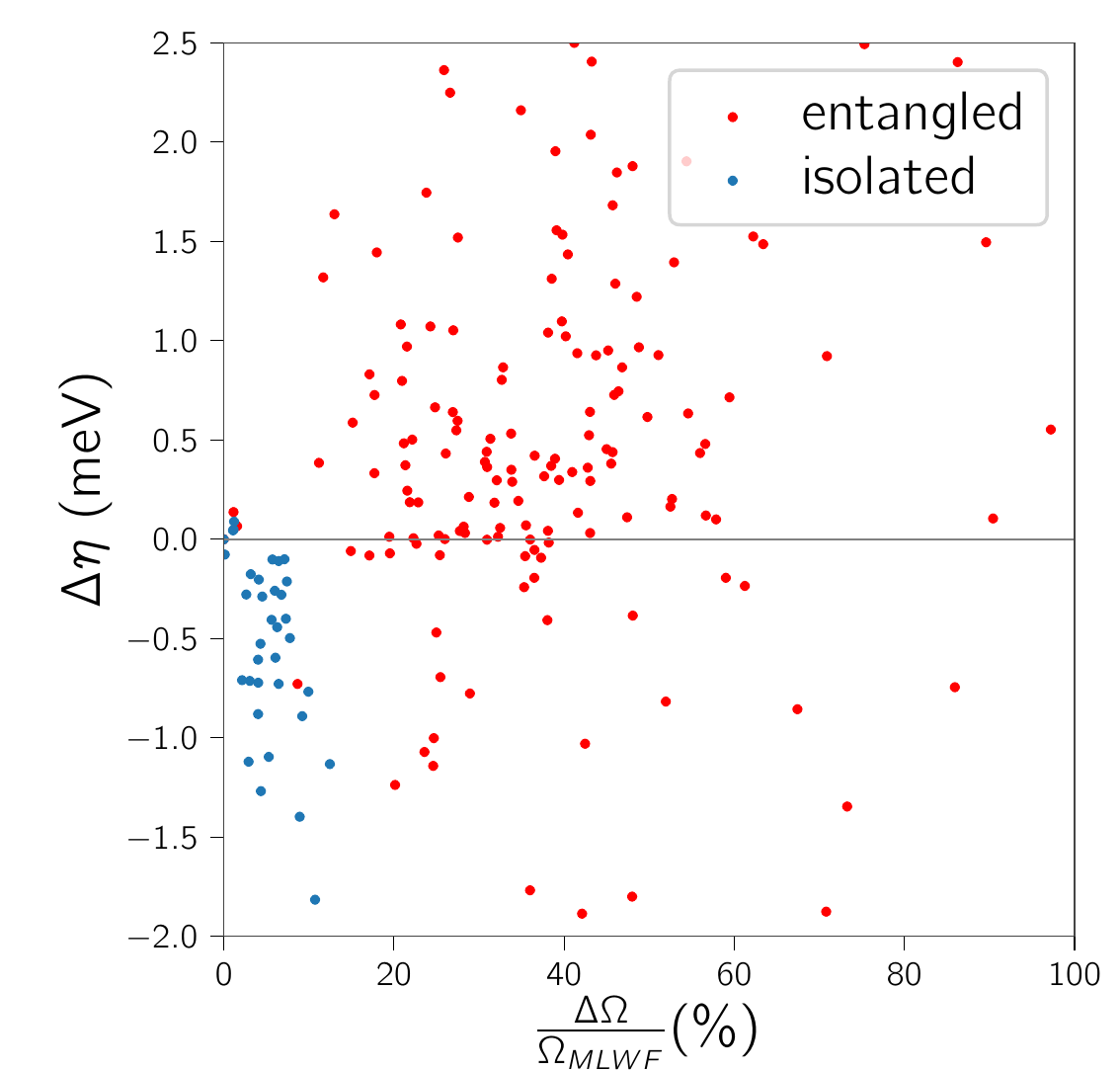}
    \caption{\titlefig{$\Delta \eta$  versus  $\Delta \Omega/\Omega^{\rm MLWF}$ scatter plot (valence and few conduction bands} The dataset consists of all 200+81 materials, with entangled bands (red dots, 148 out of 200 visible in the axes range) and with isolated bands (blue dots, 64 out of 81 visible) showing $\Delta \eta$  versus  $\Delta \Omega/\Omega^{\rm MLWF}$, that is the quantitative deviation between SCDM+MLWF and SCDM-only in terms of band structures and total spreads, respectively. Maximally-localising Wannier functions give potentially more accurate interpolated bands for valence bands only, whereas for entangled bands the trend is reversed.}
    \label{fig_scatter_met}
\end{figure}    

We now look at how the difference in spread due to the MLWF procedure correlates with the difference in the quality of the interpolated band structures. Although the correlation is not as strong as in the case of isolated bands, it can be seen (Fig.~\ref{fig_scatter_met}) that the trend is almost reversed: reducing the spread tends to worsen the quality of the band interpolation. In fact, the majority of systems ($71\%$, 142/200) show a positive change in $\Delta \eta$, meaning that SCDM-only provides better interpolation. 
The main reason behind this effect is that, in the selection of the optimal manifold $\mathcal{S}(\bfk)$, the SCDM algorithm might include contributions from higher energy states. The subsequent MLWF step does not use information on the target band structure. Therefore, 
while mixing the states via the $U$ matrix to minimise the spread, such spurious contributions can be distributed on the lower-energy states and, as a consequence, worsen the interpolation quality. However, we emphasise that in most cases, even when the MLWF algorithm increases the value of $\eta$, it does so only marginally: in 182 out of 200 systems (91\%) the MLWF scheme either increases $\eta$ by less than 5~meV or reduces it. More in detail, 163 out of these 182 materials show a variation $|\Delta\eta|$ within only 5.0~meV, and only one system among these exhibits $\eta^{\rm MLWF}>20$~meV. Moreover, for the remaining 19 (out of 182) systems the MLWF procedure improves the bands interpolation, notably yielding $\eta^{\rm MLWF}<20$~meV for all of them.
Finally, for the remaining 18 systems ($9\%$), the MLWF scheme worsens the results with $|\Delta \eta |>5$~meV and only in 6 cases the interpolation quality is quite poor ($\eta^{\rm MLWF} >20$ meV). In all these cases, a possible reason for failure might be related to the choice of columns in the SCDM algorithm, which is performed only at $\Gamma$ (see discussion in~\nameref{sec:scdm-k}), for materials where the relative order of electronic states at $\Gamma$ and at the BZ boundary is inverted. In this situation, spurious contributions might enter into the QR decomposition as discussed above.

We have presented an approach to generate a set of maximally localised Wannier functions in an automated way that has the advantage of being simple, robust and applicable also in the more general case of so-called entangled bands. 
The high sensitivity of iterative minimisation algorithms to the initial conditions, which was a long-standing problem in particular for the entangled-band case, is overcome by employing the selected columns of the density matrix\cite{DL_2015_SCDM,DL_2018_SIAM} (SCDM) algorithm to automatically choose the initial subspace.
For the Wannierisation of isolated bands, SCDM is a parameter-free method, whereas for entangled bands two real numbers $\mu$ and $\sigma$ must be specified, whose appropriate choice is critical for the success of the method, in addition to the target  dimensionality of the manifold to be described (i.e., the number of Wannier functions). We have proposed and validated a protocol to choose these parameters by leveraging information encoded in the projectability of the Bloch states on pseudo-atomic orbitals.
We found that the SCDM method works very well for band-structure interpolations, but does not perform as well for other kind of applications where, for instance, a specific symmetry character of the WFs is desirable.

To make the method available to any researcher, we have implemented the SCDM algorithm in \pwtowannier{}, part of the open-source \QE{} distribution, and added corresponding functionality to the open-source \Wannier{} code. 
We have also discussed how the full procedure is implemented as AiiDA\cite{Pizzi_AiiDA} workflows, encoding the knowledge that is needed to perform all steps (DFT simulations, selection of the parameters, Wannierisation) into an automated software. This enables MLWFs to be obtained and used to calculate material properties by providing the crystal structure of a material as the only input.
Furthermore, we are distributing publicly and freely all codes and 
workflows discussed in this work within a virtual machine\cite{MaterialsCloudArchiveEntry}
preconfigured with the open source codes AiiDA, \QE{} and 
\Wannier{}. This VM  allows anyone to explore and reproduce 
straightforwardly the present results without the need to install or 
configure anything, and without the need of implementing again workflows 
and algorithms, in the true spirit of Open Science. In addition, 
interested researchers are not constrained to re-run the calculations 
performed in this work, but can perform their own simulations, either 
with different parameters or on new materials. To the best of our 
knowledge, this is the first time that such level of reproducibility is 
offered accompanying a scientific paper in the field of DFT simulations.

We have demonstrated the robustness of the present approach by carrying out high-throughput calculations on a dataset of 200 bulk crystalline materials, of which 81 are insulators, spanning a wide chemical and structural space.
The main metric we used to assess the results is the so-called band distance\cite{sssp_paper}, quantifying the difference between the Wannier-interpolated band structures and the corresponding direct DFT band structures. In particular, we obtain excellent interpolations: for entangled bands, 97\% of the materials show an average bands distance $\eta < 20$~meV and 72\% show $\eta < 5$~meV. 
For the insulating subset, when limiting to valence bands only, 93\% show $\eta < 2$~meV.

We believe that this work is a significant step forward towards completely automated high-throughput calculations of advanced materials properties exploiting Wannier functions.

\section*{Methods}\label{sec6:implementation}
AiiDA~\cite{Pizzi_AiiDA} is a python materials' informatics platform to automate, manage and coordinate simulations and workflows, and to encourage sharing of both the resulting data and the workflow codes used to generate them. While general in its design, its plugins cover many materials science codes, including \QE{}\cite{aiida-qe} and \Wannier{}\cite{aiida-w90}. 

Our implementation of the SCDM method inside the open-source code \QE{} makes it available to any researcher. Moreover, our protocol for the choice of the SCDM parameters discussed in~\nameref{sec:mu-sigma-protocol} describes an effective procedure to automatically compute the Wannier functions of any material.
However, the actual computation starting only from the crystal coordinates is non-trivial. The choice of numerical parameters (cutoffs, $k-$point grid density, convergence parameters) requires some prior knowledge and experience. Moreover, the full simulation for each material involves a complex sequence of steps, requiring a user to run over 10 different executables.
Therefore, we have implemented the full procedure as AiiDA workflows, making it thus possible to repeat seamlessly the calculations for many different materials with minimal effort.

\begin{figure*}[htbp]
\begin{adjustbox}{angle=0}
  \includegraphics[height=12cm]{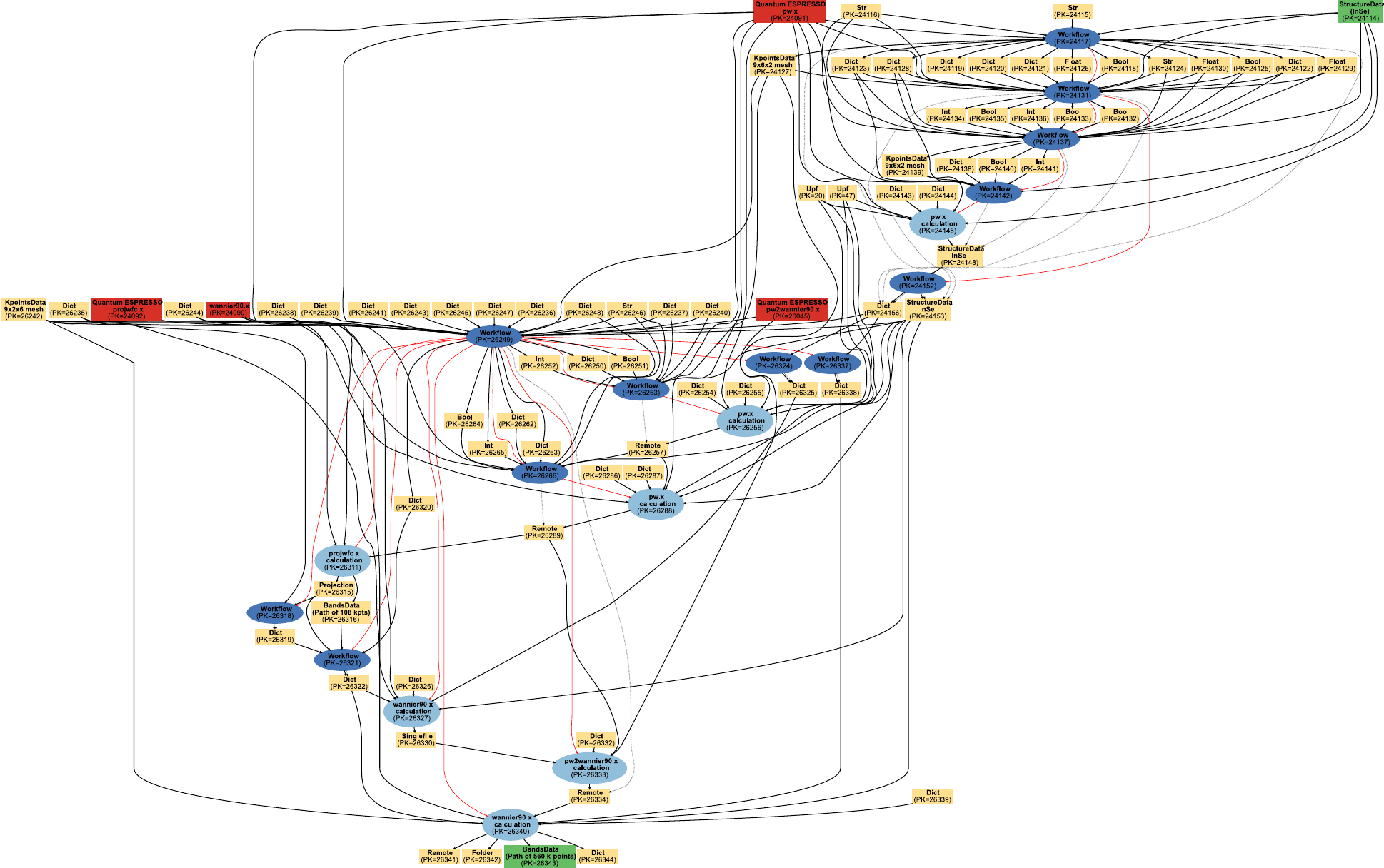}
\end{adjustbox}
\caption{\titlefig{Provenance graph automatically generated by AiiDA} The graph has been generated by running a \Wannier{} calculation using \QE{} as the input code for an InSe crystal, top green node (link labels have been removed for clarity). Red arrows represent caller-called relationships between a workflow and a subworkflow or a calculation; continuous lines connect calculations on a supercomputer (light blue ellipses) to their inputs and to the outputs they create, while dotted lines connect workflows (dark blue ellipses) to the data they return. Other data nodes are represented as yellow rectangles. In the top-right part of the graph, a set of workflows drive variable-cell relaxations of the initial structure via \QE{}; the central part contains the self-consistent, non-self-consistent and band-structure \QE{} calculations; in the bottom-left part are located the calculations computing the projection of the wavefunctions on a localised atomic basis set. At the bottom of the graph, we can find the \Wannier{} calculation, producing a set of output nodes that includes the Wannier-interpolated band structure (bottom green node).}
\label{fig:aiida-graph}
\end{figure*}

Furthermore, AiiDA keeps track of the provenance of the data generated in the simulations in a fully automated way, in the form of a directed graph (see Fig.~\ref{fig:aiida-graph} for an example of the provenance tracked for one material), where nodes can be calculations, workflows or data.
This means that any researcher accessing the AiiDA database can inspect not only the final data, but also explore which calculation generated it, its relevant (raw and parsed) outputs and the complete set of its input parameters, and see how these input data were, in turn, obtained as output of previous calculations, traversing the graph up to the original input crystal structure.

The AiiDA workflows that we have written start by calling existing subworkflows available in the AiiDA-quantumespresso\cite{aiida-qe} plug-in that, given a crystal structure, perform a variable-cell atomic relaxation to obtain the converged DFT charge density. These workflows also contain useful heuristics and recovery mechanisms to reach convergence in case of common problems (e.g., by changing the diagonalisation algorithm) as well as automatic selection of parameters, including pseudopotentials and cutoffs from the SSSP library~\cite{sssp_paper}.
Once the charge density is computed, the workflow first standardises the cell using the symmetry-detection library {\tt spglib}~\cite{spglib_arxiv} and the {\tt seekpath}~\cite{seekpath_2017} library that, in addition, provide a standardised band-structure path. Then, it proceeds along two parallel branches: on one side, it computes the DFT band structure along the suggested path. In parallel, it computes the Wannier functions: if first computes wavefunctions on a full uniform grid using a non-self-consistent \QE{} calculation, and then computes the PDOS, the projectabilities, and fits them to obtain the $\mu$ and $\sigma$ parameters for the SCDM. Using these data, it prepares the \Wannier{} input file and runs it in pre-processing mode to generate the input file needed by the code interfacing \QE{} with \Wannier{} (\pwtowannier{}). The latter is then run to compute quantities needed by \Wannier{}, including the $\mathbf{A}^{(\bfk)}$ matrices obtained with the SCDM method. Finally, the workflow drives the execution of \Wannier{} to compute the (maximally-localised) Wannier functions and produce the output quantities of interest (spreads, interpolated band structure on the same path of the DFT code, plots of the Wannier functions, etc.).

In an effort to improve the verification and dissemination of computational
results, and in order to make the present work available to all, 
we are distributing all codes and workflows discussed here 
within a preconfigured virtual machine (VM)\cite{MaterialsCloudArchiveEntry} based on the Quantum 
Mobile VM available on the Materials Cloud\cite{MaterialsCloudQuantumMobile}. The relevant
quantum codes (\QE{}, \Wannier{}) and the informatics'
platform AiiDA come pre-installed and configured in the VM, ready to run 
through the workflows described above. A simple README file guides new 
users in the installation of the VM and in the execution of the 
workflow, to compute---with essentially no user intervention---the 
interpolated band structure of a material of choice.


\section*{Data availability}\label{sec7:data}
All data generated for this work can be obtained by downloading the publicly available
Virtual Machine (VM) on the Materials Cloud (doi:\href{https://doi.org/10.24435/materialscloud:2019.0044/v2}{10.24435/materialscloud:2019.0044/v2}).
The VM contains the AiiDA workflow, the structures of the $\sim$ 200 materials (in
XSF format) and  the  simulation  codes  (\QE{}  and  \Wannier{}).   The  latter  have  been
pre-installed  and,  once  configured,  the  VM  is  ready  to  be  used.   Inside,  a  README  file
explains in detail how to retrieve all data.  In addition, the VM contains also the Ansible
scripts to regenerate the VM from scratch.

\section*{Code availability}\label{sec8:codes} 
All codes used for this work are open-source and hence available to any researcher.
In particular the latest stable version of
\Wannier{} can be downloaded at:

\href{http://www.wannier.org/download}{http://www.wannier.org/download}.

The latest stable version of
\QE{}
can be found at:

\href{https://www.quantum-espresso.org/download}{https://www.quantum-espresso.org/download}.

Likewise, for the AiiDA code the latest stable version can be found at:

\href{http://www.aiida.net/download}{http://www.aiida.net/download}.

\acknowledgements
V.V. acknowledges support from the European Union's Horizon 2020 research and innovation program under grant agreement no. 676531 (project E-CAM). G.P., A.M. and N.M. acknowledge support by the NCCR MARVEL of the Swiss National Science Foundation and the European Union's Centre of Excellence MaX ``Materials design at the Exascale'' (grant no.~824143). G.P., A.M. and N. M. acknowledge PRACE for awarding us simulation time on Piz Daint at CSCS (project ID 2016153543) and Marconi at CINECA (project ID 2016163963). V.V. and A.A.M. acknowledge support from the Thomas Young Centre under grant TYC-101. 
J.R.Y. is grateful for computational support from the UK national high performance computing service, ARCHER, for which access was obtained via the UKCP consortium and funded by EPSRC grant ref EP/P022561/1.
V.V. acknowledges Prof. Mike Payne for support, and Prof. Lin Lin and Dr. Anil Damle for useful discussions. G.P. acknowledges Dr. Francesco
Aquilante for useful discussions. A.M. acknowledges Prof. Ivo Souza for useful comments on the manuscript. We acknowledge Norma Rivano for testing the virtual machine and Dr. Sebastiaan P. Huber for the implementation of the AiiDA--\QE{} workflow for geometry relaxations.

\section*{\textbf{Competing interests}:} The authors declare no competing Financial or non-Financial

\section*{Author contributions}\label{sec9:contributions} 
V.V.  implemented  and  tested  the  SCDM  method on selected materials,  G.P.  and  A.M.  developed  the automation protocols and the workflows, run the high-throughput simulations and generated the Virtual Machine.  N.M., A.A.M. and J.R.Y supervised the project. All authors analysed the results and contributed to writing the manuscript.

\section*{Additional information}\label{sec10:info} 
\paragraph*{\textbf{Supplementary material}} accompanies the paper on the npj Computational Materials website
interests.

\bibliographystyle{naturemag}
\bibliography{biblio_SCDM} 

\end{document}


\title{Supplementary Information for:\\Automated high-throughput Wannierisation}

\author{Valerio Vitale}
\altaffiliation{\corr}
 \affiliation{Cavendish Laboratory, Department of Physics, University of Cambridge, 19 JJ Thomson Avenue Cambridge UK}
\affiliation{Departments of Materials and Physics, and the Thomas Young Centre for Theory and Simulation of Materials, Imperial College London, London SW7 2AZ, UK}

\author{Giovanni Pizzi}
\affiliation{Theory and Simulation of Materials (THEOS) and National Centre for Computational Design and Discovery of Novel Materials (MARVEL), \'Ecole Polytechnique F\'ed\'erale de Lausanne, Lausanne, Switzerland}

\author{Antimo Marrazzo}
\affiliation{Theory and Simulation of Materials (THEOS) and National Centre for Computational Design and Discovery of Novel Materials (MARVEL), \'Ecole Polytechnique F\'ed\'erale de Lausanne, Lausanne, Switzerland}

\author{Jonathan R. Yates}
\affiliation{Department of Materials, University of Oxford, Parks Road, Oxford OX1 3PH, UK}

\author{Nicola Marzari}
\affiliation{Theory and Simulation of Materials (THEOS) and National Centre for Computational Design and Discovery of Novel Materials (MARVEL), \'Ecole Polytechnique F\'ed\'erale de Lausanne, Lausanne, Switzerland}

\author{Arash A. Mostofi}
\affiliation{Departments of Materials and Physics, and the Thomas Young Centre for Theory and Simulation of Materials, Imperial College London, London SW7 2AZ, UK}
\date{\today}

\maketitle

\section*{Supplementary Methods}
\subsection*{SCDM implementation in \QE{}}\label{ssec6:scdm_implementation}
To implement the SCDM method one needs the wavefunctions from the {\it ab initio} code represented on a real space grid, see Eq.~(14) in the main text. 
Since these are not directly accessible to \Wannier, we decided to implement the method in one of the open-source DFT-to-\Wannier{} interface packages available. In particular, we have chosen the \pwtowannier{} FORTRAN code, distributed with the open-source \QE{} suite\cite{giannozzi_qe_2017}. Our SCDM implementation is available since the v6.3 release of \QE{}.
It includes the extension of the method to $k$-points and to entangled bands, and it is parallelised using the Message Passing Interface (MPI).

To compute the $A^{(\bfk)}_{mn}$ projection matrices using SCDM,
the {\tt auto\_projections}
keyword must be set to {\tt .true.} in the \Wannier{} input file. 
In addition, the following keywords should be defined in the \pwtowannier{} input file: {\tt scdm\_proj}, {\tt scdm\_entanglement}, {\tt scdm\_mu} and {\tt scdm\_sigma}.
In particular, {\tt scdm\_proj} is a boolean flag to enable the SCDM method. {\tt scdm\_entanglement} is a string defining the functional form of the $f(\varepsilon)$ function in Eq.~(15) in the main text. 
 In the cases described in this paper, the value is either {\tt isolated} (isolated bands) or {\tt erfc} (entangled bands with the $f(\varepsilon)$ of Eq.~(16) in the main text). 
An additional choice we implemented is {\tt gaussian}, see Ref.~[\onlinecite{DL_2018_SIAM}] for its functional form. Finally, {\tt scdm\_mu} and {\tt scdm\_sigma} (not needed if {\tt scdm\_entanglement} is {\tt isolated}) define, respectively, the values of $\mu$ and $\sigma$ (in eV) in Eq.~(16) in the main text. 

In \pwtowannier, the QRCP factorisation of the $\Psi^\dag_{\bfk=\Gamma}$ matrix is obtained through the \myfont{LAPACK} routine {\tt ZGEQP3}. Presently the factorisation is performed on a single MPI process (since {\tt ZGEQP3} is not available in the parallel ScaLAPACK routines) and the resulting permutation matrix $\Pi$ is broadcast to all processes. 
After the computation of the non-orthogonal SCDM functions, a L\"owdin orthogonalisation is performed. This step is not needed when providing the $A^{(\bfk)}_{mn}$ matrices to \Wannier{}, since the same orthogonalisation is performed by the code before the start of the minimisation. However, having the orthogonalisation step also in the \pwtowannier{} interface
allows users to directly employ the SCDM functions without further processing, if needed.    

As a final note, we emphasise that when ultrasoft pseudopotentials are employed, 
the $\psink{n}$ wavefunctions satisfy a generalised orthogonality condition with a non-trivial metric $\hat{S}$ being a function of the core augmentation charges\cite{US_Vanderbilt}. In this case the $\unk$ stored by \QE{} are not orthonormal, resulting in $\Psi$ being non unitary. However, in practice this usually has only a marginal effect on the results. Indeed, as we have shown, the algorithm manages to find good Wannier functions also when employing ultrasoft pseudopotentials and therefore no adaptation has been applied for the ultrasoft case. 

\subsection*{\label{appendix:QRCP-defs}Properties of the QRCP factorisation}
We recall in this section the properties of the $Q$, $R$, and $\Pi$ matrices obtained from a QRCP decomposition, in the general case where the matrix to decompose is rectangular.
For definiteness, we consider the decomposition of a rectangular $\Psi^\dagger$ matrix of shape $J\times n_G$.

The QRCP decomposition can be written as:
\begin{equation}
\Psi^\dagger \Pi = Q R
\end{equation}
where the matrices have the following properties:
\begin{enumerate}
    \item $Q$ is a $J\times J$ unitary matrix, i.e., it has orthonormal columns: $Q^\dagger Q=\boldsymbol 1_J$;
    \item $\Pi$ is a $n_G\times n_G$ permutation matrix (permuting the columns of $\Psi^\dagger$);
    \item $R$ is an upper-triangular rectangular matrix of shape $J\times n_G$, with diagonal elements sorted with decreasing absolute value: $|R_{11}| \geq |R_{22}| \geq \ldots \geq |R_{JJ}|$ (this order is ensured thanks to the action of the $\Pi$ matrix).
\end{enumerate}

\subsection*{\label{appendix:qrcp-on-factorised-matrix}QRCP column selection of $P$ from the column selection of $\Psi^\dagger$}
\emph{We consider a $n_G\times n_G$ matrix $P$ that can be written in the following form $P=\Psi\Psi^\dagger$, with $\Psi$ being a $n_G\times J$ matrix ($J<n_G$) with orthonormal columns, i.e. $\Psi^\dagger\Psi = \boldsymbol{1}_J$.
We want to show that if we consider the following QRCP decomposition for $\Psi^\dagger$:
\begin{equation}
    \label{eq:psidagger-decomp}
    \Psi^\dagger \Pi = Q R,
\end{equation}
then we can construct a QRCP decomposition for $P$ having the same permutation matrix $\Pi$:
\begin{equation}
    \label{eq:P-decomp}
    P \Pi = Q' R'.
\end{equation}
}

Let us start by multiplying Eq.~\eqref{eq:psidagger-decomp} on the left by $\Psi$:
\begin{equation}
    \label{eq:partial-proof-B2-1}
    P\Pi \equiv \Psi\Psi^\dagger \Pi = (\Psi Q) R \equiv Q' R,
\end{equation}
where we have defined $Q'\equiv\Psi Q$.

Let us first verify that $Q'$ has orthonormal columns:
\begin{equation}
    (Q')^\dagger Q' = (Q^\dagger \Psi^\dagger) (\Psi Q) = Q^\dagger Q = \boldsymbol{1}_J,
\end{equation}
where we have used the orthonormality of the columns of $\Psi$ (by hypothesis) and of $Q$ (since it is the output of a QRCP algorithm, see point 1 in~\nameref{appendix:QRCP-defs}).

Let us now define the following $n_G\times n_G$ matrices:
\begin{equation}
Q'' \equiv \left(\begin{array}{c|c}
Q' & \tilde Q
\end{array}\right), \qquad R'' \equiv 
\left(\begin{array}{c}
    R \\ \hline \boldsymbol 0_{(n_G-J)\times n_G}
\end{array}\right),
\end{equation}
where $\left(\begin{array}{c|c} A & B \end{array}\right)$ means horizontal concatenation of matrix $A$ with matrix $B$, where $A$ and $B$ must have the same number of rows, and $\left(\begin{array}{c} A \\ \hline B \end{array}\right)$ means vertical concatenation ($A$ and $B$ must have same number of columns). the additional columns $\tilde Q$ of $Q''$ are chosen to complete the columns of $Q'$ to an orthonormal basis of $\mathbb{R}^{n_G}$ (always possible) and $R''$ extends $R$ with $(n_G-J)$ additional rows of zeros.

We want now to prove that $P\Pi = Q''R''$ is a 
QRCP decomposition of $P$. Indeed, by multiplying by blocks the two matrices $Q''$ and $R''$, we get $Q''R'' = Q'R + \Tilde Q \boldsymbol 0 = Q'R = P\Pi$ by virtue of Eq.~\eqref{eq:partial-proof-B2-1}. Moreover, $Q''$ is a unitary matrix by construction, and $R''$ is clearly an upper-triangular matrix since $R$ is according to point 3 of \nameref{appendix:QRCP-defs}, and the diagonal elements are still sorted in decreasing magnitude order since the additional elements are all zero. Therefore, we have shown that the same permutation matrix $\Pi$ obtained by applying the QRCP to $\Psi^\dagger$ is a valid QRCP permutation matrix also for $P$. 

A different, equivalent approach to show the same result is to observe that the (complex) scalar product ${\bf v}_1\cdot{\bf v}_2 \equiv ({\bf v}_1^*)^T {\bf v}_2$ between columns of $P$ is the same as the scalar product of the columns of $\Psi^\dagger$.
Indeed, we first note that, as it can be easily proven from its explicit expression Eq.~(12) in the main text,
 $P$ is a projector and it holds that $P^2=P$ and $P^\dagger = P$. Therefore, we have $P^\dagger P = P$. But the elements of $P^\dagger P$ are nothing else than the scalar products of the columns of $P$, and therefore $P_{ij} = {\bf p}_i\cdot {\bf p}_j$, with ${\bf p}_i$ indicating the $i$-th column of $P$.
At the same time, from the definition of $P=\Psi\Psi^\dagger = (\Psi^\dagger)^\dagger \Psi^\dagger$ we immediately notice that the elements of $P$ are also the scalar products of the columns of $\Psi^\dagger$, i.e. the complex conjugate $\psi^*_i$ of the wavefunctions of the system, proving our statement that 
\begin{equation}
P_{ij} = {\bf p}_i\cdot{\bf p}_j = (\braket{\psi_i|\psi_j})^*.
\label{eqn:scalar-product-P-psi}
\end{equation}

\subsection*{\label{appendix:QRCP-looks-for-most-orthogonal}Geometrical interpretation of the column selection in the QRCP algorithm}
QRCP is a greedy algorithm, where the $\Pi$ matrix is constructed by picking the columns one by one to obtain the condition $|R_{11}| \geq |R_{22}| \geq \ldots \geq |R_{JJ}|$.
In the case of the QRCP decomposition of a $P$ matrix, the first column to be picked $({\bf p}_1)_i \equiv (P\Pi)_{i1}$ is chosen as the one with largest norm. This can be easily proven by noting that
\begin{equation*}
    (P\Pi)_{i1} = (QR)_{i1} = Q_{i1} R_{11},
\end{equation*}
because of the triangular form of $R$. Moreover, since the columns of $Q$ have unit norm, then $\|{\bf p}_1\| = |R_{11}|$ by construction (see point 3 of~\nameref{appendix:QRCP-defs}) is the largest possible.

More generally, the $j$-th column ${\bf p}_j$ is chosen to maximise the norm of the  component ${\bf p}_j^\perp$ orthogonal to the subspace $\mathcal S_{j-1}$ spanned by the previous $(j-1)$ columns. To prove this, let us first write ${\bf p}_j = {\bf p}_j^{\|} + {\bf p}_j^\perp$, where ${\bf p}_j^{\|}$ is the projection of ${\bf p}_j$ within $\mathcal S_{j-1}$.
We first note (again due to the triangular form of $R$) that in general the first $j$ columns of $Q$ also span the space $\mathcal S_j$ and, moreover, they are a orthonormal basis set for $\mathcal S_j$ since the $Q$ columns are orthonormal. Furthermore ${\bf p}_j$ is, by definition, in the $\mathcal S_j$ subspace.
Therefore, we can write the $j$-th column in this basis set of $\mathcal S_j$ as
\begin{equation*}
({\bf p}_j)_i = (P\Pi)_{ij} = (QR)_{ij} = \sum_{m=1}^j Q_{im}R_{mj},
\end{equation*}
and, thanks to the orthonormality of the $\{{\bf q}_m\}$ basis (${\bf q}_i$ being the $i$-th column of $Q$), we have 
\begin{equation*}
\left({\bf p}_j^{\|}\right)_i = \sum_{m=1}^{j-1} Q_{im}R_{mj}, \quad 
\left({\bf p}_j^{\perp}\right)_i = Q_{ij}R_{jj},
\end{equation*}
or equivalently in vector form ${\bf p}_j^{\perp} = {\bf q}_j R_{jj}$.

Therefore, the norm of this orthogonal component is simply $\|{\bf p}_j^{\perp}\| = |R_{jj}|$ that, again, is chosen by the algorithm to have maximal value (in order to have decreasing diagonal elements of $R$), therefore proving our intuitive explanation of the QRCP  column selection.

To give a more physical interpretation of the column selection in terms of the charge density or wavefunctions, we observe that from Eq.~\eqref{eqn:scalar-product-P-psi} we know that the square modulus of the $i$-th column of $P$ is $\|{\bf p}_i\|^2 = P_{ii}$, and the diagonal element of $P$ is simply $\rho({\bf r}_i)$, i.e. the charge density at the discretised grid point ${\bf r}_i$. Therefore, the algorithm will choose the first column ${\bf p}_{\Pi(1)}$ as the one corresponding to the point in space ${\bf r}_i$ with maximal charge density (i.e., the projection of a delta-like function centred on ${\bf r}_i$).

The second (and following) columns, that are projections of delta-like functions on other grid points, will then be chosen (as discussed before) so as to maximise the orthogonality of this projection with respect to the subspace defined by all previous ones.
For instance, for the second vector ${\bf p}_{\Pi(2)}$, the norm of its orthogonal component to ${\bf p}_{\Pi(1)}$ can be shown to be 
\begin{align}
    \|{\bf p}^\perp_{\Pi(2)}\|^2 &= \|{\bf p}_{\Pi(2)}\|^2 - \frac{|{\bf p}_{\Pi(1)}\cdot{\bf p}_{\Pi(2)}|^2}{\|{\bf p}_{\Pi(1)}\|^2} =\nonumber \\
    &= \rho({\bf r}_{\Pi(2)}) - \frac{|P_{\Pi(1)\Pi(2)}|^2}{\rho({\bf r}_{\Pi(1)})},
\end{align}
and therefore choosing $\Pi(2)$ to maximise it (at fixed chosen $\Pi(1)$) is equivalent to maximising
\begin{equation}
    \label{eq:SCDM-quantity-to-maximize}
    \rho({\bf r}_{\Pi(2)}) - \frac{|P_{\Pi(1)\Pi(2)}|^2}{\rho({\bf r}_{\Pi(1)})}.
\end{equation}

\subsection*{\label{appendix:equivalence-cholesky}Equivalence of the SCDM method with the Cholesky orbitals}
We want to show here that the algorithm to obtain the Cholesky orbitals of Aquilante \etal{} \cite{Aquilante_2006} provides the same selection of columns as the QRCP prescribed by the SCDM method.

As also explained in Ref.~[\onlinecite{Aquilante2011}], the following algorithm can be employed in order to obtain the $k$-th selected column $\Pi(k)$:
\begin{enumerate}
\item Define an initial matrix $P^{(0)}=P$ being the density matrix of the system.
\item At every step $k\geq 1$, choose $\Pi(k)$ as the index of the column where the matrix $P^{(k-1)}$ has maximum diagonal element. Also, we define the $k$-th Cholesky vector ${\bf c}_k$ as the $\Pi(k)$-th column of $P^{(k-1)}$, rescaled by the inverse square root of the corresponding diagonal element:
\begin{equation}
    \label{eq:cholesky-vector}
    ({\bf c}_k)_j = \frac 1 {\sqrt{P^{(k-1)}_{\Pi(k)\Pi(k)}}}[P^{(k-1)}]_{j\Pi(k)}.
\end{equation}
\item Define the matrix $P^{(k)}$ for the next iteration as follows:
\begin{equation}\label{eq:cholesky-iteration}
    P^{(k)} = P^{(k-1)} - {\bf c}_k \cdot {\bf c}_k^\dagger
\end{equation}
(where ${\bf c}_k \cdot {\bf c}_k^\dagger$ indicates a matrix product).
\item Iterate the previous two points until the needed number of selected columns is obtained.
\end{enumerate}

We can now show that this approach is equivalent to the selection of columns of the QRCP algorithm. In particular, in the first step, the Cholesky approach selects the column corresponding to the largest diagonal element of $P$, which is exactly the same choice as the QRCP algorithm, as discussed in~\nameref{appendix:QRCP-looks-for-most-orthogonal}.

At the second step ($k=1$), substituting Eq.~\eqref{eq:cholesky-vector} in Eq.~\eqref{eq:cholesky-iteration} and using $P^{(0)} = P$, we have
\begin{align}
    P^{(1)}_{ij} &= P_{ij} - [{\bf c}_{\Pi(1)} \cdot {\bf c}_{\Pi(1)}^\dagger]_{ij} = \nonumber \\
    &= P_{ij} - \frac{ P_{i\Pi(1)}P_{j\Pi(1)}^* }{P_{\Pi(1)\Pi(1)}} = P_{ij} - \frac{ P_{i\Pi(1)}P_{j\Pi(1)}^* }{\rho({\bf r}_{\Pi(1)})}.
\end{align}

In particular, we can notice now that the diagonal elements $P^{(1)}_{jj}$ of $P^{(1)}$ can be written as
\begin{align}
    P^{(1)}_{jj} = P_{jj} - \frac{ |P_{\Pi(1)j}|^2 }{\rho({\bf r}_{\Pi(1)})},
\end{align}
(where we have used $P^\dagger = P$) and therefore the choice of $j=\Pi(2)$ based on the largest diagonal element of $P^{(1)}$, as prescribed by the Cholesky algorithm, is equivalent to the QRCP choice maximising Eq.~\eqref{eq:SCDM-quantity-to-maximize}.

Finally, we note that the $\Pi(1)-$th column of $P^{(1)}$ is composed only by zeros (and analogously for the $\Pi(1)-$th row since the $P^{(i)}$ matrices are Hermitian), since
\begin{align}
    P^{(1)}_{i\Pi(1)} = P_{i\Pi(1)} - \frac{ P_{i\Pi(1)}P_{\Pi(1)\Pi(1)}^* }{P_{\Pi(1)\Pi(1)}} = P_{i\Pi(1)} - P_{i\Pi(1)} = 0
\end{align}
(where we have used the fact that the diagonal elements of $P$ are real). This fact, in addition to providing numerical stability to the Cholesky-orbital algorithm by forcing these elements to be numerically zero, allows us to ``remove'' the zero row and column from $P^{(1)}$ and repeat the reasoning by induction for all subsequent Cholesky vectors, working with smaller and smaller matrices.

Equivalently, one could understand more intuitively the result by noting that the Cholesky vectors of Eq.~\eqref{eq:cholesky-vector} are normalised to 1 because of Eq.~\eqref{eqn:scalar-product-P-psi}.
Therefore, Eq.~\eqref{eq:cholesky-iteration} constructs a new projection operator $P^{(k)}$ projecting on the subspace of the span of $P^{(k-1)}$ that is also orthogonal to ${\bf c}_k$, and then the Cholesky algorithm selects the largest vector in this subspace, that is exactly what the QRCP algorithm also does, as discussed in~\nameref{appendix:QRCP-looks-for-most-orthogonal}.

\section*{Supplementary Note 1}
\begin{figure}[h!]
     \centering
     \subfloat[]{\includegraphics[width=5.cm,trim={60pt 60pt 80pt 60pt},clip,rotate=90]{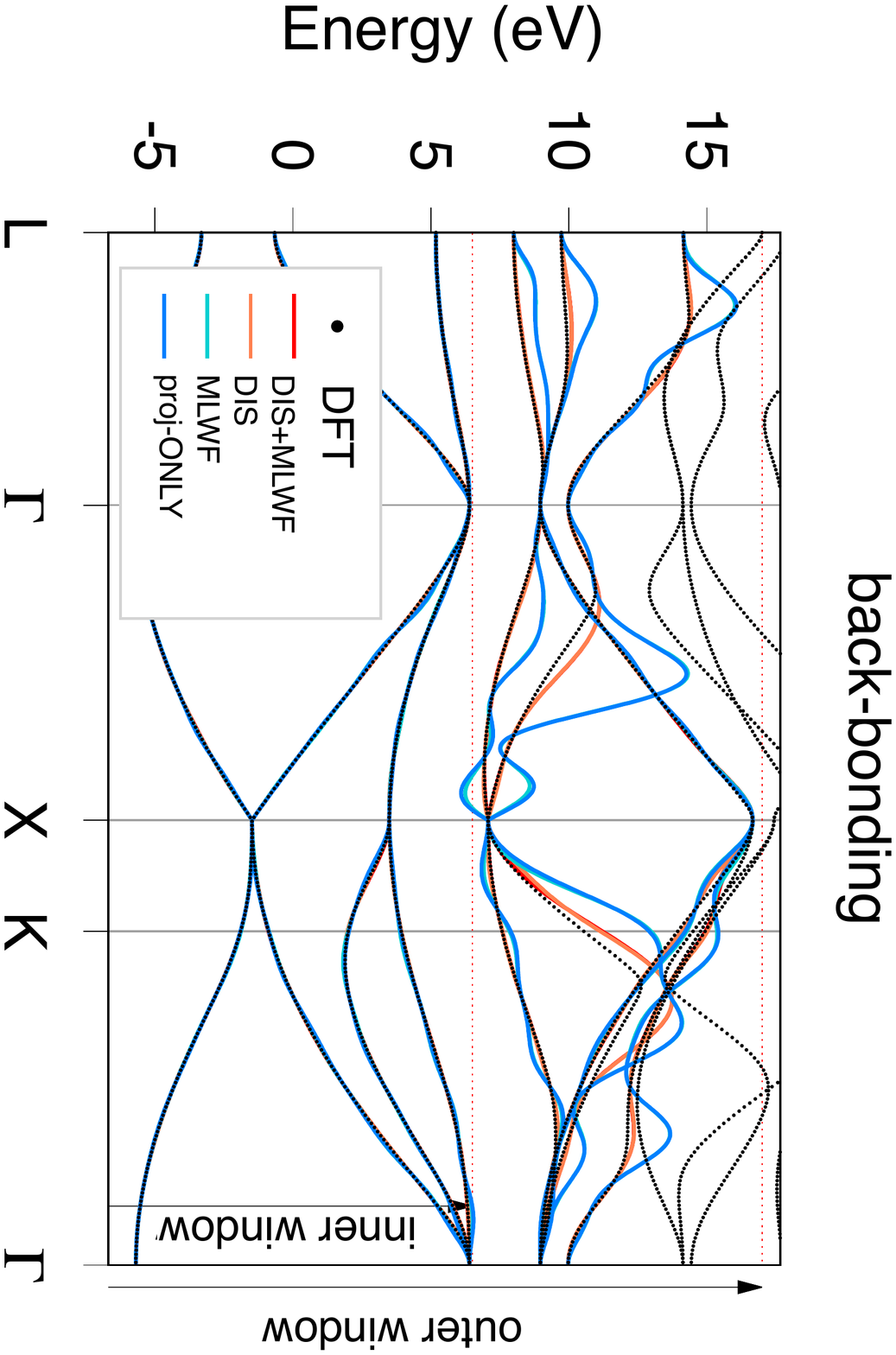}}
     \subfloat[]{\includegraphics[width=5.cm,trim={60pt 60pt 80pt 60pt},clip,rotate=90]{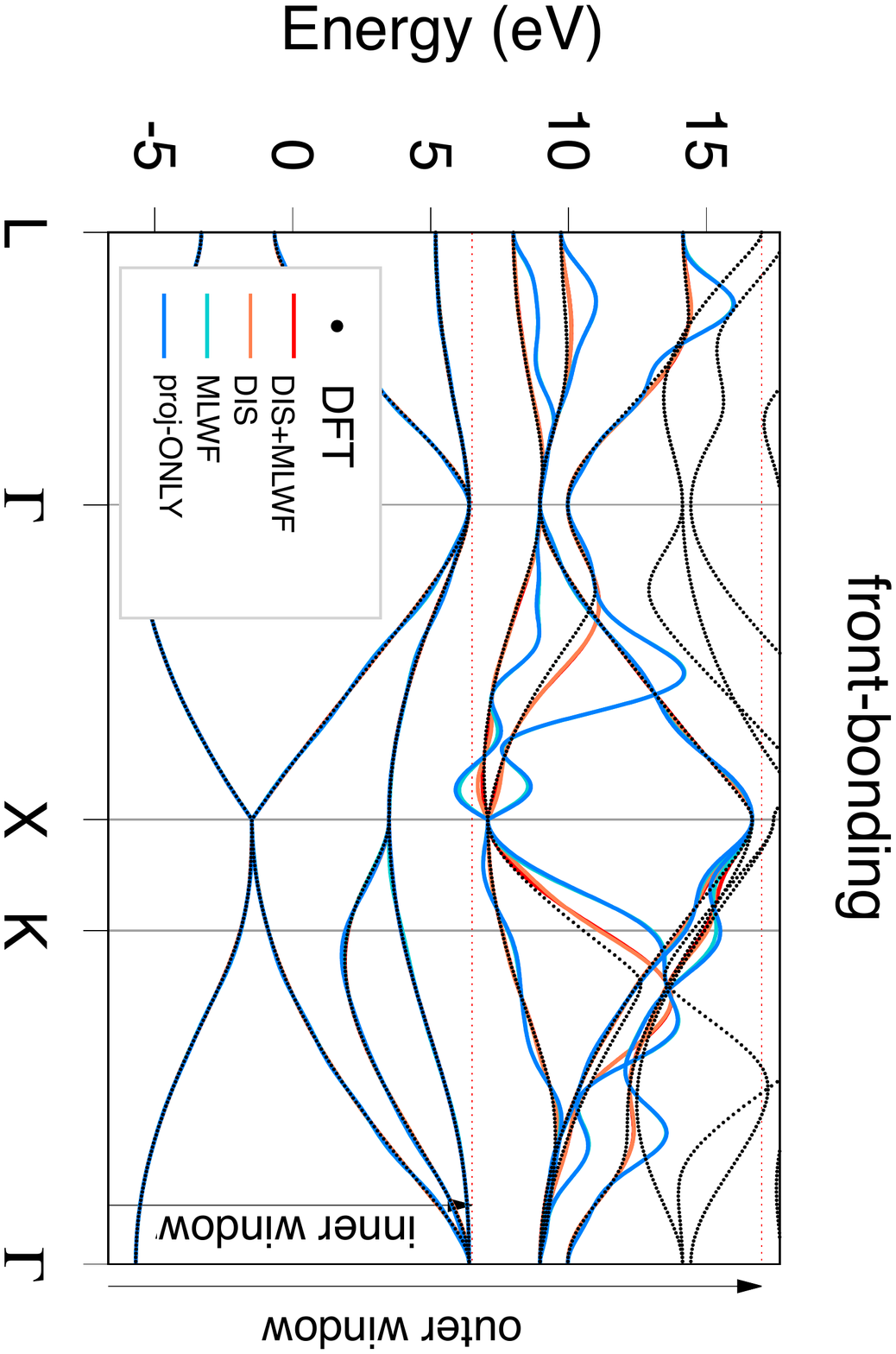}}\\
     \subfloat[]{\includegraphics[width=5.cm,trim={60pt 60pt 80pt 60pt},clip,rotate=90]{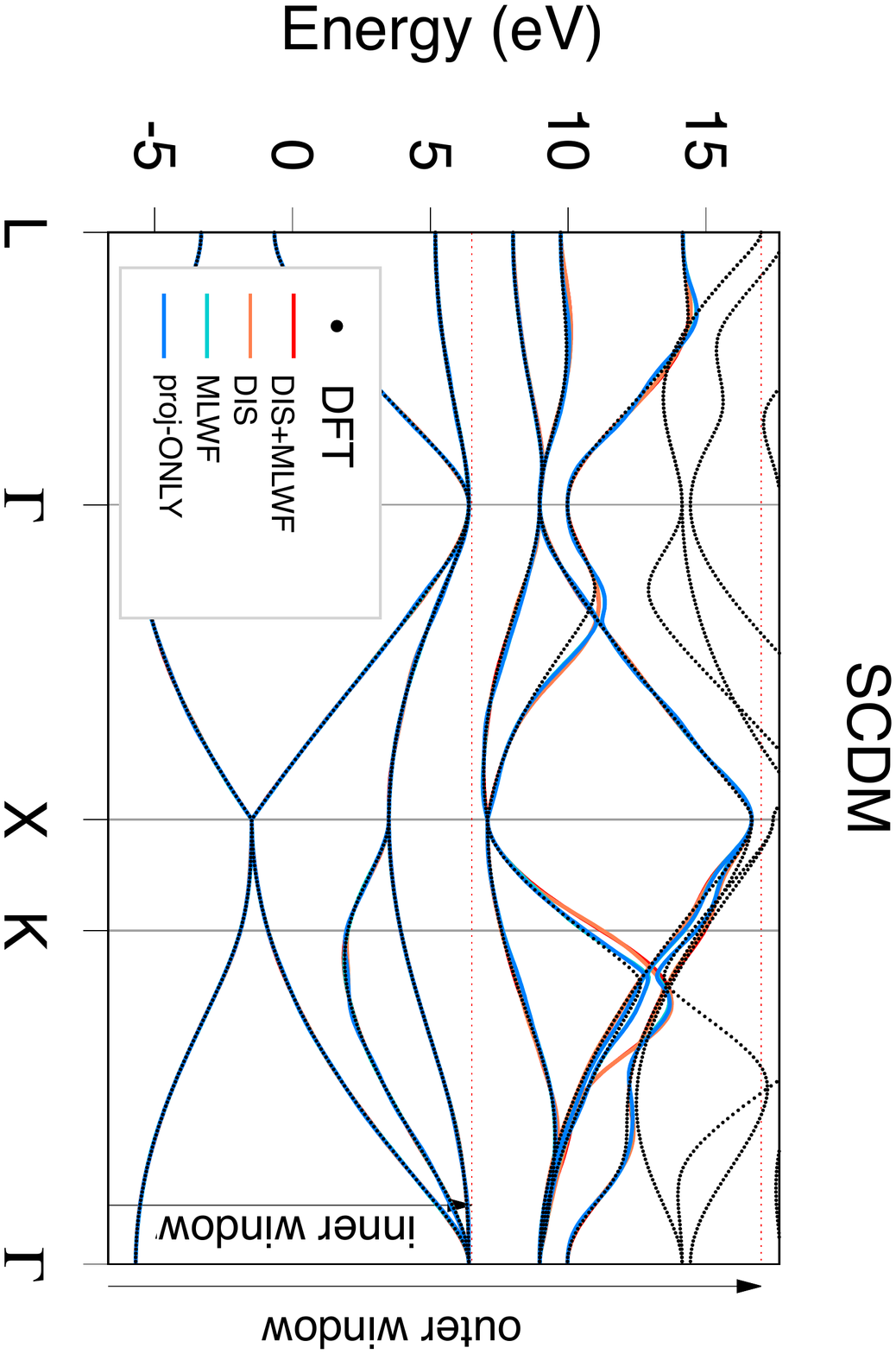}}
     \caption{Wannier interpolated valence bands and four low-lying conduction bands in silicon from three different set of initial projections: a) eight $sp^3$ in the {\it back-bonding} configuration; b) eight $sp^3$ in the {\it front-bonding} configuration and c) from SCDM with $\mu=10$~eV and $\sigma=2$~eV. For each plot the interpolation from four minimisation schemes are shown: 1) full minimisation of $\Omega$ ({\tt DIS+MLWF}), with $\varepsilon\tinysub{outer}=17$~eV and $\varepsilon\tinysub{inner}=6.5$~eV for the disentanglement step (solid red). 2) Minimisation of $\omi$ only "disentanglement" ({\tt DIS}) with $\varepsilon\tinysub{outer}=17$~eV and $\varepsilon\tinysub{inner}=6.5$~eV (solid coral). 3) Minimisation of $\omt$ only ({\tt MLWF}) in the projected subspace (solid turquoise). 4) No minimisation ({\tt proj-ONLY}) (solid blue). The DFT band-structure is also shown for reference (dotted black). It is worth clarifying that regardless of the initial projections, after a full minimisation---solid red line in all three panels---the Wannier interpolation is extremely good, particularly for the valence manifold. In fact, the three methods give almost indistinguishable results.}
     \label{fig3.3}
     \end{figure}

In~\ref{fig3.3} we report the Wannier-interpolated valence bands and four low-lying conduction bands in silicon for three different set of initial projections (two explicit sets of projections and one using the SCDM method).

    \clearpage

\section*{Supplementary Note 2}
In~\ref{table:z_valence_projections} we report, for every element appearing in at least one structure used in this work, the number of valence electrons and the atomic pseudo-orbitals included in the pseudopotential files used in the simulations discussed in this work.

    \begin{table*}[b!]
        \centering
    \begin{tabular}{lllp{2cm}lll}
        Symbol & $Z_{\text{val}}$ & Pseudo-orbitals & & Symbol & $Z_{\text{val}}$ & Pseudo-orbitals \\ \hline
        H & 1 & 1s                    &  &           Br & 7 & 1s, 1p \\
        He & 2 & 1s                   &  &           Kr & 8 & 1s, 1p \\
        Li & 3 & 1s, 1p, 2s           &  &           Rb & 9 & 1s, 1p, 2s \\
        Be & 4 & 1s, 1p, 2s           &  &           Sr & 10 & 1s, 1p, 1d, 2s, 2p \\
        B & 3 & 1s, 1p                &  &           Y & 11 & 1s, 1p, 1d, 2s, 2p \\
        C & 4 & 1s, 1p                &  &           Zr & 12 & 1s, 1p, 1d, 2s, 2p \\
        N & 5 & 1s, 1p                &  &           Nb & 13 & 1s, 1p, 1d, 2s \\
        O & 6 & 1s, 1p                &  &           Mo & 14 & 1s, 1p, 1d, 2s \\
        F & 7 & 1s, 1p                &  &           Ru & 16 & 1s, 1p, 1d, 2s \\
        Ne & 8 & 1s, 1p               &  &           Rh & 17 & 1s, 1p, 1d, 2s \\
        Na & 9 & 1s, 1p, 2s           &  &           Pd & 18 & 1s, 1p, 1d, 2s \\
        Mg & 2 & 1s, 1p               &  &           Ag & 19 & 1s, 1p, 1d, 2s \\
        Al & 3 & 1s, 1p               &  &           Cd & 12 & 1s, 1p, 1d \\
        Si & 4 & 1s, 1p               &  &           In & 13 & 1s, 1p, 1d \\
        P & 5 & 1s, 1p                &  &           Sn & 14 & 1s, 1p, 1d \\
        S & 6 & 1s, 1p                &  &           Sb & 15 & 1s, 1p, 1d \\
        Cl & 7 & 1s, 1p               &  &           Te & 6 & 1s, 1p
     \end{tabular}
    \caption{\label{table:z_valence_projections}List of number of valence electrons included in the pseudopotential ($Z_\text{val}$) and atomic pseudo-orbitals included in the pseudopotential file (1s refers to an atomic $s$ pseudo-orbital without radial nodes, $2p$ to an atomic $p$ pseudo-orbital with one radial node, \ldots)}
     \end{table*}
     
    \begin{table*}[t!]
        \centering
     \begin{tabular}{lllp{2cm}lll}
        Symbol & $Z_{\text{val}}$ & Pseudo-orbitals & & Symbol & $Z_{\text{val}}$ & Pseudo-orbitals \\ \hline
        Ar & 8 & 1s, 1p               &  &           I & 7 & 1s, 1p \\
        K & 9 & 1s, 1p, 2s, 2p        &  &           Xe & 18 & 1s, 1p, 1d \\
        Ca & 10 & 1s, 1p, 1d, 2s      &  &           Cs & 9 & 1s, 1p, 1d, 2s, 2p \\
        Sc & 11 & 1s, 1p, 1d, 2s      &  &           Ba & 10 & 1s, 1p, 2s \\
        Ti & 12 & 1s, 1p, 1d, 2s      &  &           Hf & 12 & 1s, 1p, 1d, 2s \\
        V & 13 & 1s, 1p, 1d, 2s       &  &           Ta & 13 & 1s, 1p, 1d, 2s, 2p \\
        Cr & 14 & 1s, 1p, 1d, 2s      &  &           W & 14 & 1s, 1p, 1d, 2s, 2p \\
        Mn & 15 & 1s, 1p, 1d, 2s, 2p  &  &           Re & 15 & 1s, 1p, 1d, 2s, 2p \\
        Fe & 16 & 1s, 1p, 1d, 2s, 2p  &  &           Os & 30 & 1s, 1p, 1d, 1f, 2s, 2p \\
        Co & 17 & 1s, 1p, 1d, 2s, 2p  &  &           Ir & 15 & 1s, 1p, 1d, 2p \\
        Ni & 18 & 1s, 1p, 1d, 2s, 2p  &  &           Pt & 16 & 1s, 1p, 1d, 2p \\
        Cu & 19 & 1s, 1p, 1d, 2s, 2p  &  &           Au & 19 & 1s, 1p, 1d, 2s \\
        Zn & 20 & 1s, 1p, 1d, 2s, 2p  &  &           Hg & 20 & 1s, 1p, 1d, 2s \\
        Ga & 13 & 1s, 1p, 1d          &  &           Tl & 13 & 1s, 1p, 1d \\
        Ge & 14 & 1s, 1p, 1d          &  &           Pb & 14 & 1s, 1p, 1d \\
        As & 5 & 1s, 1p               &  &           Bi & 15 & 1s, 1p, 1d \\
        Se & 6 & 1s, 1p 
    \end{tabular}
    \caption*{Table \ref{table:z_valence_projections} continued}
    \end{table*}

    \clearpage

\bibliographystyle{naturemag}
\bibliography{supp_biblio_SCDM}